\documentclass[galaxies,article,accept,moreauthors,pdftex, biblatex, 10pt,a4paper]{mdpi}

\firstpage{1}
\articlenumber{x}
\doinum{10.3390/------}
\pubvolume{xx}
\pubyear{2017}
\copyrightyear{2017}
\history{Received: 30 October 2017; Accepted: 5 December 2017; Published: date}
 \theoremstyle{mdpi}
 \newcounter{thm}
 \newcounter{ex}
 \newcounter{re}
\usepackage{booktabs}

\setitemize{parsep=6pt,itemsep=0pt,leftmargin=*,labelsep=5.5mm}
\setenumerate{parsep=6pt,itemsep=0pt,leftmargin=*,labelsep=5.5mm}
\setlist[description]{itemsep=0mm}

\usepackage{subfigure}
\makeatletter
\renewcommand{\@thesubfigure}{\normalsize(\textbf{\alph{subfigure}})}
\makeatother

\Title{Multi-Band Intra-Night Optical Variability of~BL~Lacertae}

\Author{Haritma Gaur $^{1,}$*, Alok C.\ Gupta $^{2}$, R.\ Bachev $^{3}$, A.\ Strigachev $^{3}$, E.\ Semkov $^{3}$, Paul J.\ Wiita $^{4,5}$, \linebreak M. F. Gu $^{1}$ and S. Ibryamov $^{6}$ }
\AuthorNames{Firstname Lastname, Firstname Lastname and Firstname Lastname}

\address{%
$^{1}$ \quad Key Laboratory for Research in Galaxies and Cosmology, Shanghai Astronomical Observatory,
Chinese~Academy of Sciences, 80 Nandan Road, Shanghai 200030, China; harry.gaur31@gmail.com (H.G); gumf@shao.ac.cn (M.F.G) \\ 
$^{2}$ \quad Aryabhatta Research Institute of Observational Sciences (ARIES), Manora Peak, Nainital 263002, India; acgupta30@gmail.com\\
$^{3}$ \quad Institute of Astronomy and National Astronomical Observatory, Bulgarian Academy of Sciences,
72~Tsarigradsko Shosse Blvd., 1784 Sofia, Bulgaria; blazonstone@gmail.com (R.B); \linebreak anton.strigachev@gmail.com (A.S); esemkov@astro.bas.bg (E.S)\\
$^{4}$ \quad Department of Physics, The College of New Jersey, P.O.\ Box 7718, Ewing, NJ 08628-0718, USA; wiitap@tcnj.edu\\
$^{5}$ \quad Kavli Institute for Particle Astrophysics and Cosmology, SLAC, 2575 Sand Hill Rd, Menlo Park, \linebreak CA 94025, USA \\
$^{6}$ \quad Department of Theoretical and Applied Physics, Faculty of Natural Sciences, University of Shumen, 115, Universitetska Str., 9712 Shumen, Bulgaria; sibryamov@shu.bg} 

\corres{Correspondence: harry.gaur31@gmail.com}

\abstract{We monitored BL Lacertae frequently during 2014--2016 when it was generally in a  high state. We searched for intra-day variability for 43 nights using quasi-simultaneous measurements in the B, V, R, and I bands (totaling 143
light curves); the typical sampling interval was about eight minutes. On hour-like timescales, BL Lac exhibited significant
variations during 13 nights in various optical bands. Significant spectral variations are seen during most of these nights
such that the optical spectrum becomes bluer when brighter. The amplitude of variability is usually greater for longer
observations but is lower when BL Lac is brighter. No evidence for periodicities
or characteristic variability time-scales in the light curves was found.
The color variations are mildly chromatic on long timescales.}


\begin{document}

\section{Introduction}

BL Lacertae is a prototype of the blazar class at redshift $z=0.069$ ~\cite{Miller1977}.  BL Lacs, along with many flat spectrum
radio quasars, constitute the blazar class of active galaxies. BL Lacs are differentiated by the absence or extreme
weakness of  emission lines (usually taken to be an equivalent width in the rest frame of the host galaxy of $<$5 \AA). They exhibit
strong flux and spectral variability across all of the electromagnetic spectrum  over a wide range of time-scales,
and significant and variable radio and optical polarization is common to all blazars (e.g., \cite{Wagner1995}).
However, on some occasions, BL Lacertae has shown
broad $H{\alpha}$ and $H{\beta}$ emission lines in its spectrum, thereby raising the issue of it actually being part of
its eponymous class~\cite{Vermeulen1995}.
BL Lac is classified as a low frequency peaked blazar (LBL) because of its first spectral hump  peaks in the
near IR/optical region, which is explained as the synchrotron emission from highly relativistic electrons within a helical jet
~\cite{Raiteri2009}. Optical observations are particularly helpful in studying the acceleration and cooling timescales of
the electrons in the relativistic jets.

BL Lacertae has shown strong variability in optical bands and hence is one of the favorite targets of multi-wavelength
campaigns carried out by the Whole Earth Blazar Telescope
(WEBT/GASP;~\cite{Raiteri2009,Bottcher2003,Villata2009,Raiteri2010,Raiteri2013}~and references therein). BL Lac displays intense optical variability on short and intra-day time-scales (e.g.,~\cite{Massaro1998,Tosti1999,Clements2001,Hagen2004,Agarwal2015})
as well as substantial variations in its polarization  (~\cite{Marscher2008,Gaur2014} and references therein). Previous studies have stressed the flux variations and spectral changes
(~\cite{Villata2002,Papadakis2003,Hu2006}).

In this work, we present observations of BL Lac, made at several telescopes in Bulgaria and Greece, during
the period 2014--2016 over a total of 43 nights with sufficient measurements so as to be able to properly search for intra-night
optical variability.  A key motivation for this study is to understand the nature of flux and spectral variability of BL Lac on the
shortest intra-day timescales. This study is a follow up of our earlier extensive optical monitoring of BL Lacertae covering
the period 2010--2012 (~\cite{Gaur2015a,Gaur2015b}). These additional observations of this particular blazar  help us to better constrain
its duty cycle for hourly variability in optical bands and let us study the dependences of the amplitude of variability on the
overall flux levels.  In Section \ref{sec2}, we briefly describe our observations and data reductions. Section \ref{sec3} summarizes the methods we
used to quantify variability. We present our results in Section \ref{sec4}. Sections \ref{sec5} and \ref{sec6}, respectively, give a discussion and our conclusions.

\section{Observations and Data Reduction}\label{sec2}

Our new observations of BL Lacertae began on 29 July 2014 and concluded on 30 September 2016 continuing those
taken between 10 June 2010 and 26 October 2012 that we reported in~\cite{Gaur2015a}.
 The~complete log of observations is given in Table~\ref{table1}. Our observations were made using
two telescopes in Bulgaria and  Greece.
Details about the  telescopes in Bulgaria and Greece are given in (~\cite{Gaur2012}, Table~\ref{table1}). The standard data
reduction methods we employed on all observations are described in detail in Section 3 of~\cite{Gaur2012}. Typical
seeing varied between 1--3 arcsec during these measurements.

To summarize,  standard data reduction was performed using \texttt{IRAF}\footnote{IRAF is distributed by the National

Optical Astronomy Observatories, which are operated by the Association of Universities for Research in Astronomy, Inc., Cranford, NJ, USA,
under cooperative agreement with the
National Science Foundation.}, including bias subtraction and flat-field division. Instrumental magnitudes of BL Lacertae
 as well as four comparison stars in its field~\cite{Villata1998} were produced
using the \texttt{IRAF} package \texttt{DAOPHOT}\footnote{Dominion Astrophysical
Observatory Photometry software}; normally, an aperture radius of
{ $8^{"}$ was~used.}

In these measurements of BL Lacertae's light curve (LC), we observe comparison stars B, C, and~H in the same field
and take their magnitudes from~\cite{Villata1998}. Star C is used for calibration because both its magnitude and
color were most similar to those of BL Lac during this period.
BL Lac is within a relatively bright host galaxy and we removed its contribution from the
observed magnitudes by following the method provided in~\cite{Gaur2015a}.

\begin{table}[H]
\centering
\caption {Observation log of BL Lacertae.}
\label{table1}
\noindent
\setlength{\tabcolsep}{0.020in}
\begin{tabular}{cccc} \toprule
{\bf Date of Observation} & {\bf Telescope} & {\bf Band} & {\bf Number of}  \\
{\bf dd.mm.yyyy}     &      &          & {\bf Image Frames}        \\\midrule
29.07.2014    &A      &B, V, R, I   &10, 10, 10, 10   \\
28.09.2014    &B      &B, V, R, I   &13,  17, 19, 23   \\
29.09.2014    &B      &B, V, R, I   &12,  16,  17,  19   \\
30.09.2014    &B      &B, V, R, I   &19,  20,  22,  23   \\
29.06.2015    &A      &B, V, R, I   &10,  10,  10,  10   \\
30.06.2015    &A      &B, V, R, I   &10,  10,  10,  10   \\
18.07.2015    &B      &B, R, I   &33,  33,  32   \\
19.07.2015    &B      &B, R, I   &26,  27,  27   \\
20.07.2015    &B      &B, R, I   &22,  24,  24   \\\bottomrule
\end{tabular}
\end{table}

\begin{table}[H]\ContinuedFloat
\centering
\caption {{\em Cont.}}
\label{table1}
\noindent
\setlength{\tabcolsep}{0.020in}
\begin{tabular}{cccc} \toprule
{\bf Date of Observation} & {\bf Telescope} & {\bf Band} & {\bf Number of}  \\
{\bf dd.mm.yyyy}     &      &          & {\bf Image Frames}        \\\midrule
21.07.2015    &B      &B, R, I   &20, 20, 20   \\
22.07.2015    &B      &B, R, I   &23, 24, 24   \\
23.07.2015    &B      &B, R, I   &18, 22, 20   \\
18.08.2015    &B      &B, R, I   &28, 28, 28   \\
26.08.2015    &B      &B, R, I   &17, 20, 17   \\
27.08.2015    &B      &B, R, I   &19, 21, 20   \\
28.08.2015    &B      &B, R, I   &22, 23, 21   \\
08.09.2015    &B      &B, R, I     &45, 45, 45   \\
14.09.2015    &B      &B, R, I     &23, 27, 29   \\
15.09.2015    &B      &B, R, I   &28, 28, 30   \\
03.11.2015    &B      &B, R, I     &21, 21, 22   \\
04.11.2015    &B      &B, R, I     &26, 28, 28   \\
05.11.2015    &B      &B, R, I     &30, 30, 30   \\
06.11.2015    &B      &B, R, I   &31, 33, 32   \\
07.11.2015    &B      &B, R, I     &28, 32, 32   \\
11.11.2015    &B      &B, R, I     &46, 46, 45   \\
12.11.2015    &B      &B, R, I     &43, 42, 41   \\
03.12.2015    &B      &B, R, I     &15, 16, 18   \\
14.12.2015    &B      &B, R, I     &29, 30, 30   \\
30.06.2016    &B      &V, R, I     &20, 20, 20   \\
01.07.2016    &B      &V, R, I     &14, 14, 13   \\
08.08.2016    &B      &B, V, R     &15, 15, 30   \\
09.08.2016    &B      &B, V, R     &25, 25, 51   \\
25.08.2016    &B      &B, V, R     &13, 14, 28   \\
26.08.2016    &B      &B, V, R     &25, 26, 59   \\
28.08.2016    &B      &B, V, R, I   &14, 14, 12, 13   \\
30.08.2016    &B      &B, V, R, I   &45, 54, 53, 53   \\
23.09.2016    &B      &B, V, R, I   &44, 44, 42, 23   \\
24.09.2016    &B     &B, V, R, I   &31, 30, 30, 22   \\
26.09.2016    &B      &B, V, R, I   &45, 45, 43, 42   \\
27.09.2016    &B      &B, V, R, I   &42, 42, 41, 37   \\
28.09.2016    &B      &B, V, R, I   &50, 46, 47, 49   \\
29.09.2016    &B      &B, V, R, I   &49, 49, 50, 42   \\
30.09.2016    &B      &B, V, R, I   &38, 35, 34, 36   \\ \bottomrule
\end{tabular} \\
\begin{tabular}{@{}c@{}}
\multicolumn{1}{p{\textwidth -.88in}}{\footnotesize
A: 1.3-m Skinakas Observatory, Crete, Greece;
B: 60-cm Cassegrain Telescope at Astronomical Observatory Belogradchik,~Bulgaria.}
\end{tabular}
\end{table}

\section{Variability Detection Criterion}\label{sec3}

The $F$-test is considered to be a sensible way to search for the presence of the intra-night variability and has frequently been used
to do so (i.e., \cite{Diego2010}). However, recently,~Reference \cite{Diego2014} pointed out a problem with this apparently robust method, showing
that either substantial brightness differences or significant variability in a comparison star can produce underestimation of the
source variability with respect to the dimmer or less variable star.

To avoid these problems, we employed the power-enhanced $F$-test (~\cite{Diego2014,Diego2015}). The basic idea is that
one first increases the number of degrees of freedom in the denominator of the $F$-distribution by stacking all the LCs of the
standard stars.  Next, the comparison star differential LCs are transformed so as to have the same photometric noise as the blazar.
 The key point is that, by using multiple comparison stars, the probability of false claims of intra-night variability are
reduced compared to when only a~single standard star is used. See~\cite{Diego2015} for a detailed discussion.

The smaller the $\alpha$ value, the more improbable it is that the result is produced by chance.
If $F$ exceeds the critical value, the no variability (null) hypothesis should be discarded. We use a
 stringent level of $\alpha = 0.001$ and have performed the power enhanced $F$-test on all those LCs
where at least ten points are available.

\section{Results}\label{sec4}

\subsection{Intra-Day Variability}

\begin{figure}[H]
\centering*
\vspace*{-0.6in}
\includegraphics[scale=0.18,trim=5 1 5 2, clip=true]{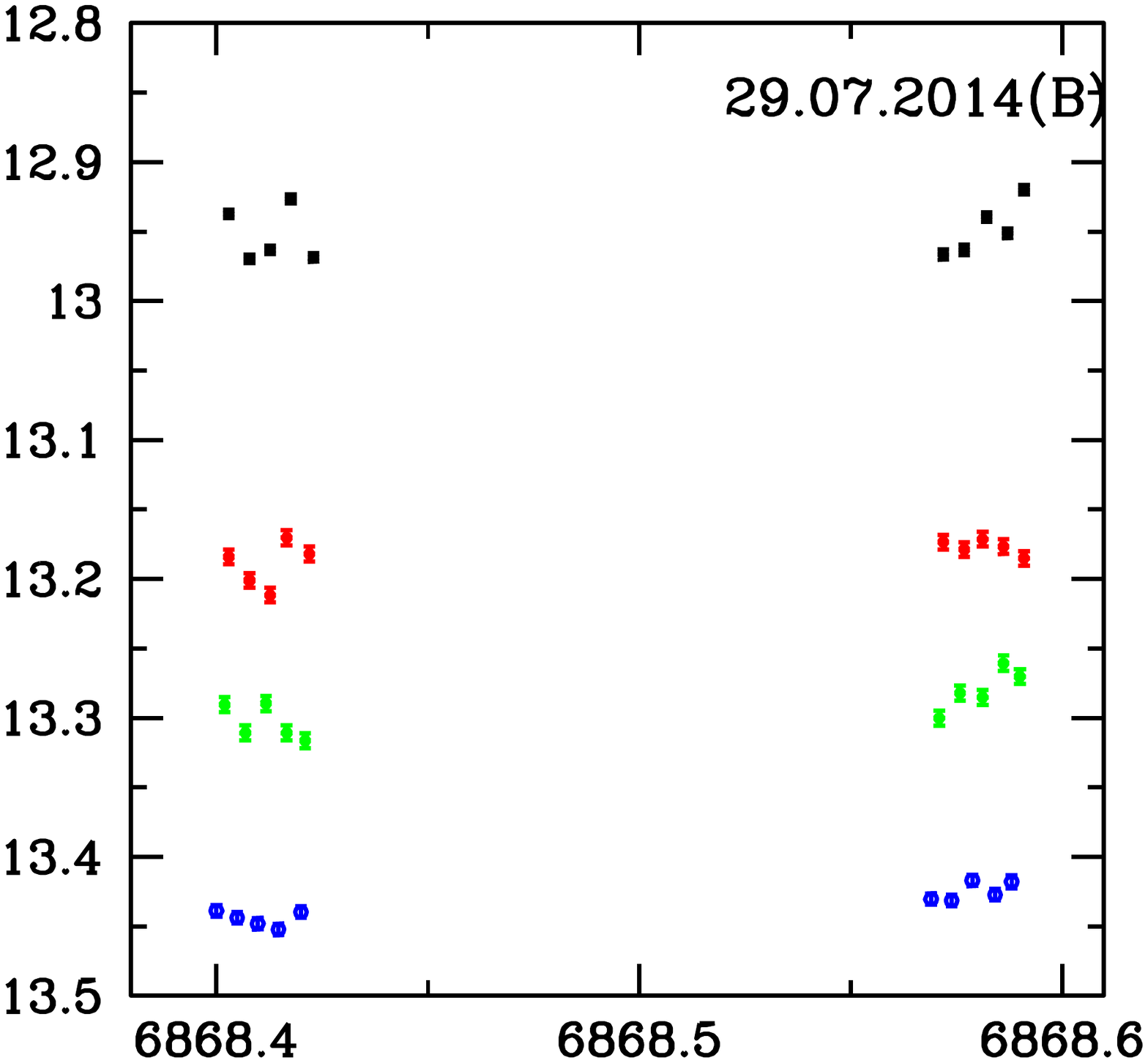}
\includegraphics[scale=0.18,trim=5 1 5 2, clip=true]{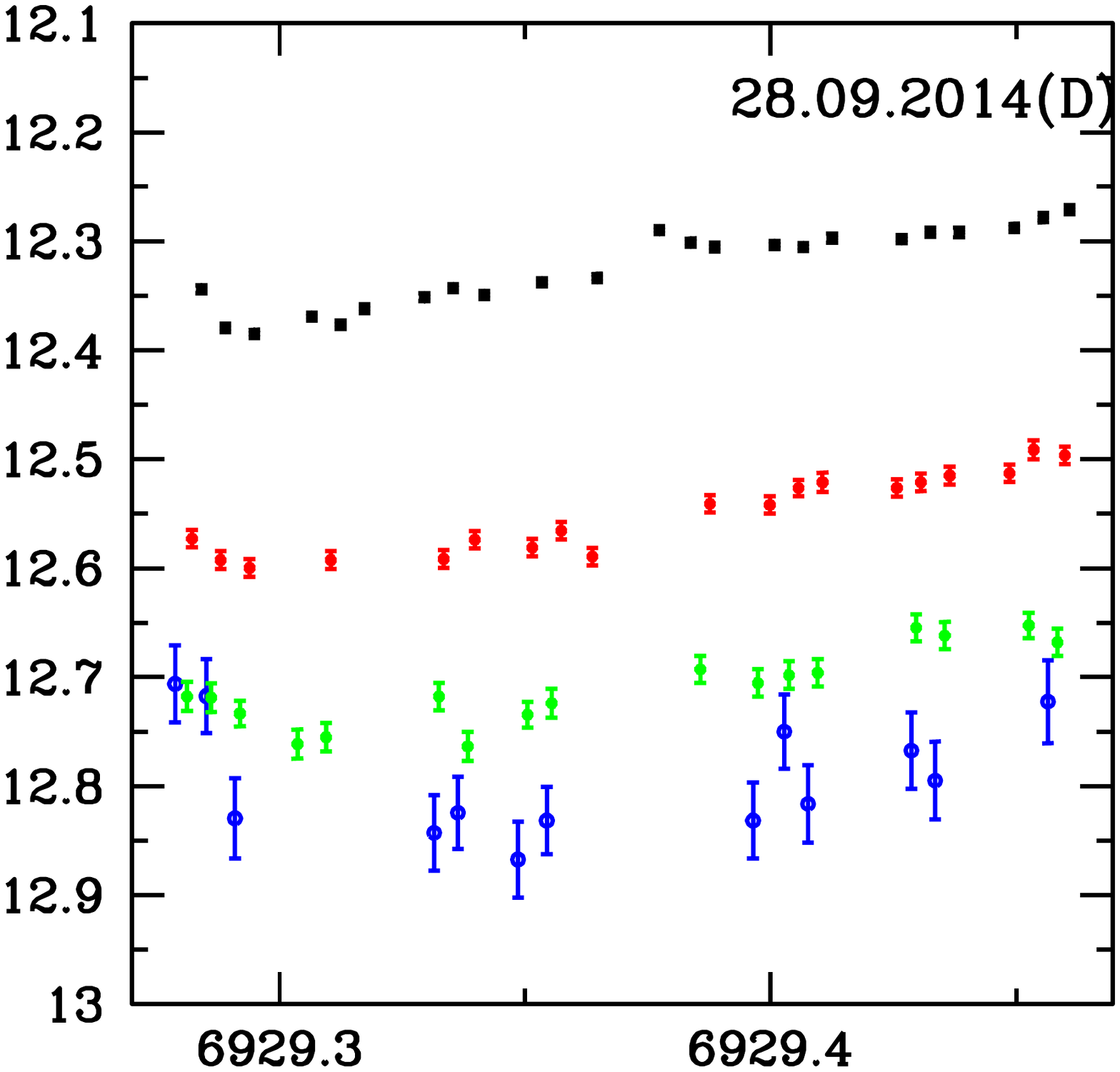}
\includegraphics[scale=0.18,trim=5 1 5 2, clip=true]{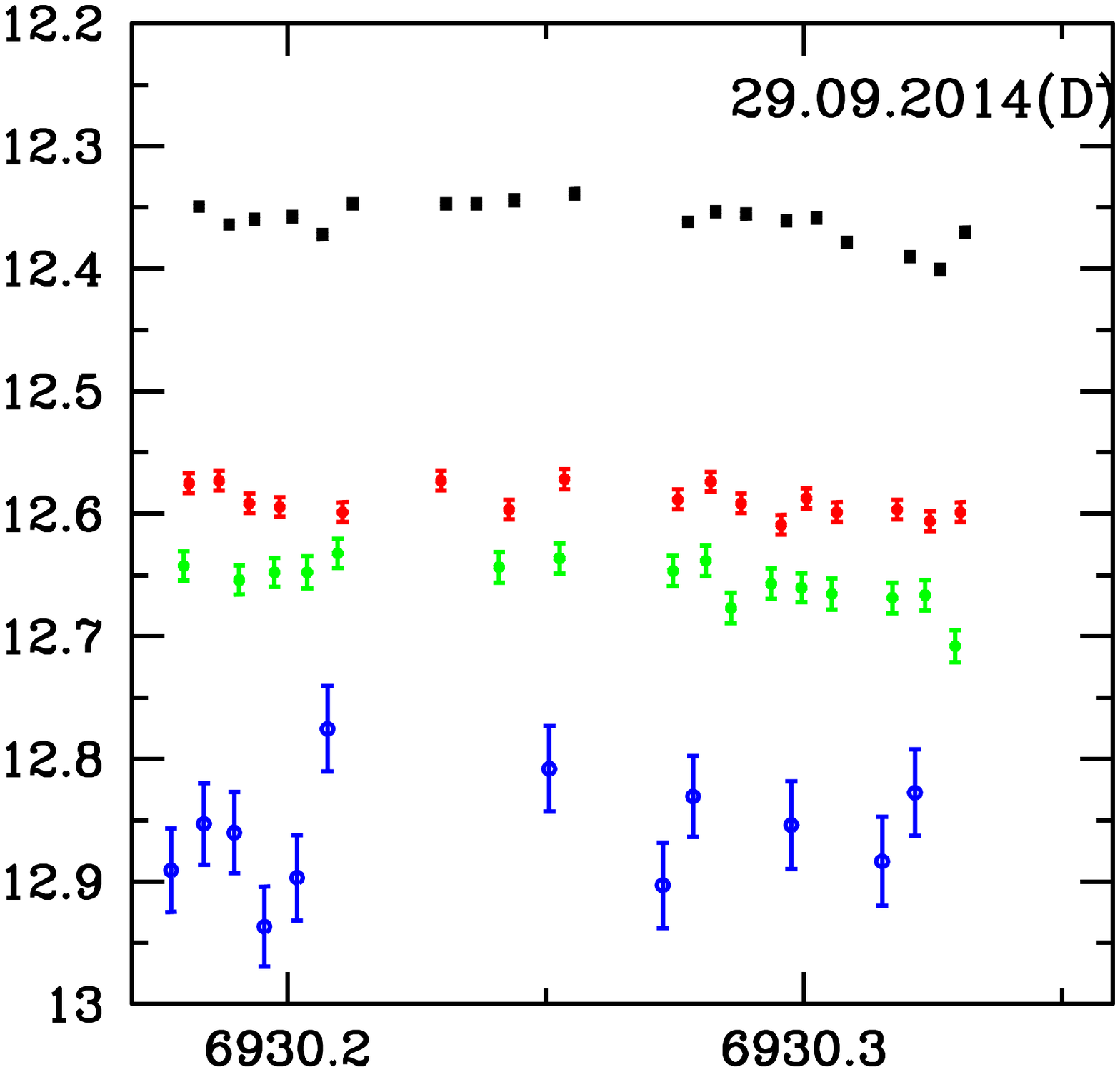}
\includegraphics[scale=0.18,trim=5 1 5 2, clip=true]{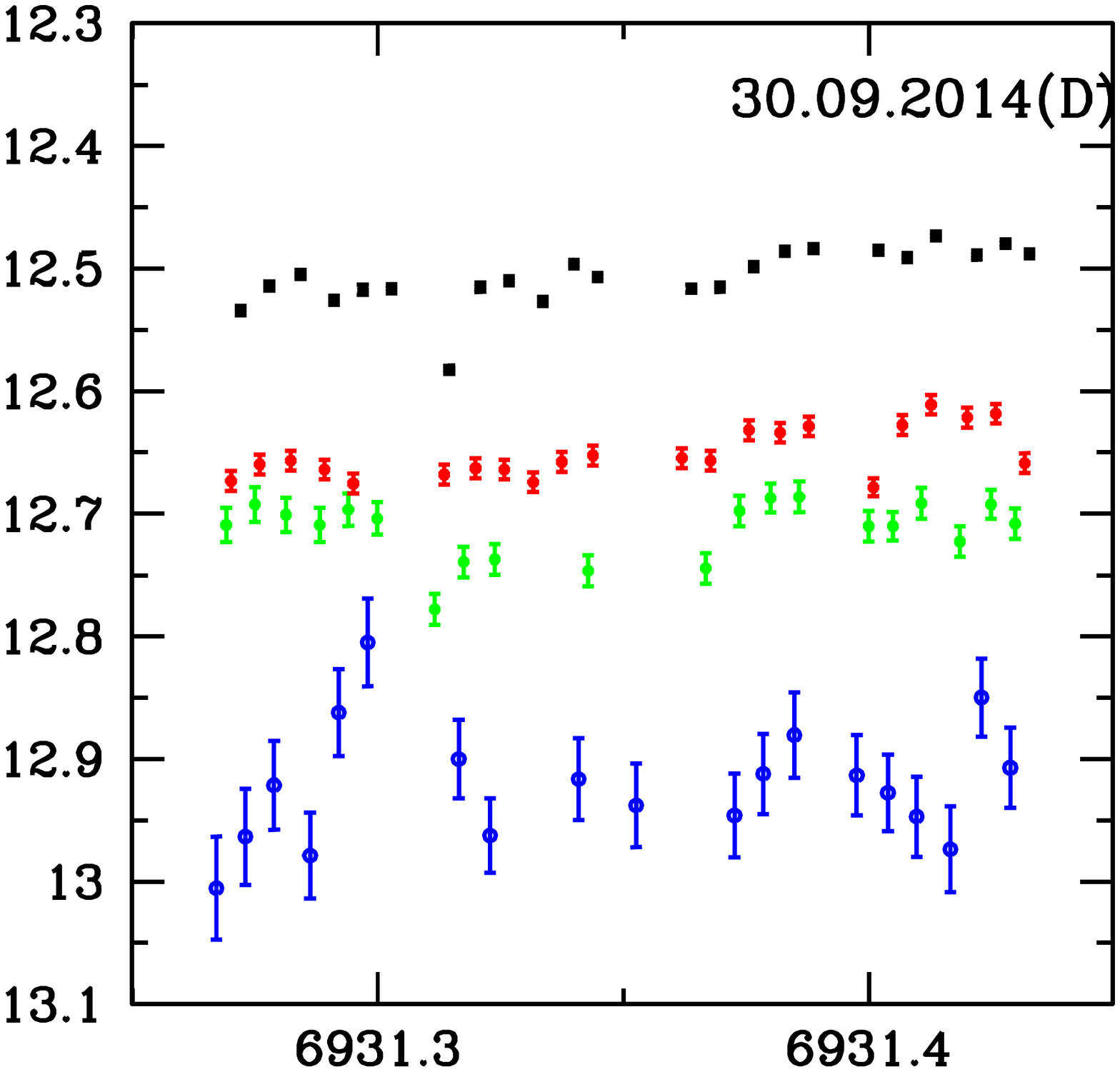}
\vspace*{-0.6in}
\includegraphics[scale=0.18,trim=5 1 5 2, clip=true]{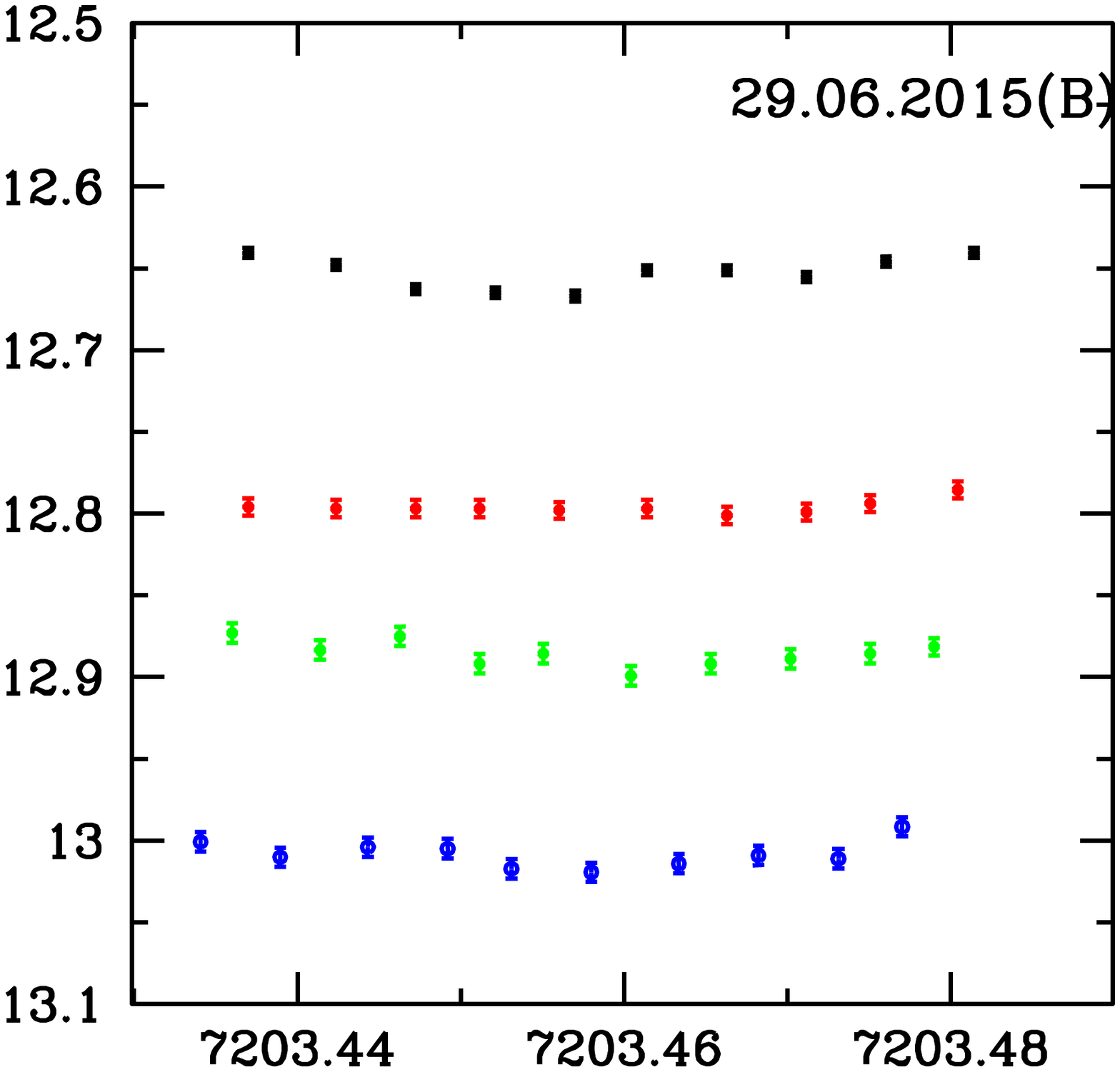}
\includegraphics[scale=0.18,trim=5 1 5 2, clip=true]{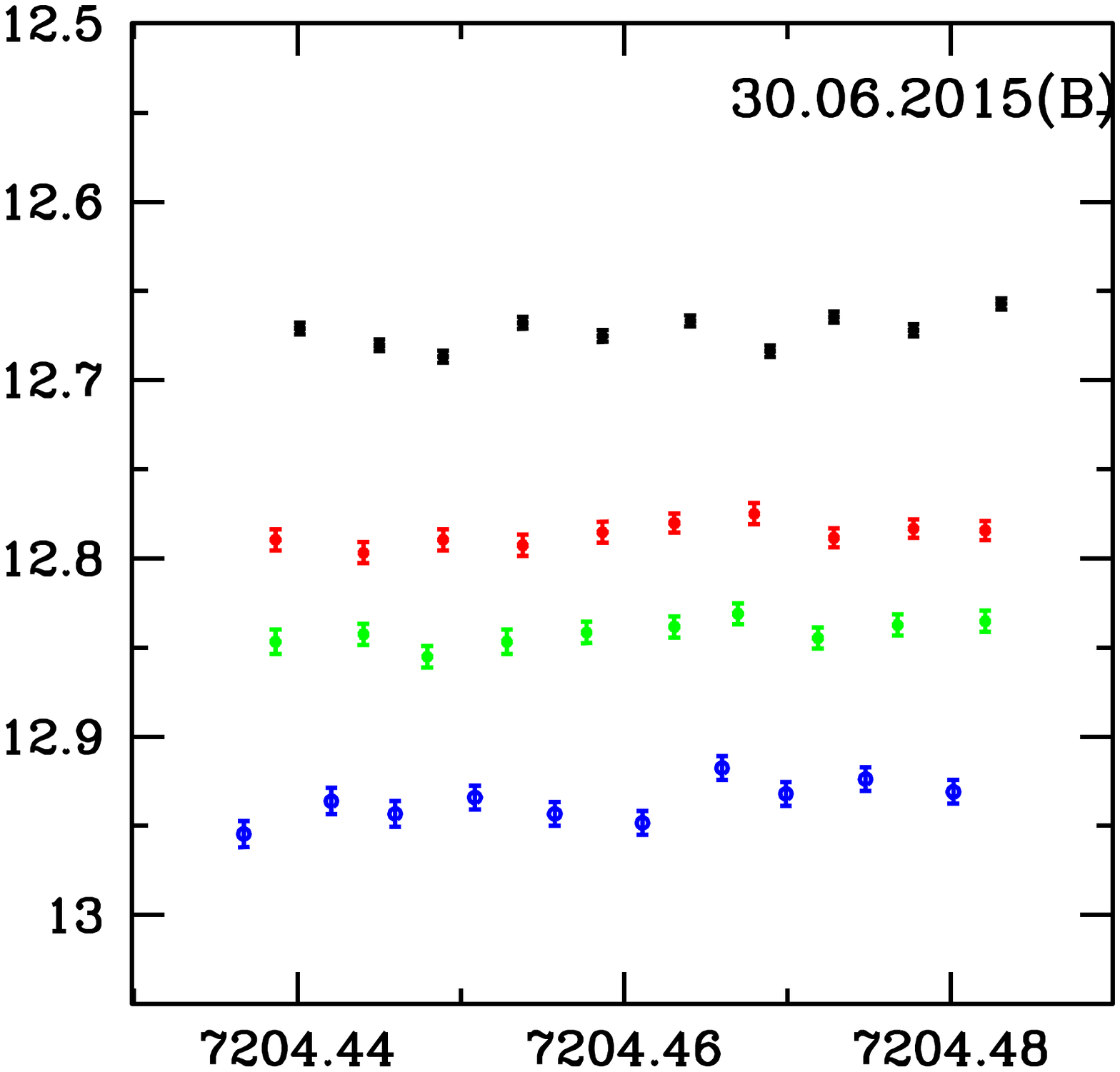}
\includegraphics[scale=0.18,trim=5 1 5 2, clip=true]{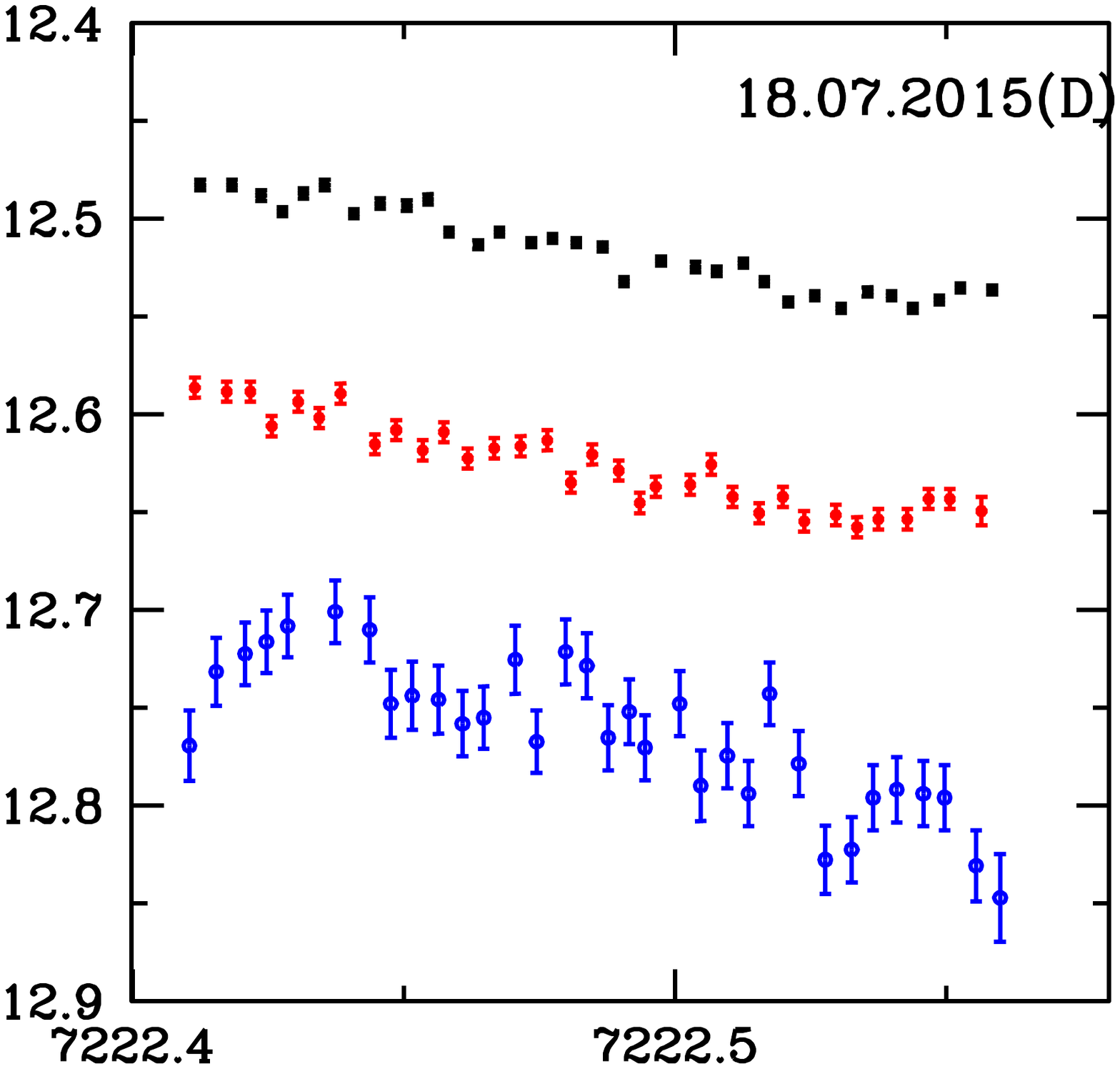}
\includegraphics[scale=0.18,trim=5 1 5 2, clip=true]{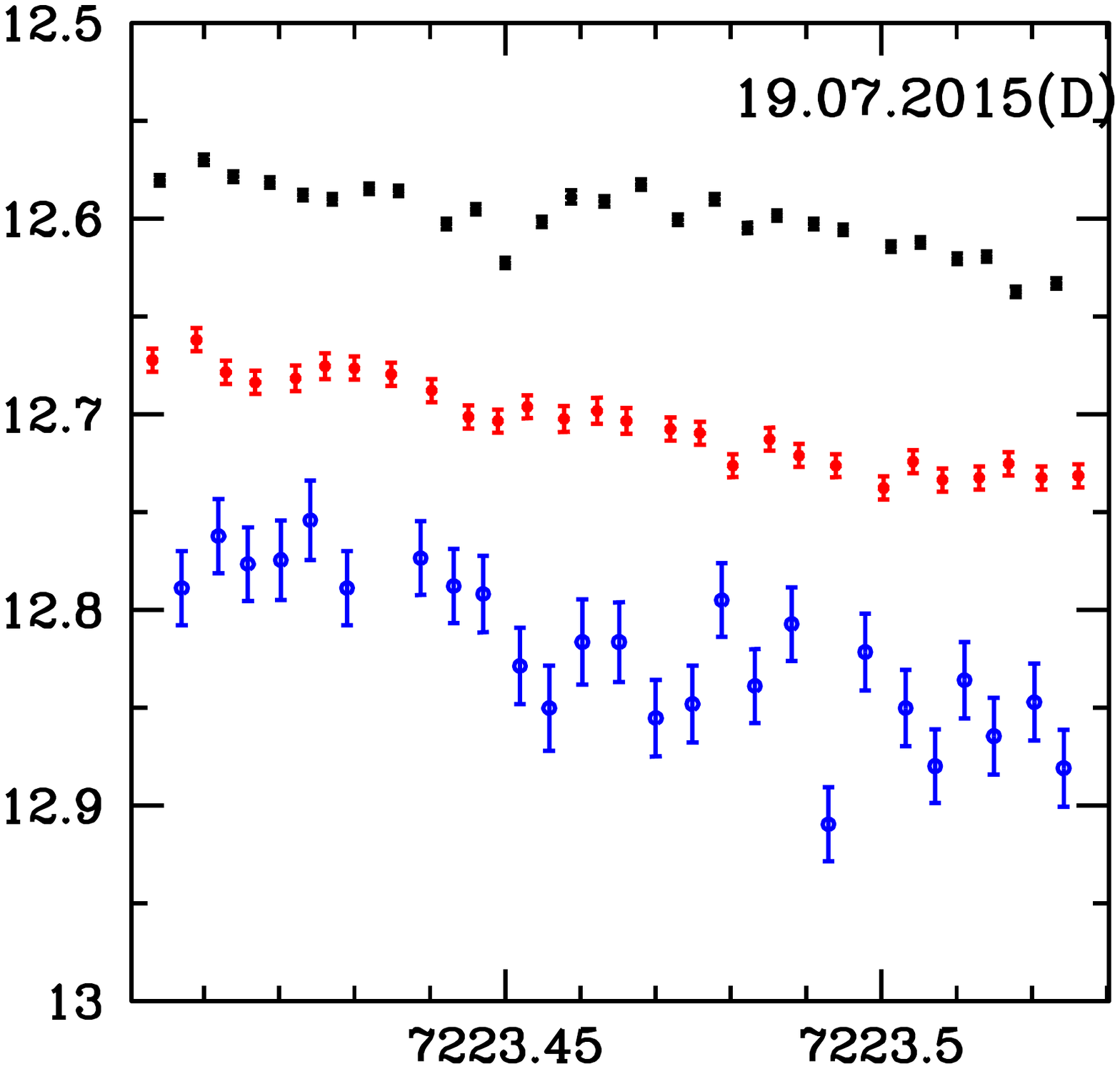}
\vspace*{-0.6in}
\includegraphics[scale=0.18,trim=5 1 5 2, clip=true]{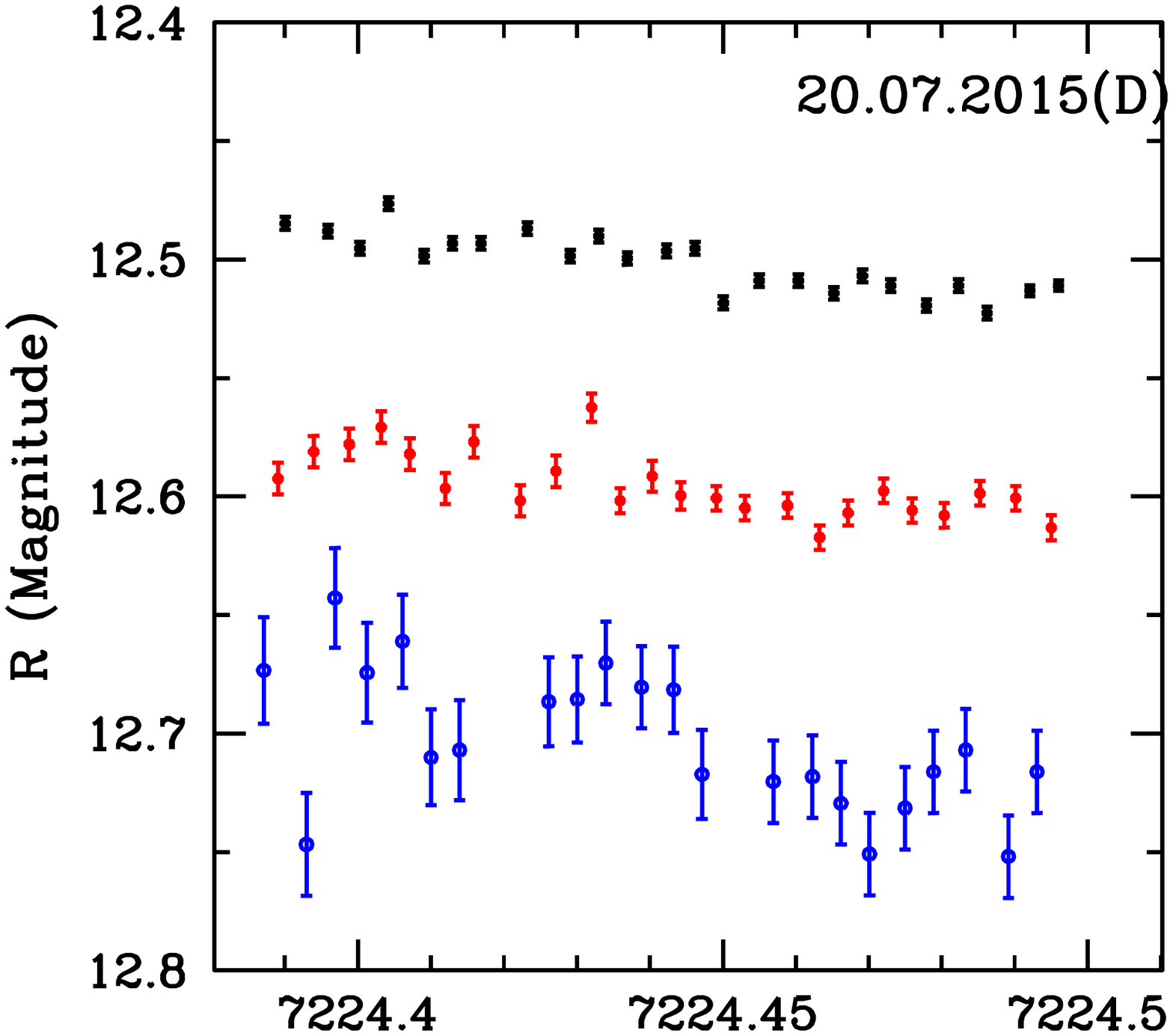}
\includegraphics[scale=0.18,trim=5 1 5 2, clip=true]{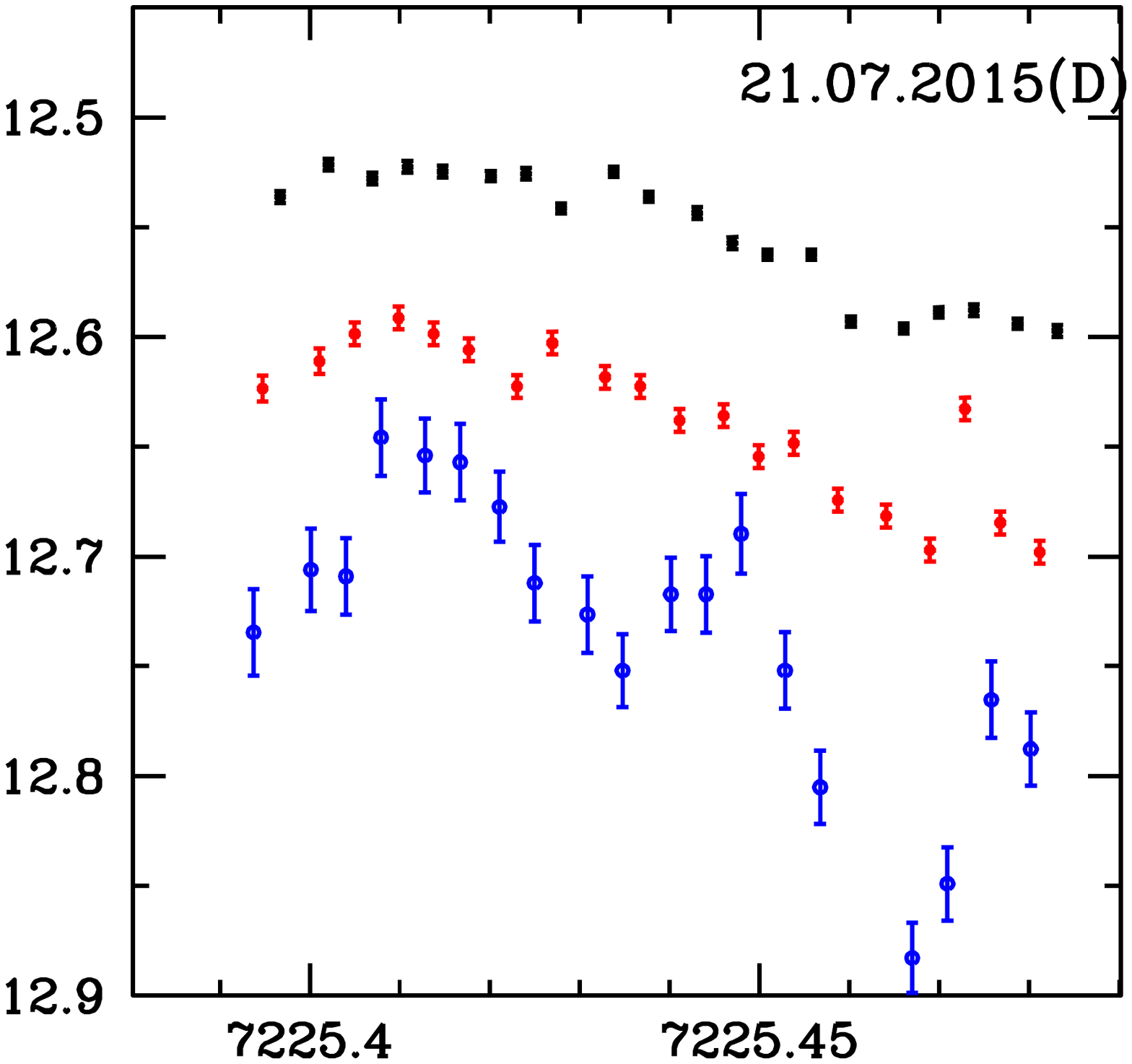}
\includegraphics[scale=0.18,trim=5 1 5 2, clip=true]{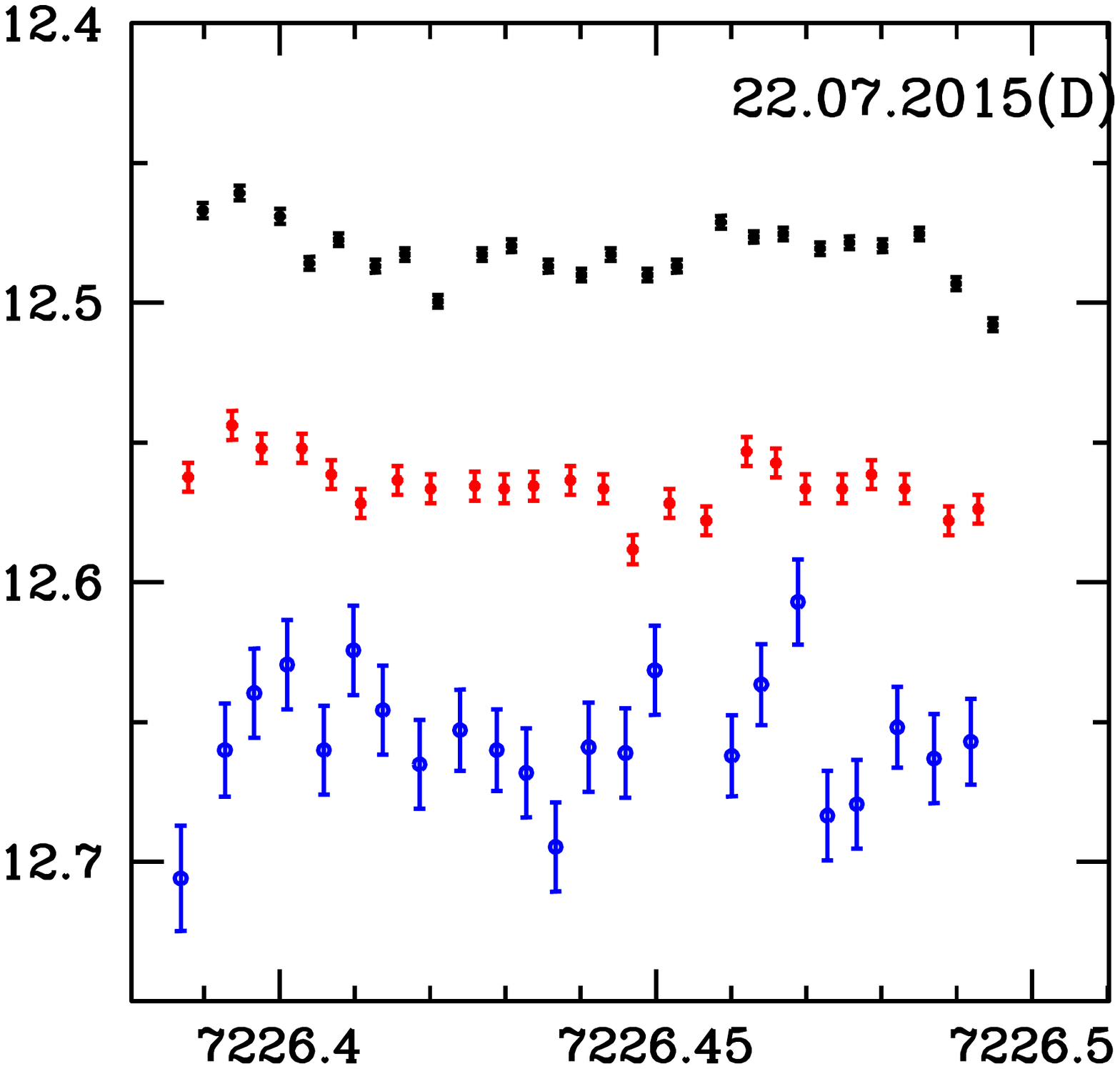}
\includegraphics[scale=0.18,trim=5 1 5 2, clip=true]{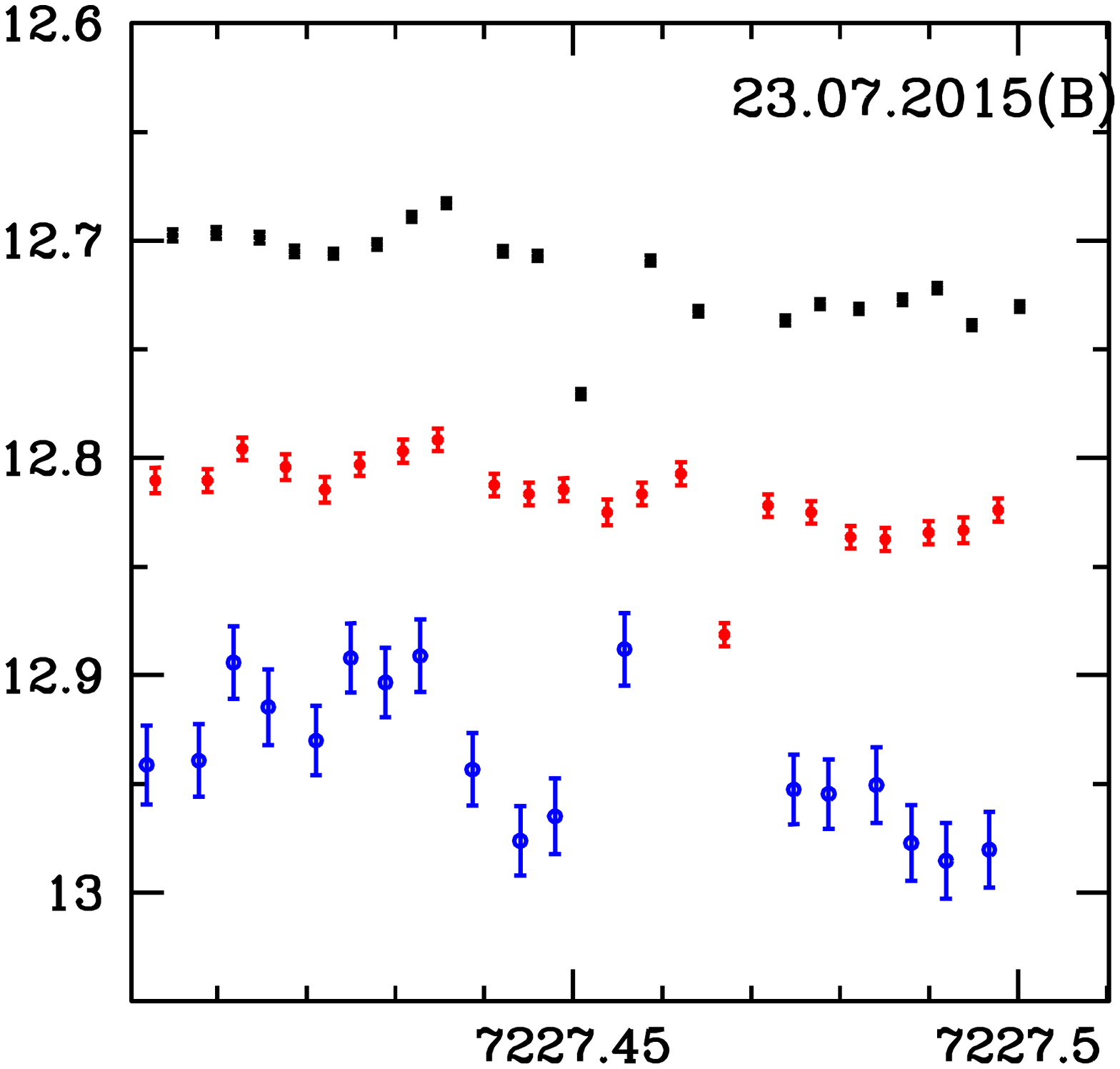}
\vspace*{-0.6in}
\includegraphics[scale=0.18,trim=5 1 5 2, clip=true]{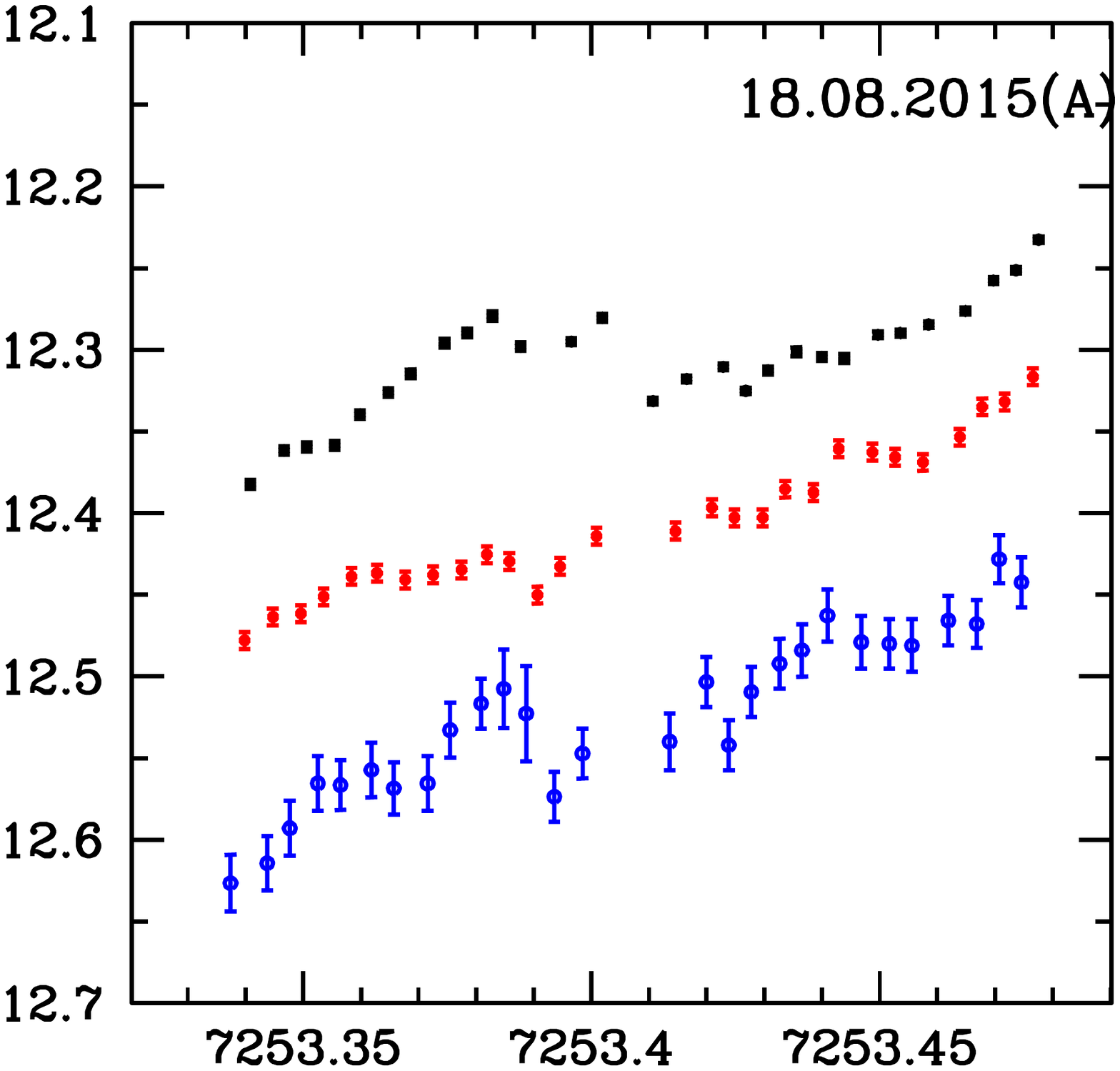}
\includegraphics[scale=0.18,trim=5 1 5 2, clip=true]{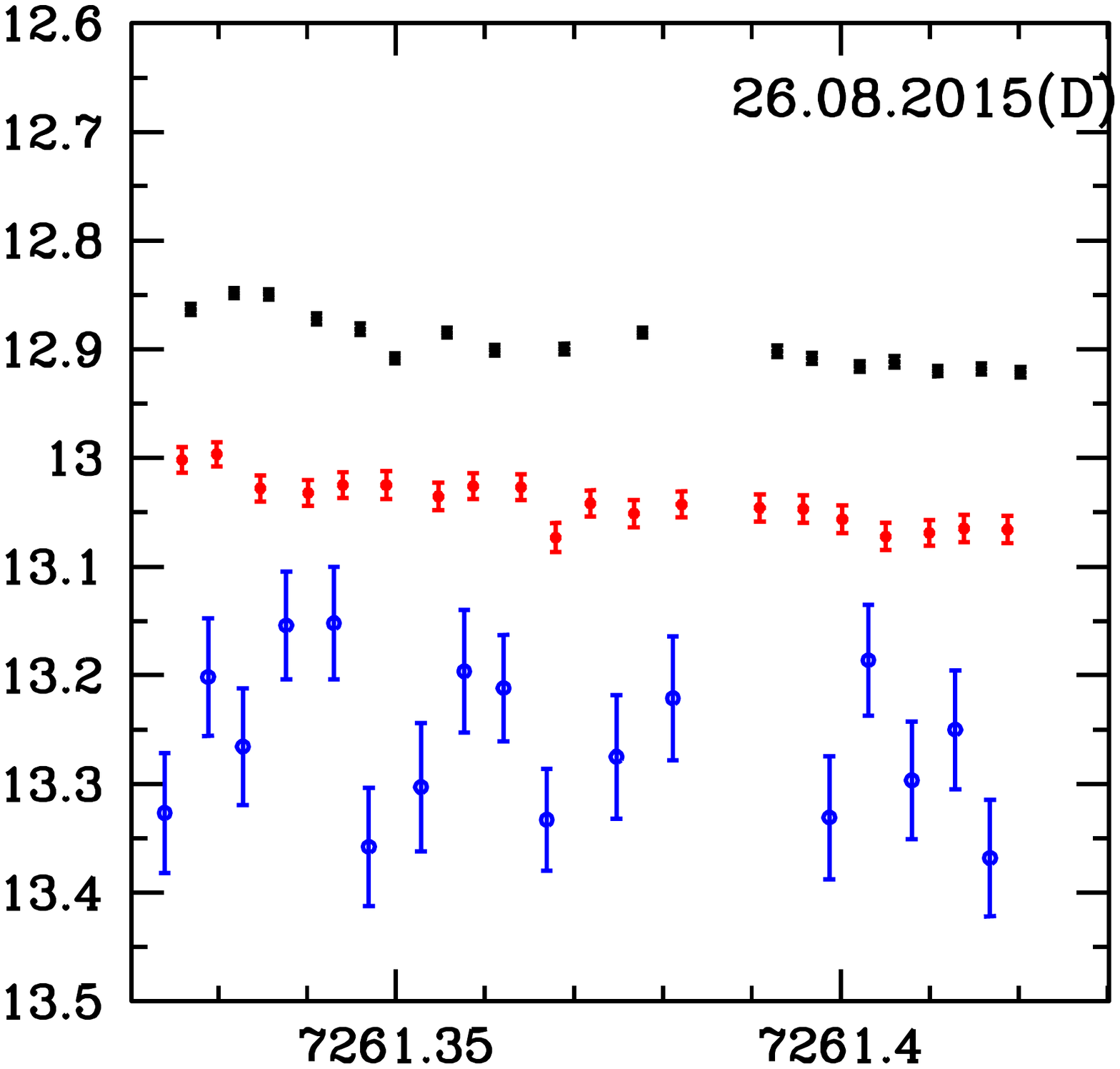}
\includegraphics[scale=0.18,trim=5 1 5 2, clip=true]{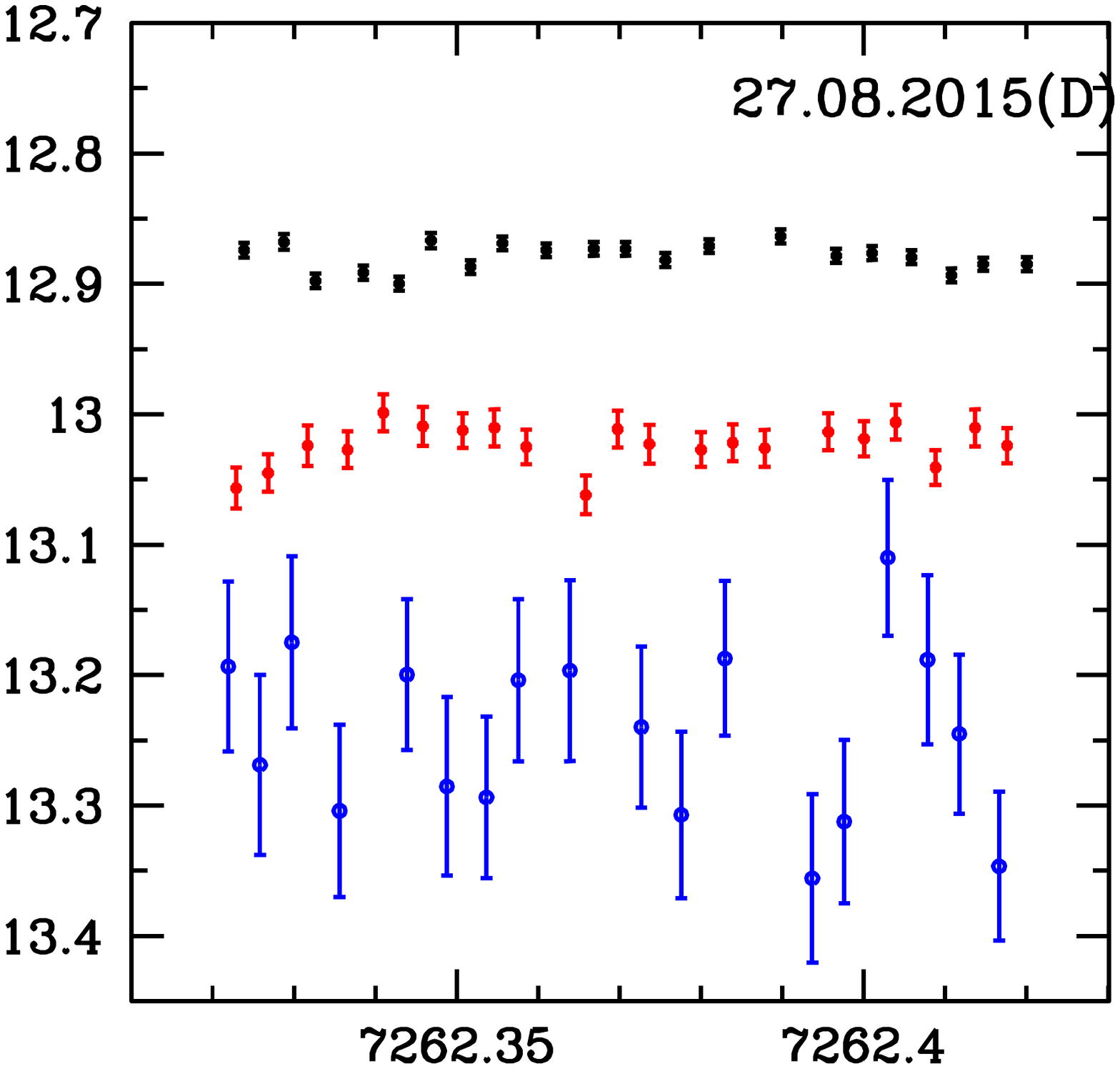}
\includegraphics[scale=0.18,trim=5 1 5 2, clip=true]{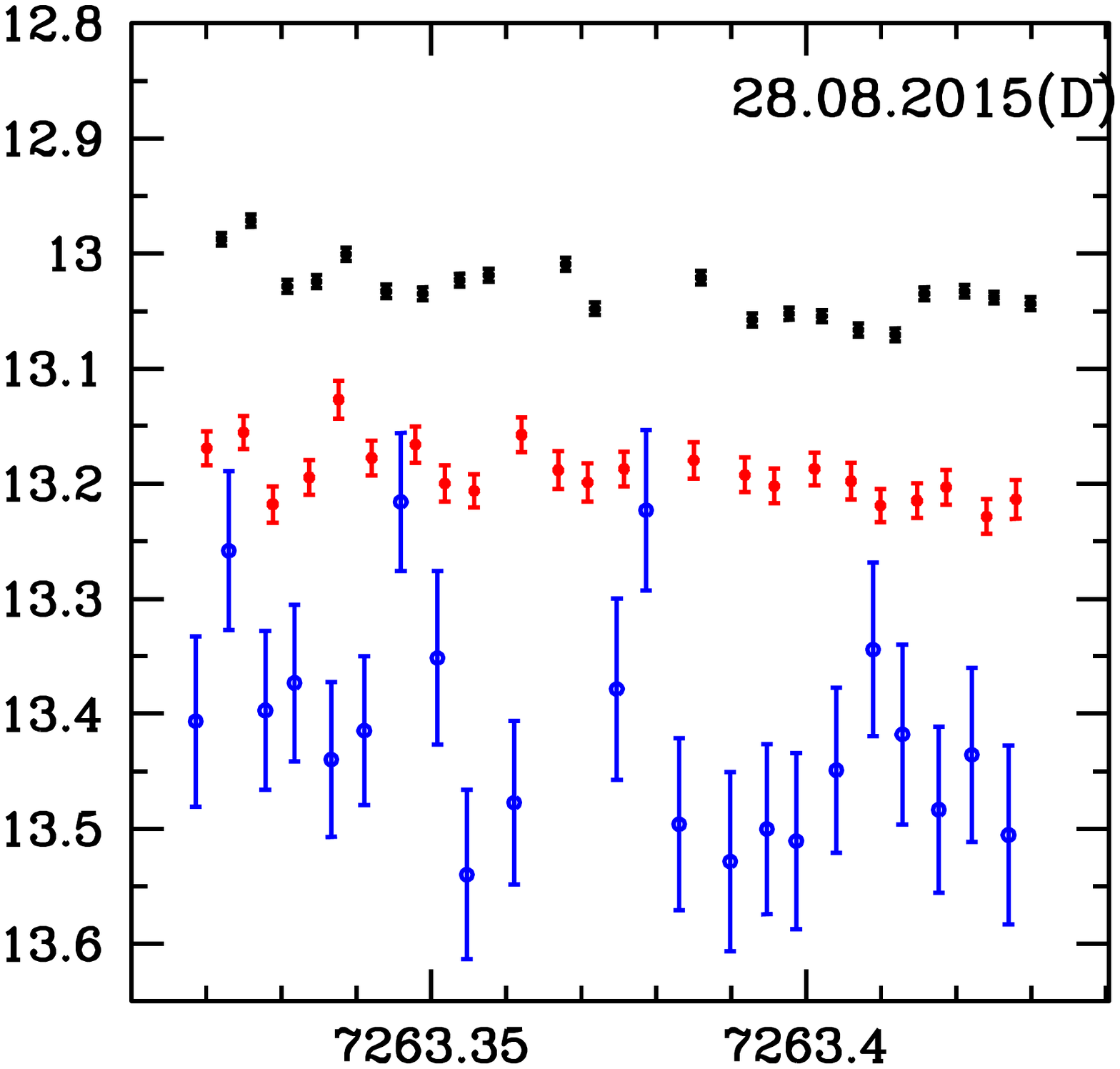}
\vspace*{-0.6in}
\includegraphics[scale=0.18,trim=5 5 5 2, clip=true]{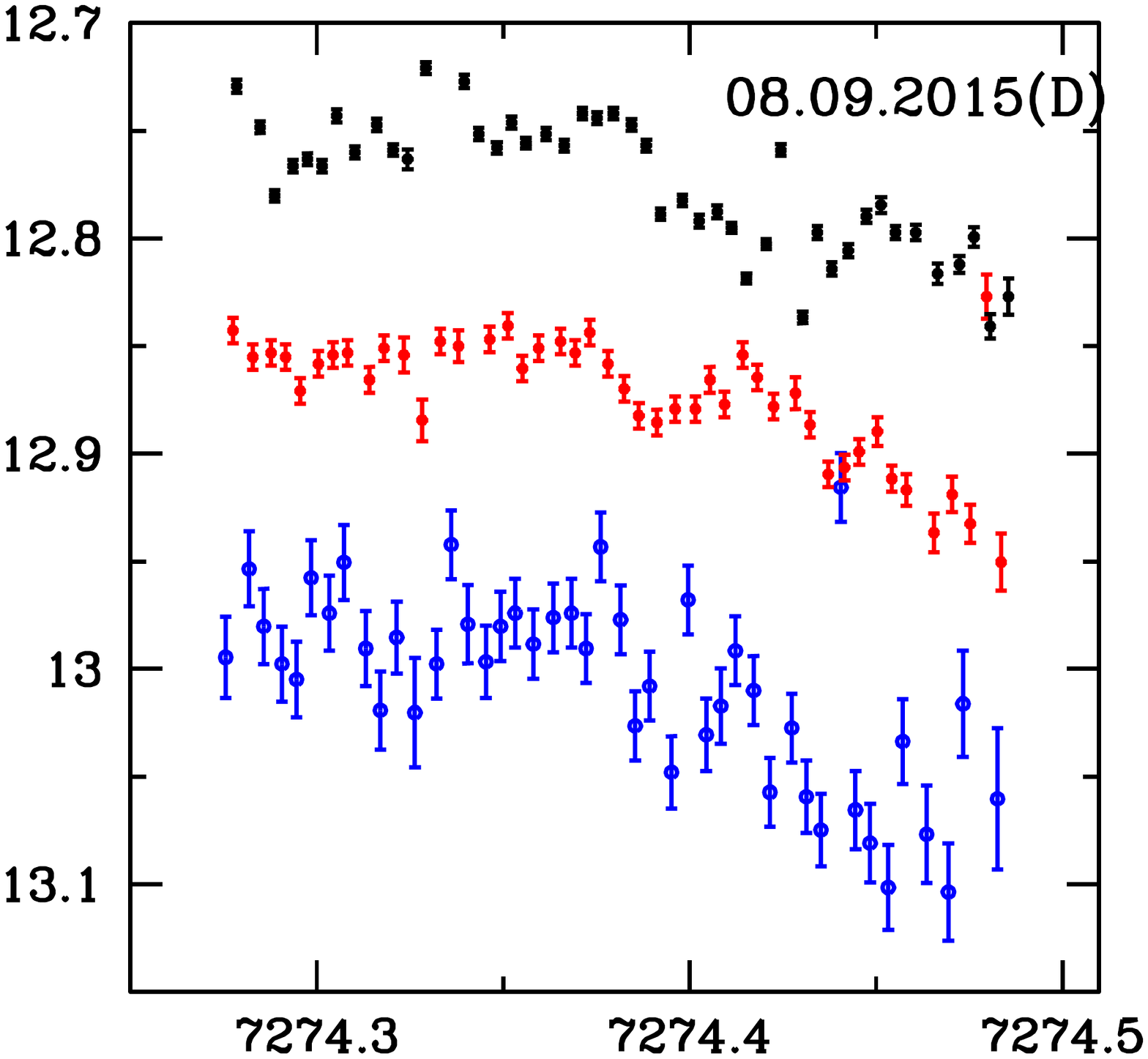}
\includegraphics[scale=0.18,trim=5 5 5 2, clip=true]{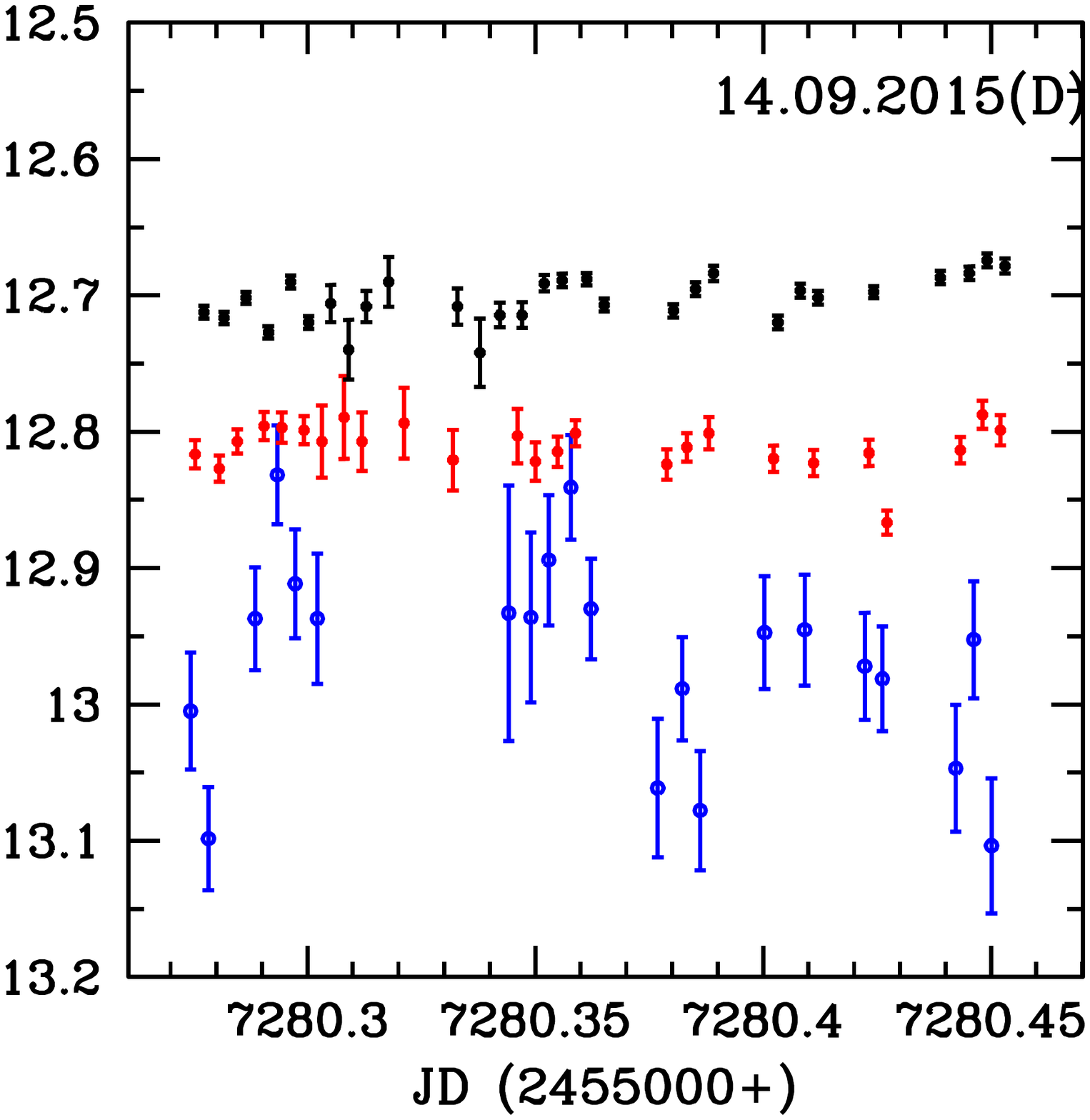}
\includegraphics[scale=0.18,trim=5 5 5 2, clip=true]{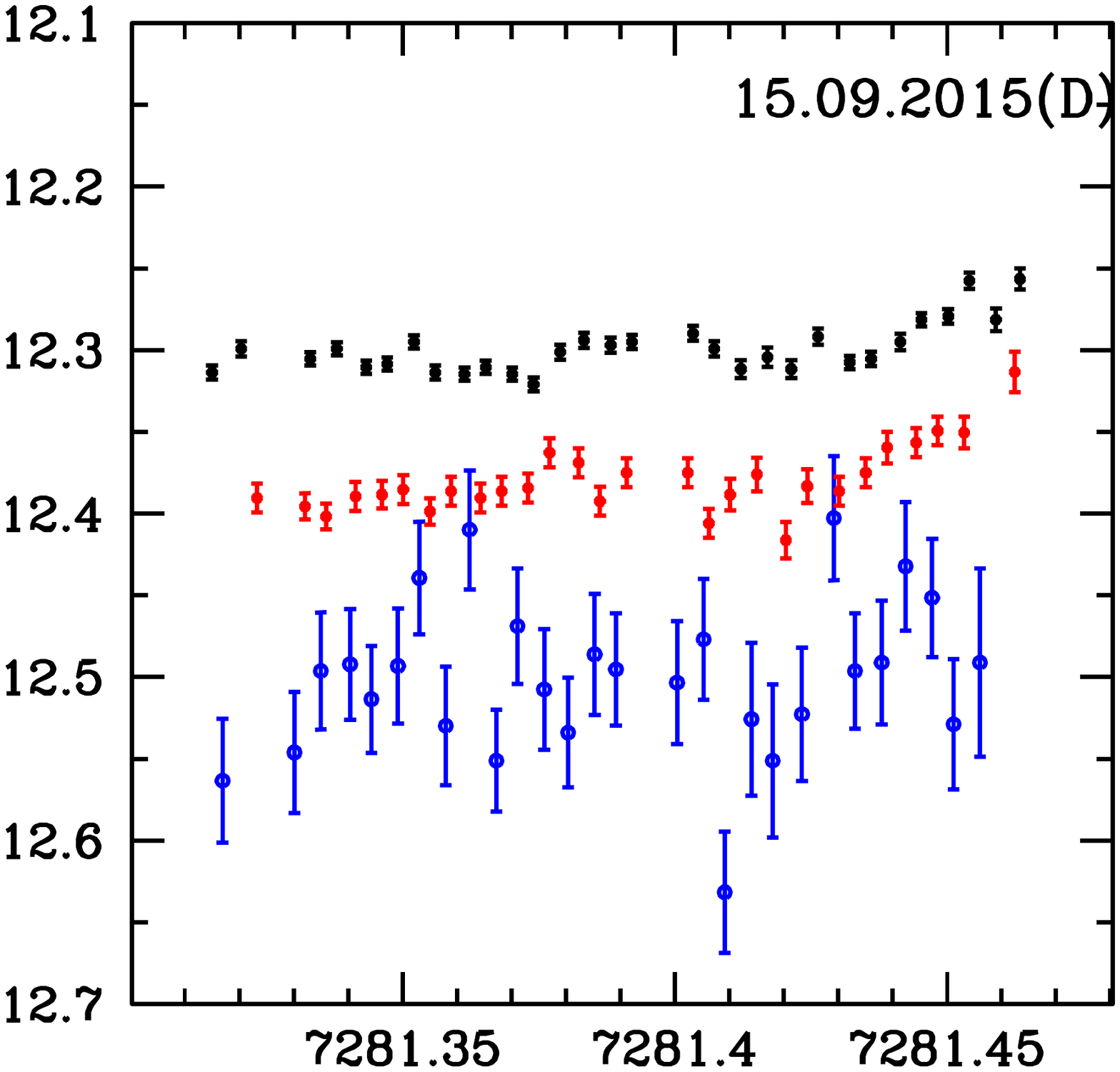}
\includegraphics[scale=0.18,trim=5 5 5 2, clip=true]{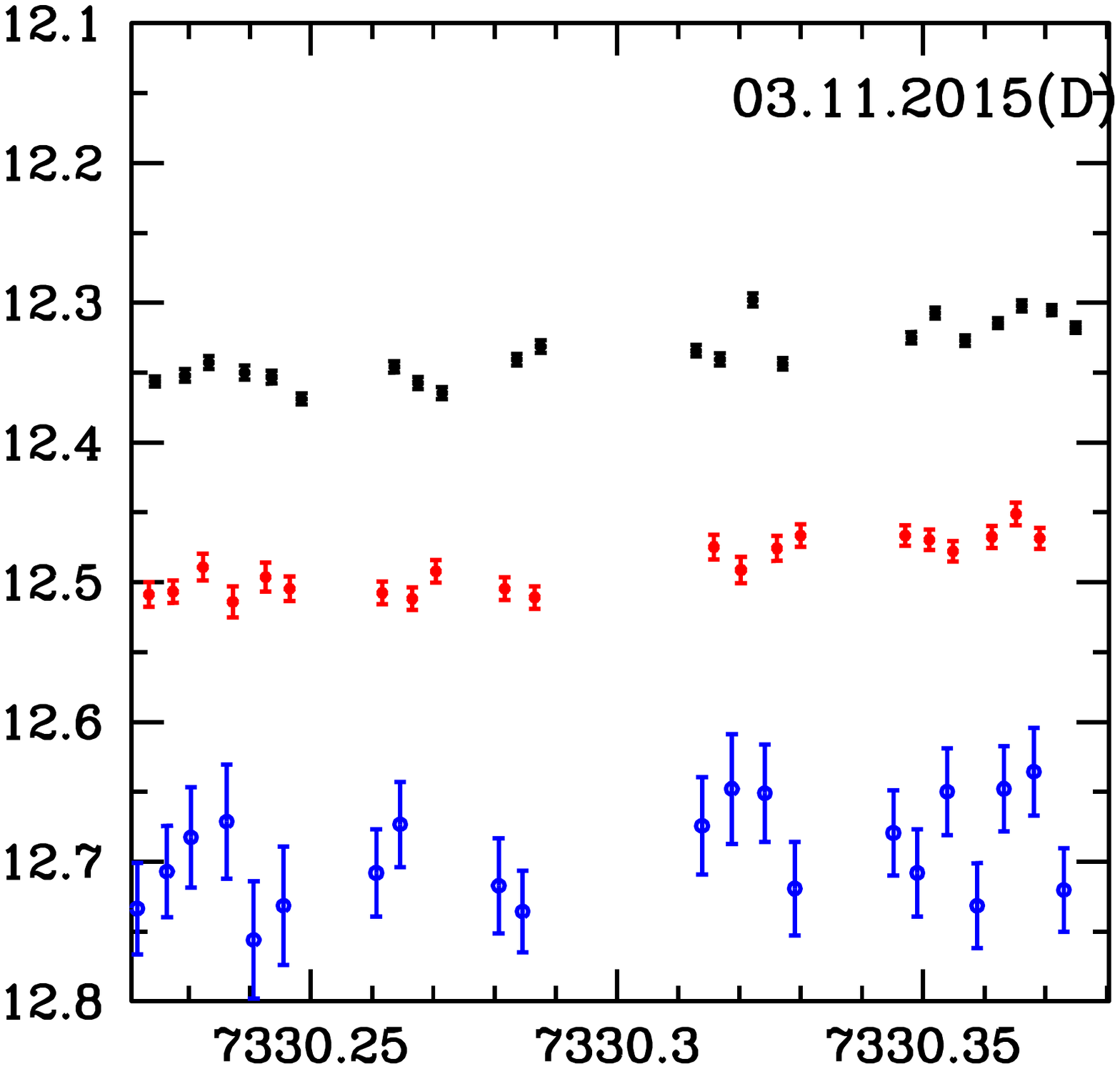}
\vspace{6pt}
\caption{{IDV LCs for BL Lacertae} from July 2014 through November 2015 in the  B (blue), V (green), R (red) and I (black) bands.  The \emph{x}-axis is JD (2455000+), and the \emph{y}-axis is the calibrated R band magnitudes in each of the panels. For clarity, the B, V and I LCs are shifted by arbitrary offsets with respect to the R-band LC.}
\label{fig1}
\end{figure}

\begin{figure}[H]
\center*
\vspace*{-0.6in}
\includegraphics[scale=0.18,trim=5 1 5 1, clip=true]{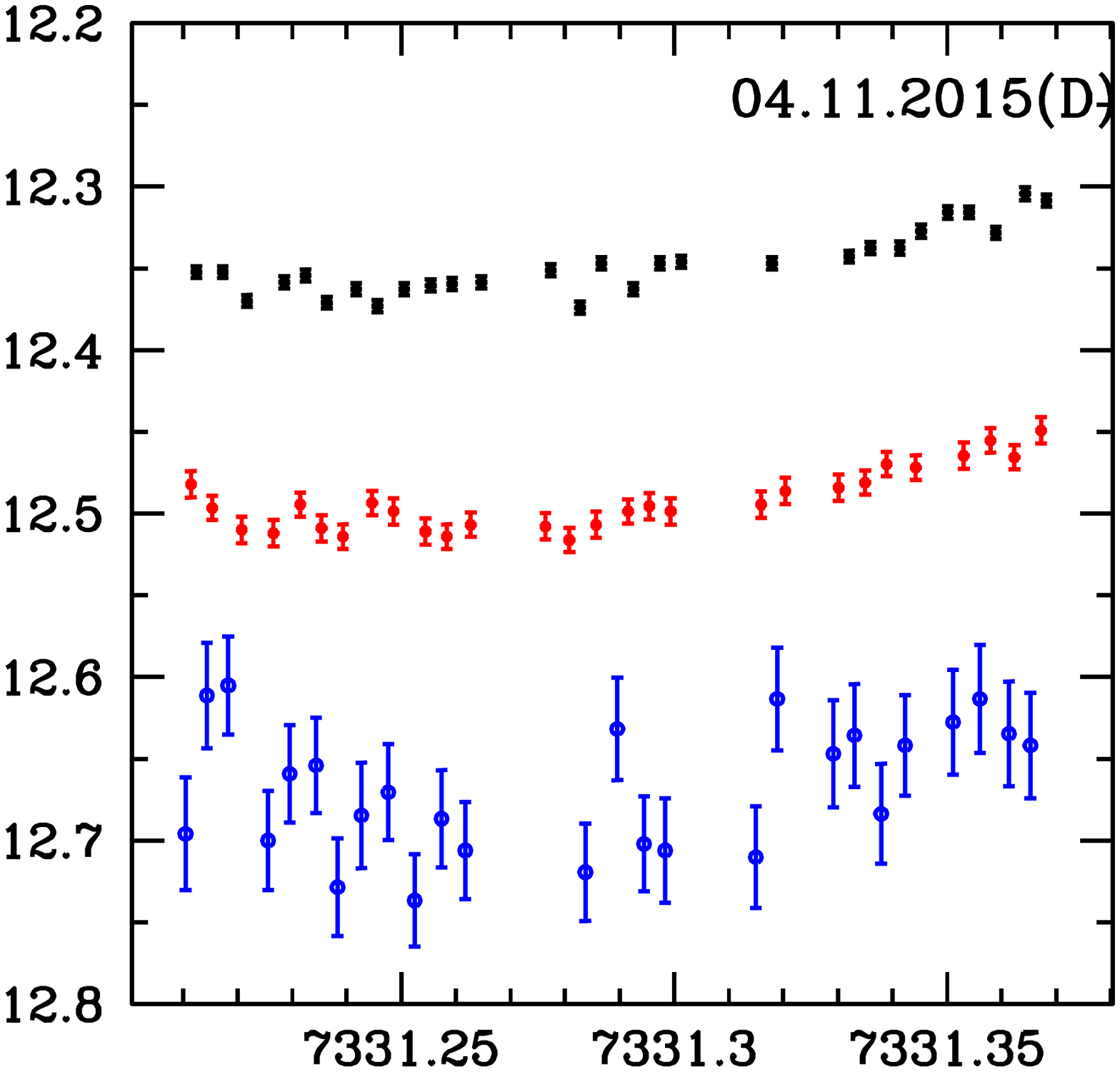}
\includegraphics[scale=0.18,trim=5 1 5 1, clip=true]{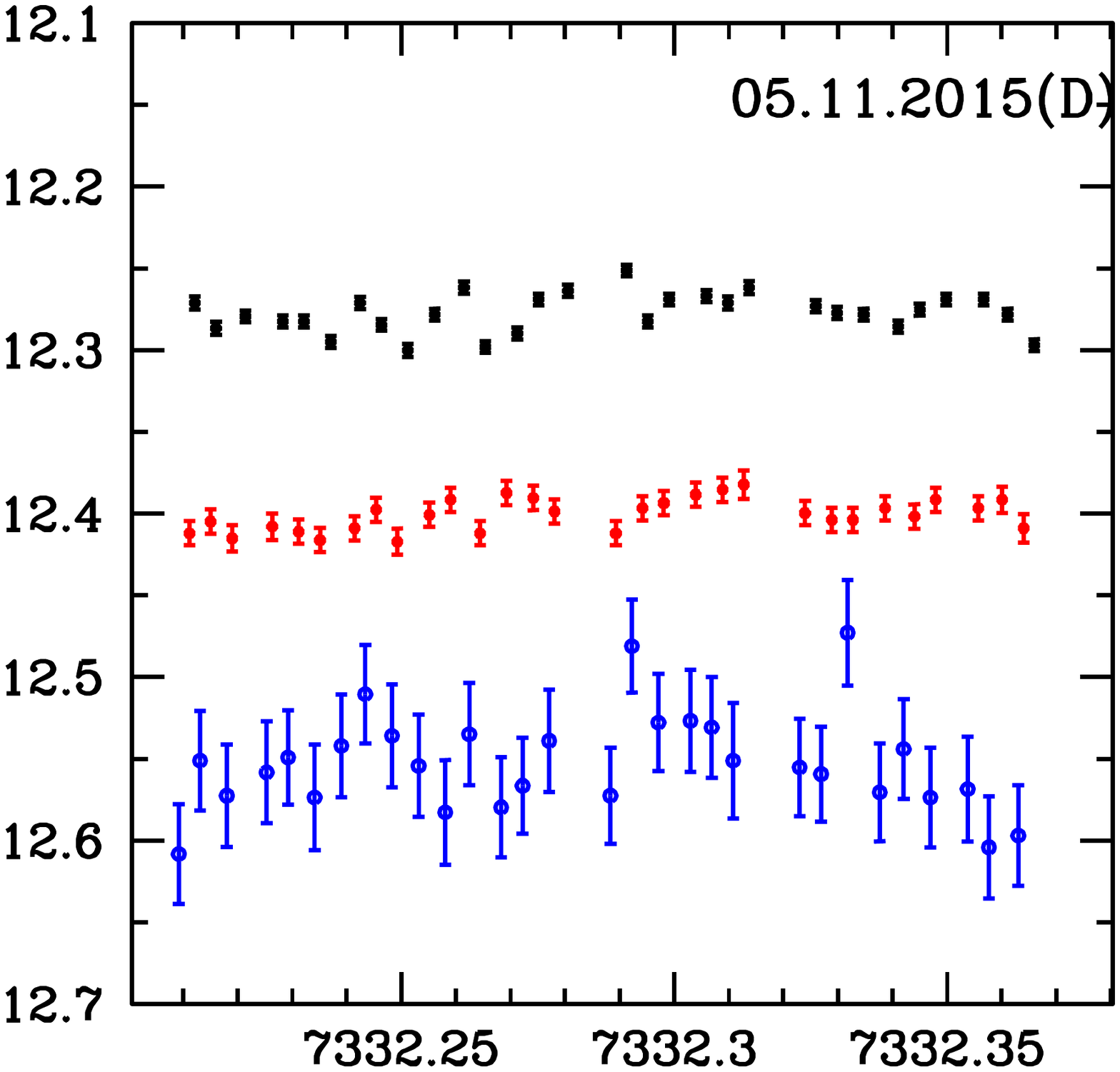}
\includegraphics[scale=0.18,trim=5 1 5 1, clip=true]{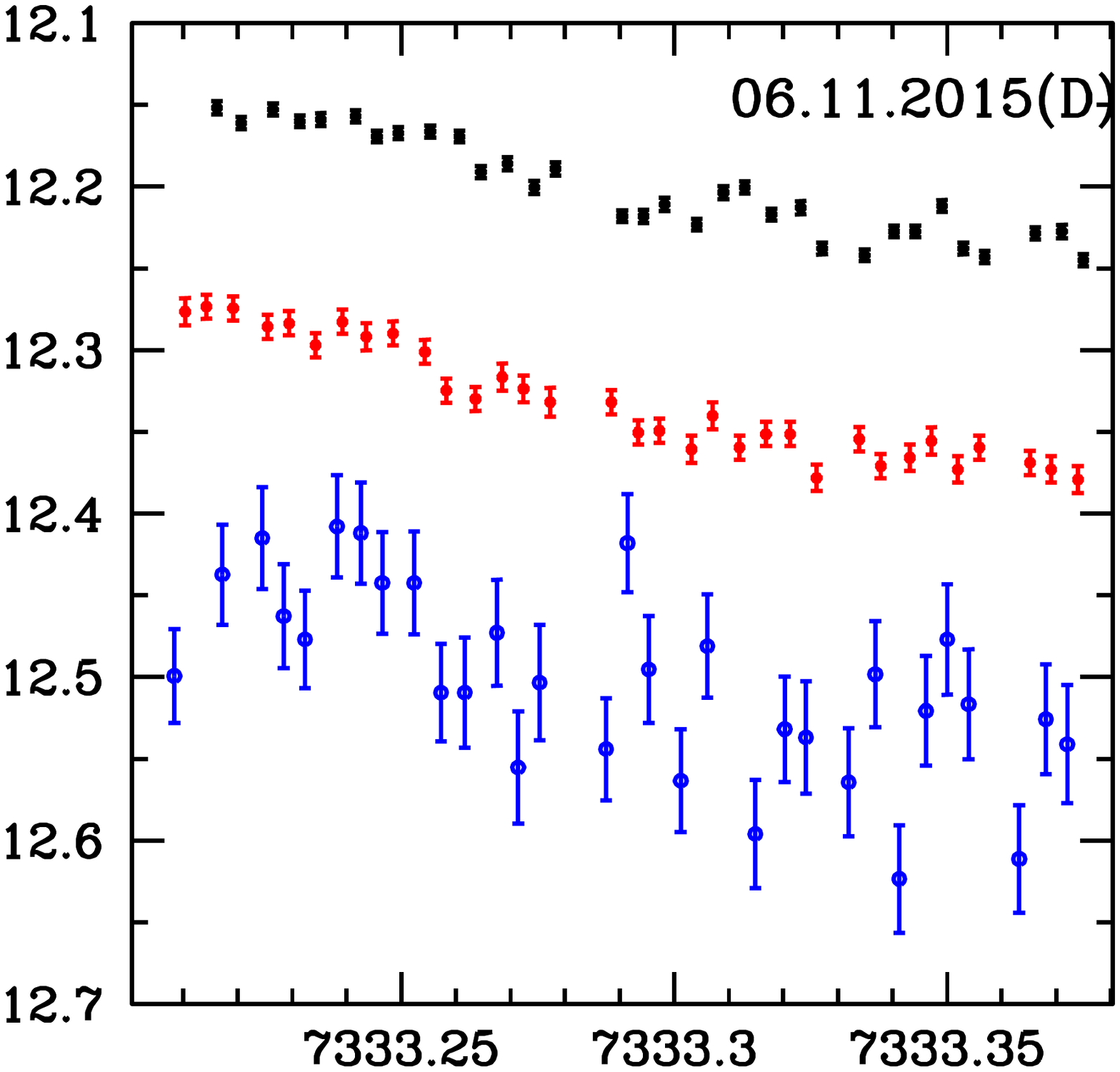}
\includegraphics[scale=0.18,trim=5 1 5 1, clip=true]{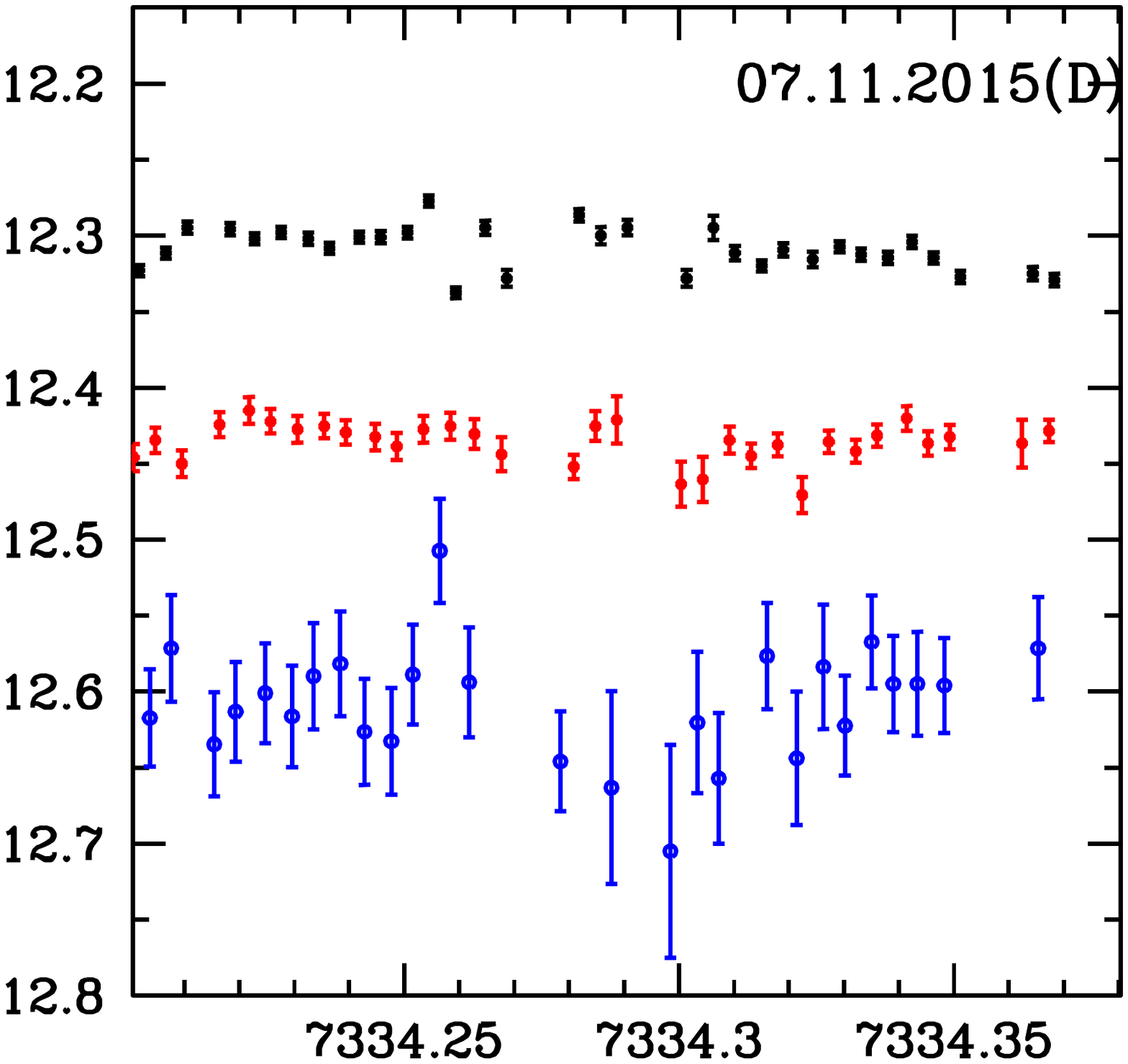}
\vspace*{-0.6in}
\includegraphics[scale=0.18,trim=5 1 5 1, clip=true]{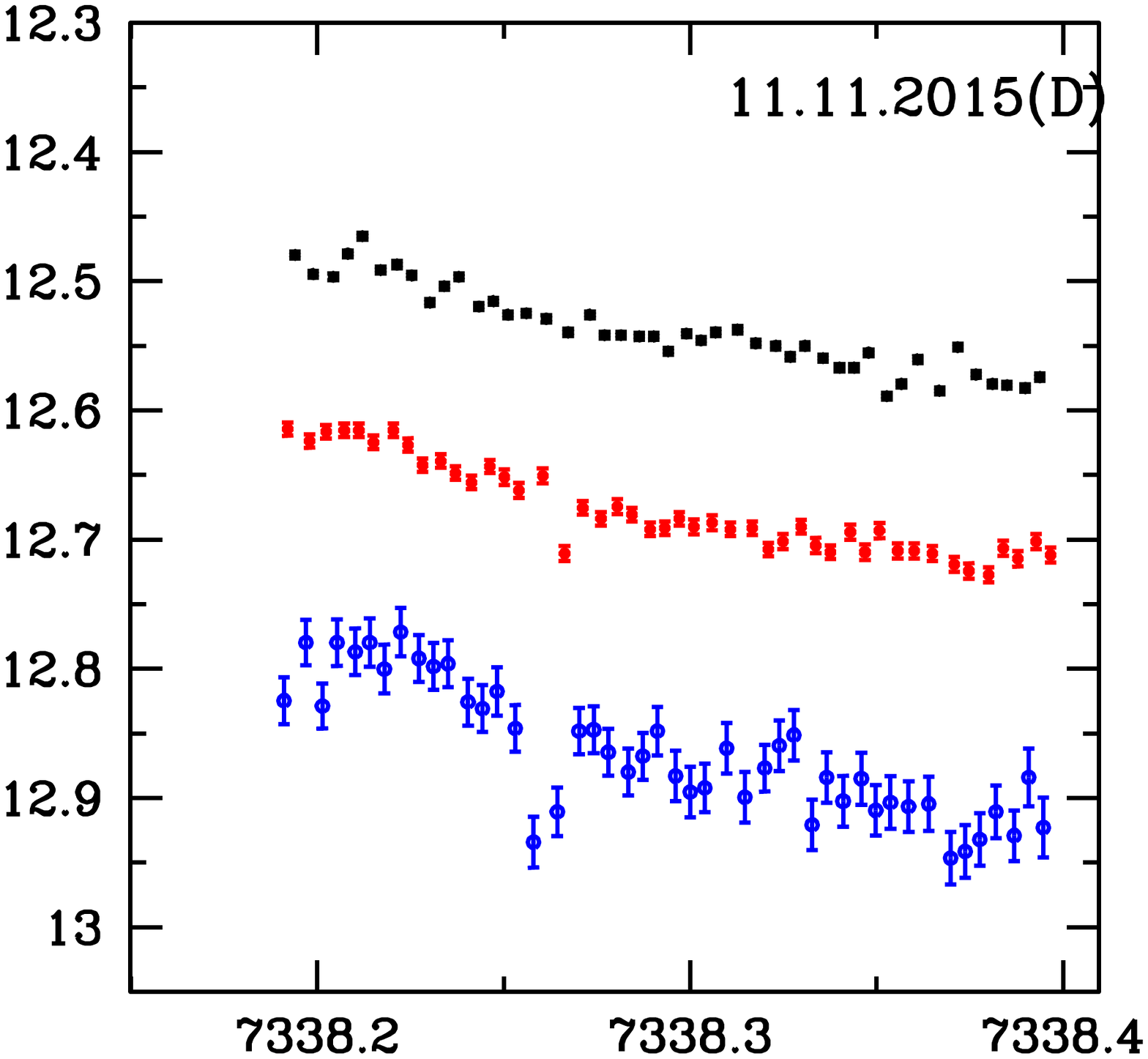}
\includegraphics[scale=0.18,trim=5 1 5 1, clip=true]{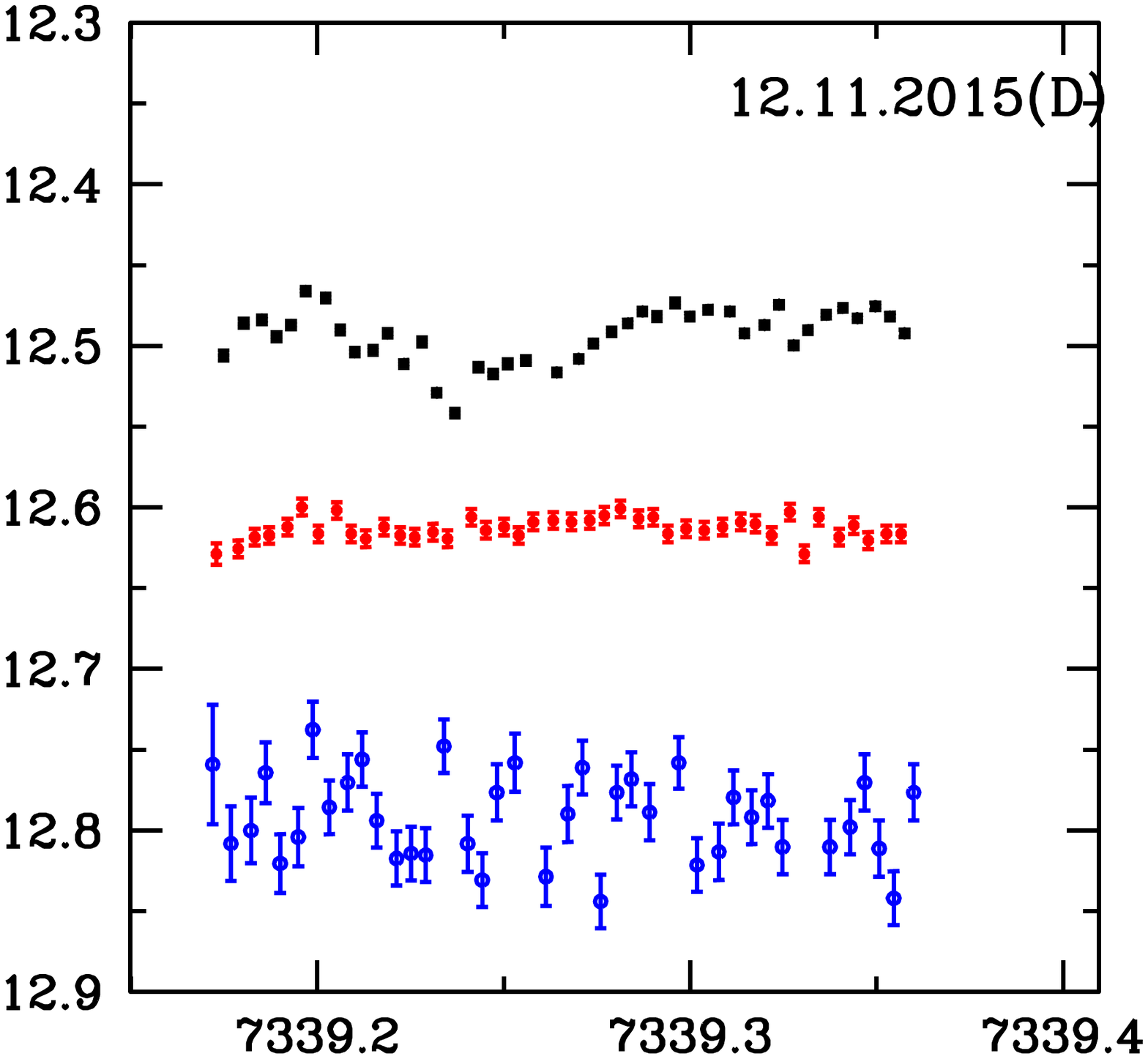}
\includegraphics[scale=0.18,trim=5 1 5 1, clip=true]{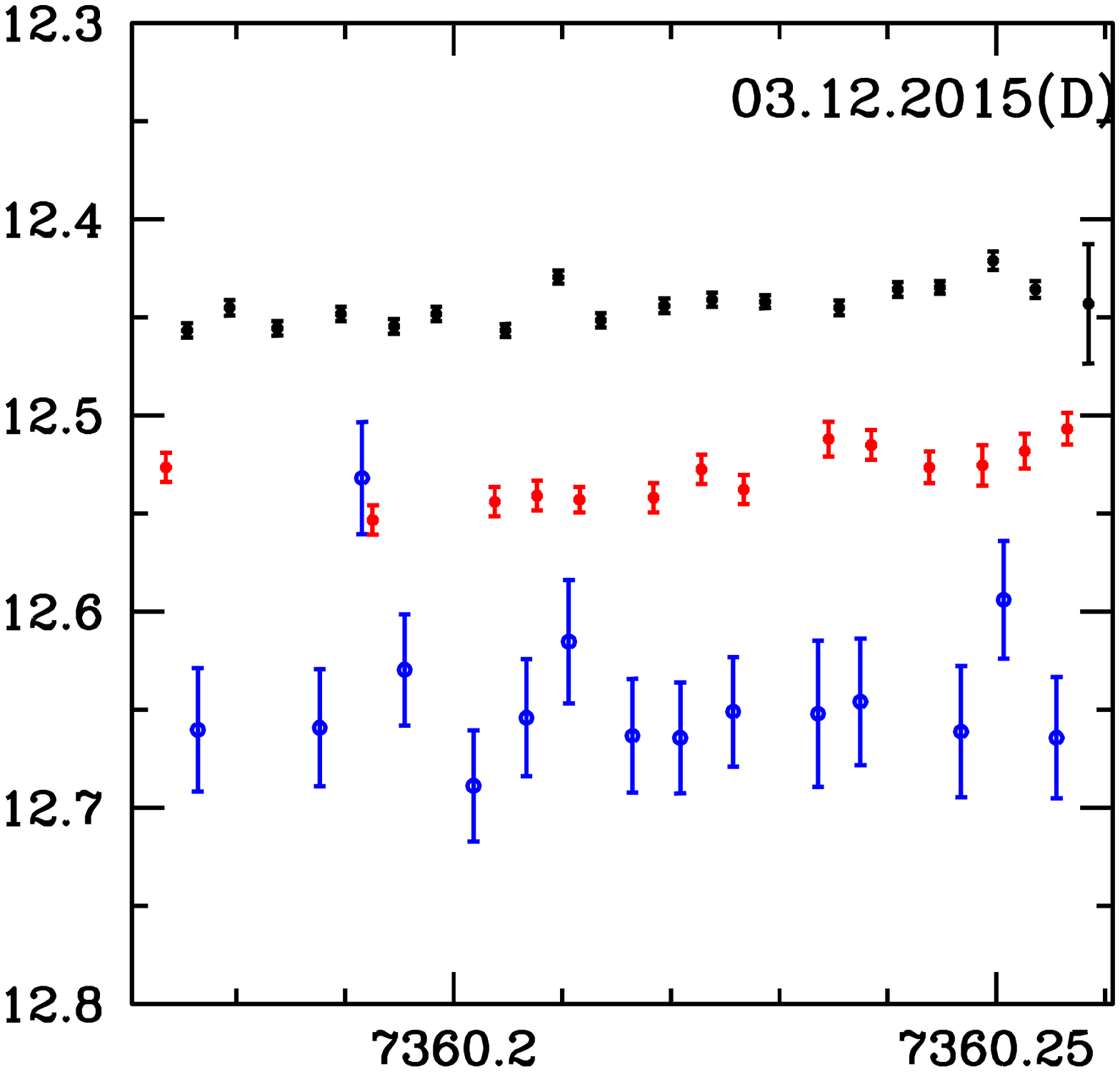}
\includegraphics[scale=0.18,trim=5 1 5 1, clip=true]{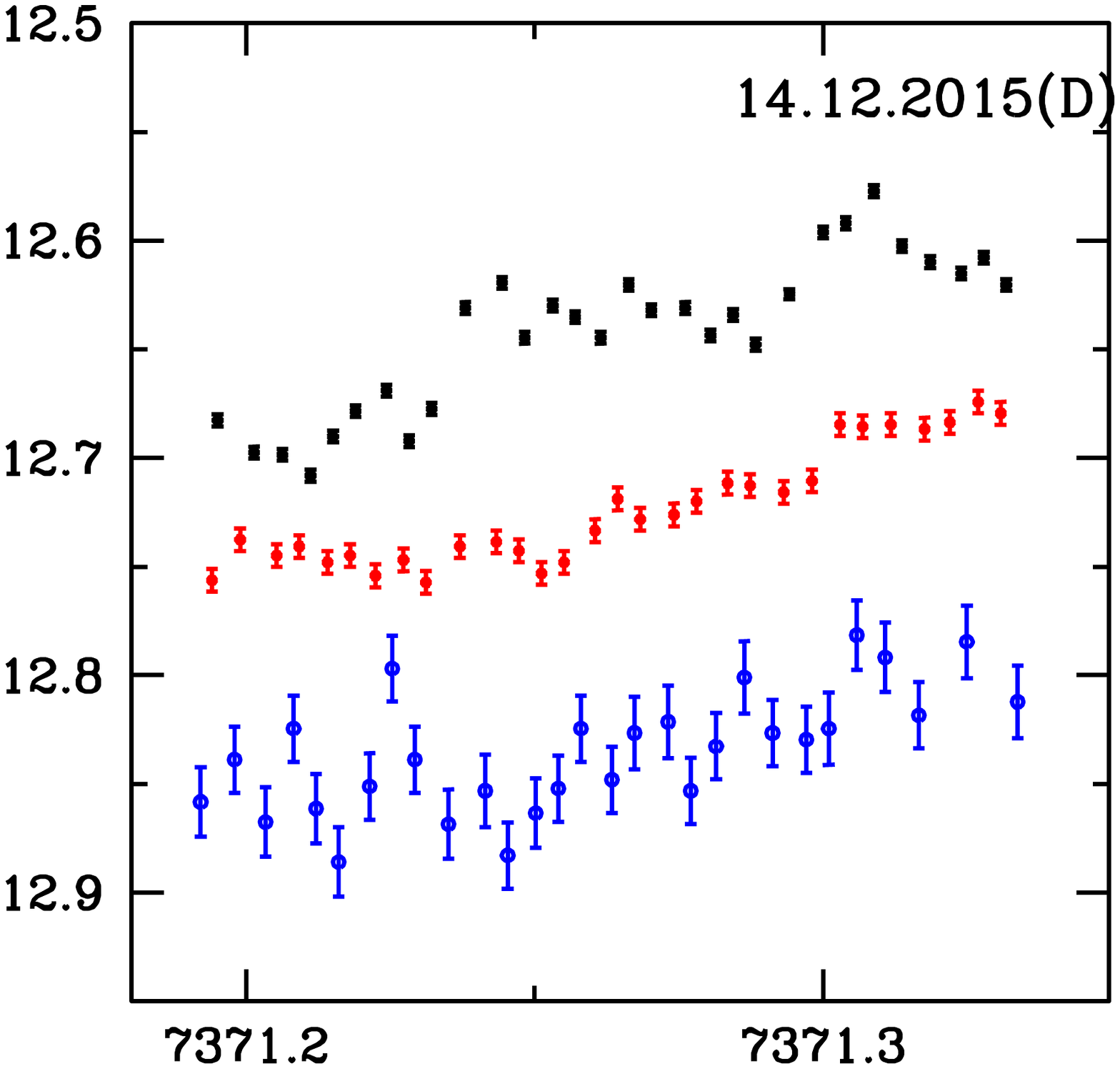}
\vspace*{-0.6in}
\includegraphics[scale=0.18,trim=5 1 5 2, clip=true]{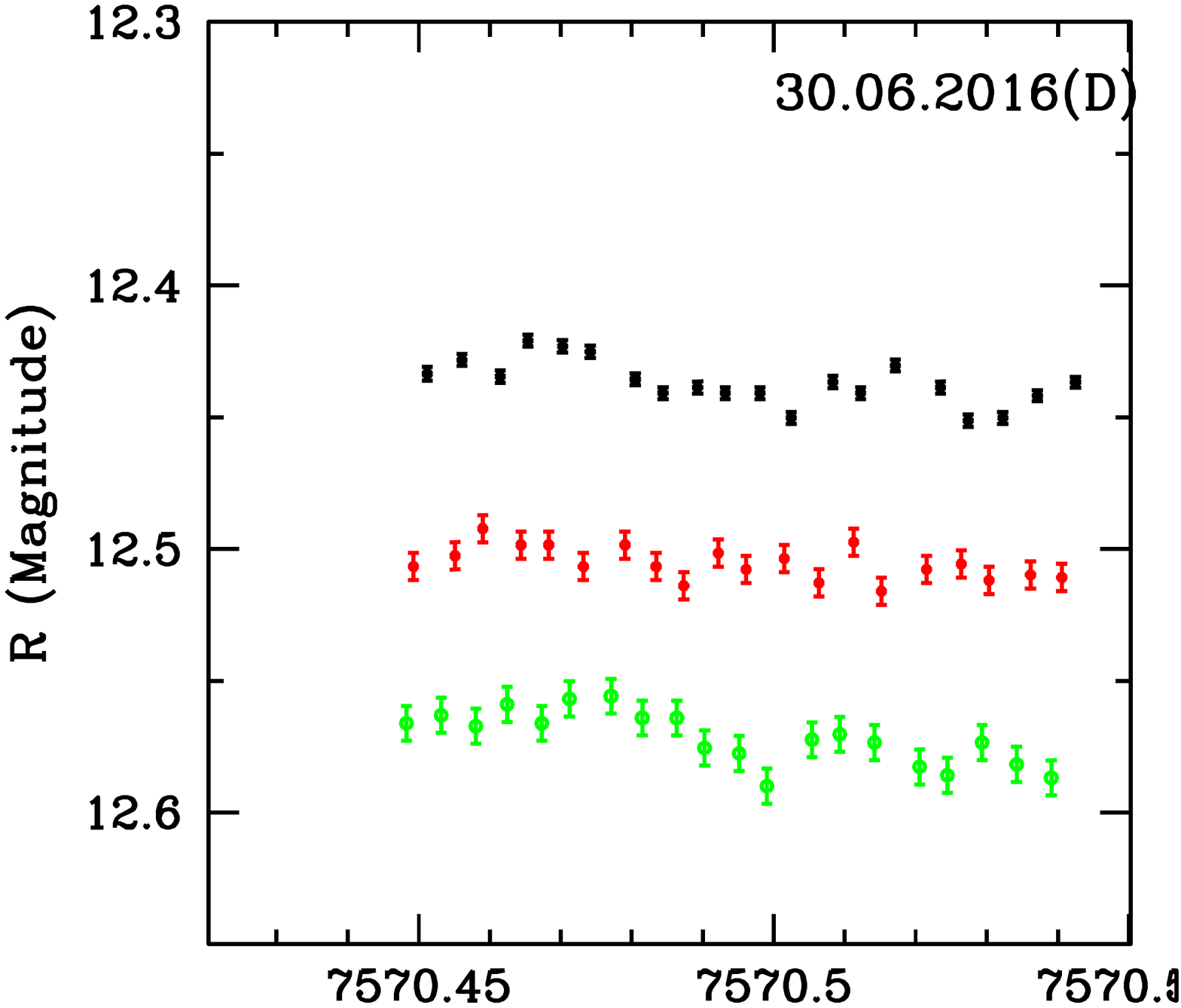}
\includegraphics[scale=0.18,trim=5 1 5 2, clip=true]{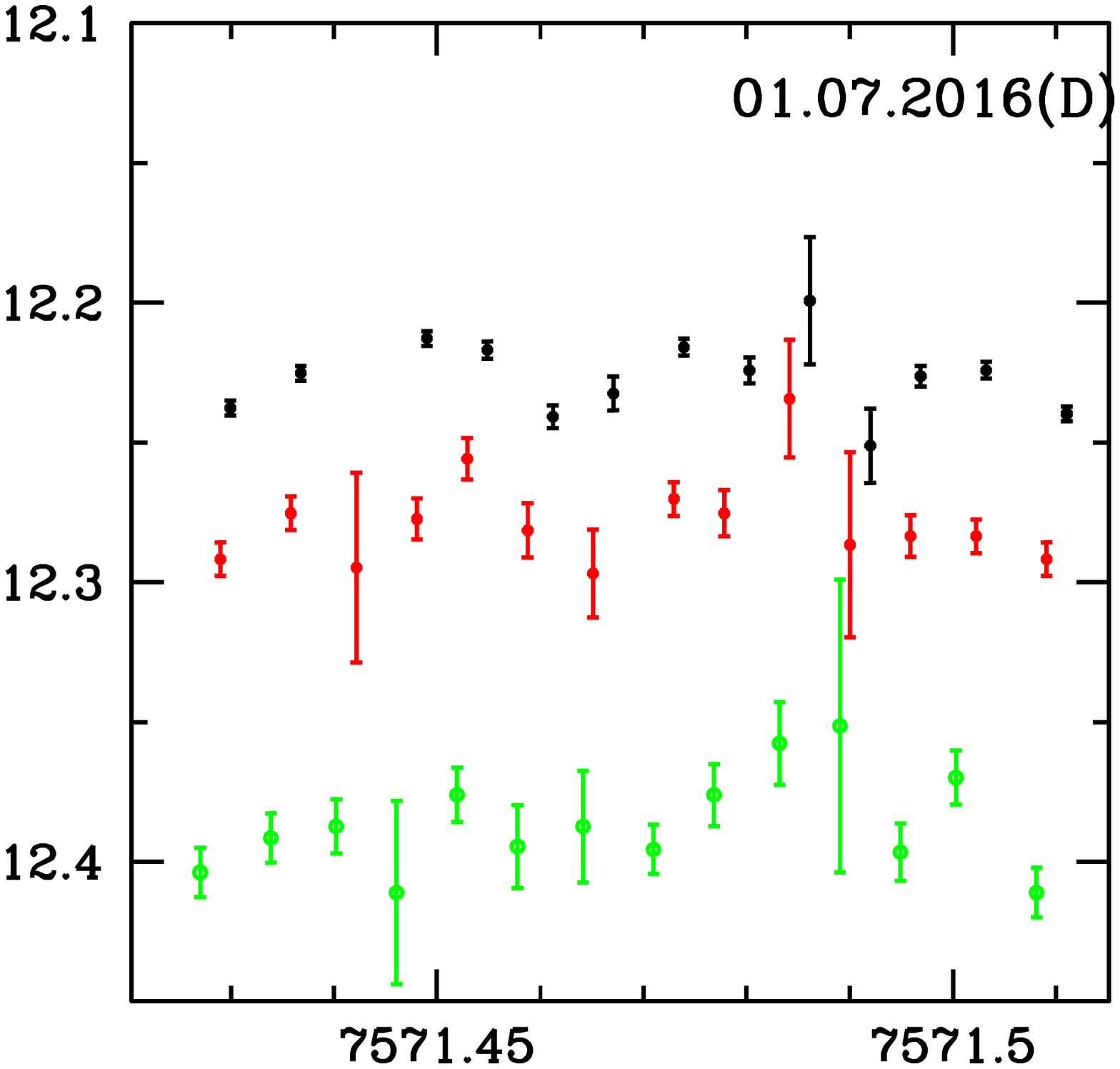}
\includegraphics[scale=0.18,trim=5 1 5 2, clip=true]{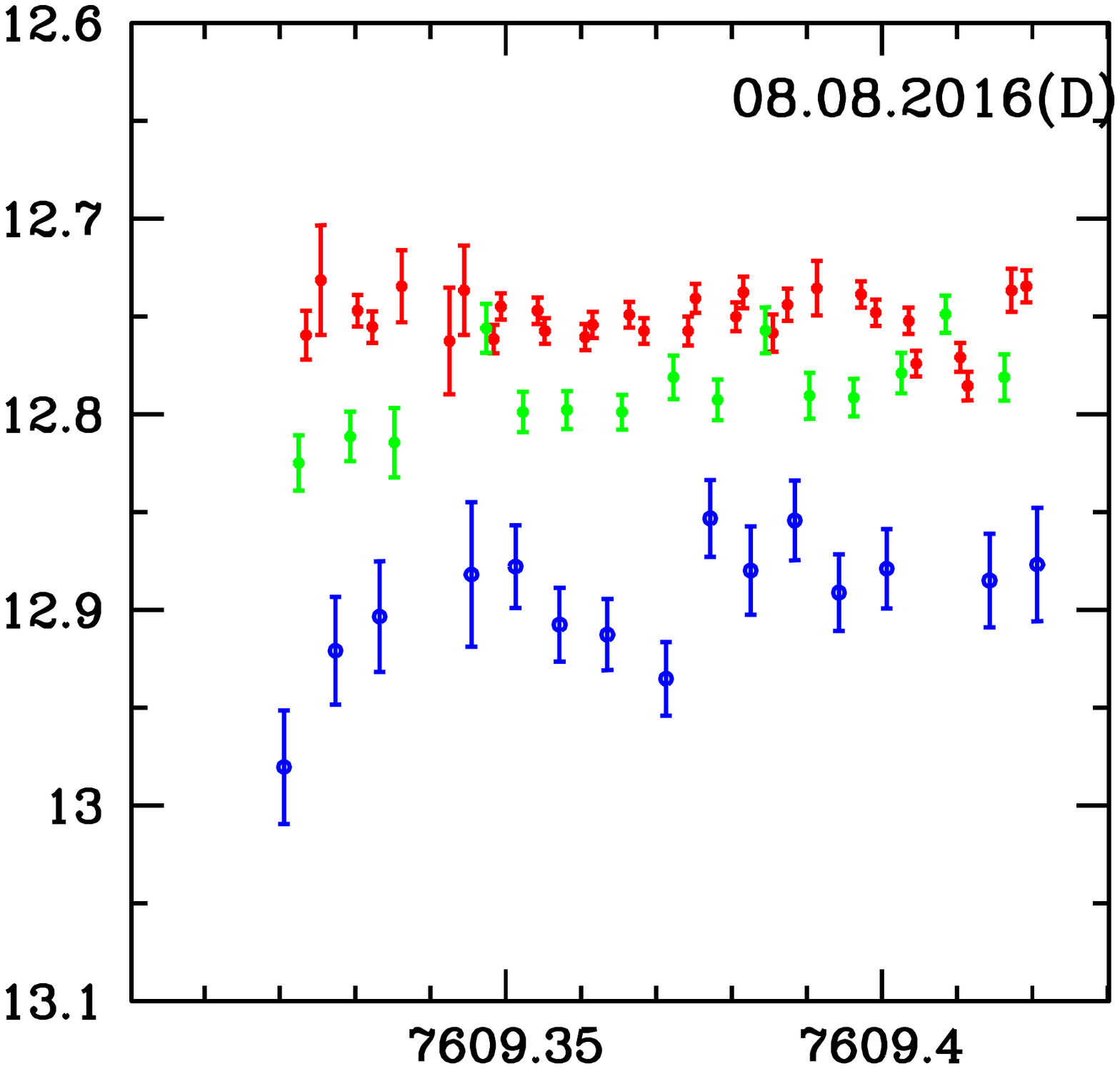}
\includegraphics[scale=0.18,trim=5 1 5 2, clip=true]{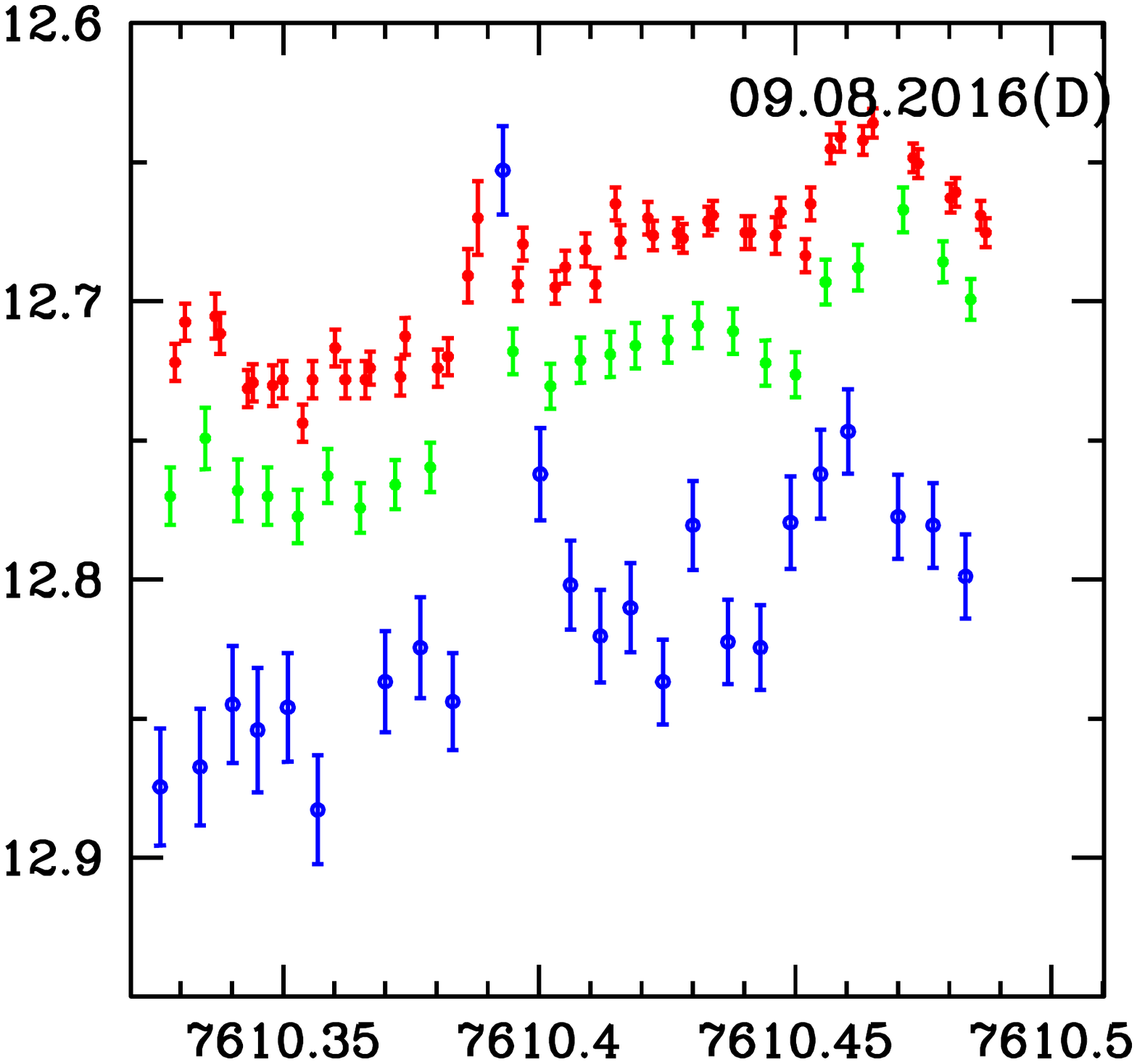}
\vspace*{-0.6in}
\includegraphics[scale=0.18,trim=5 1 5 2, clip=true]{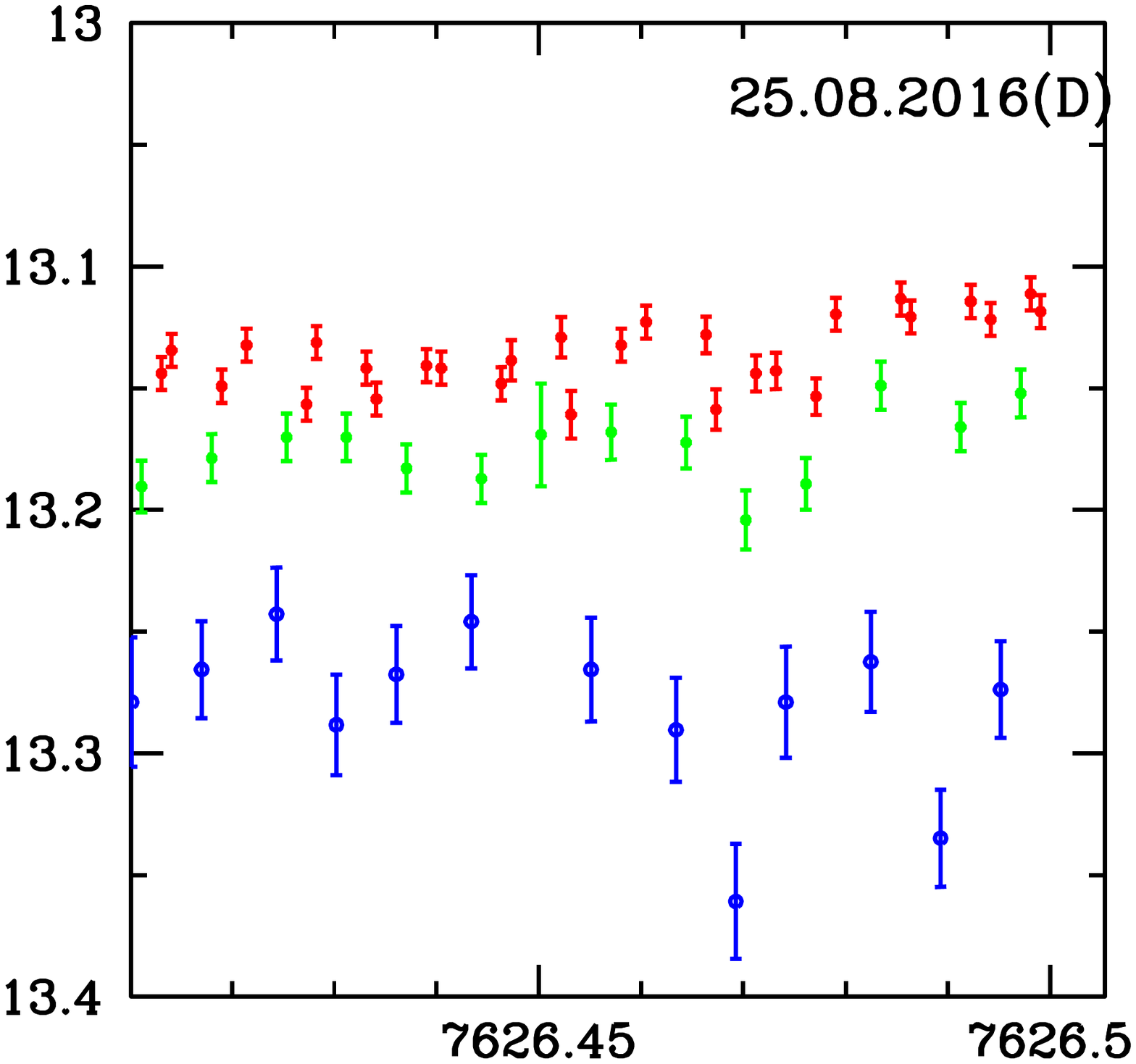}
\includegraphics[scale=0.18,trim=5 1 5 2, clip=true]{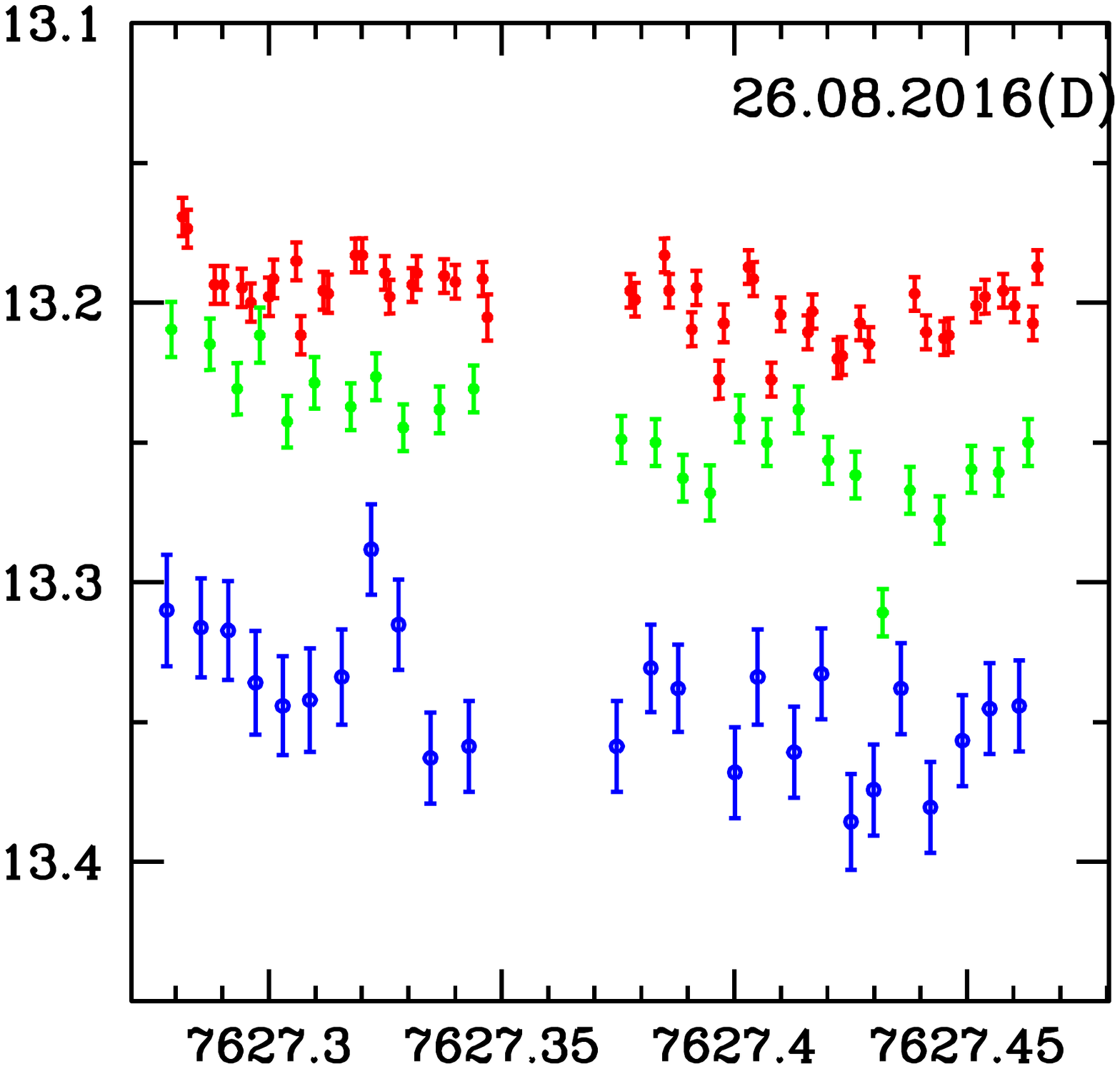}
\includegraphics[scale=0.18,trim=5 1 5 2, clip=true]{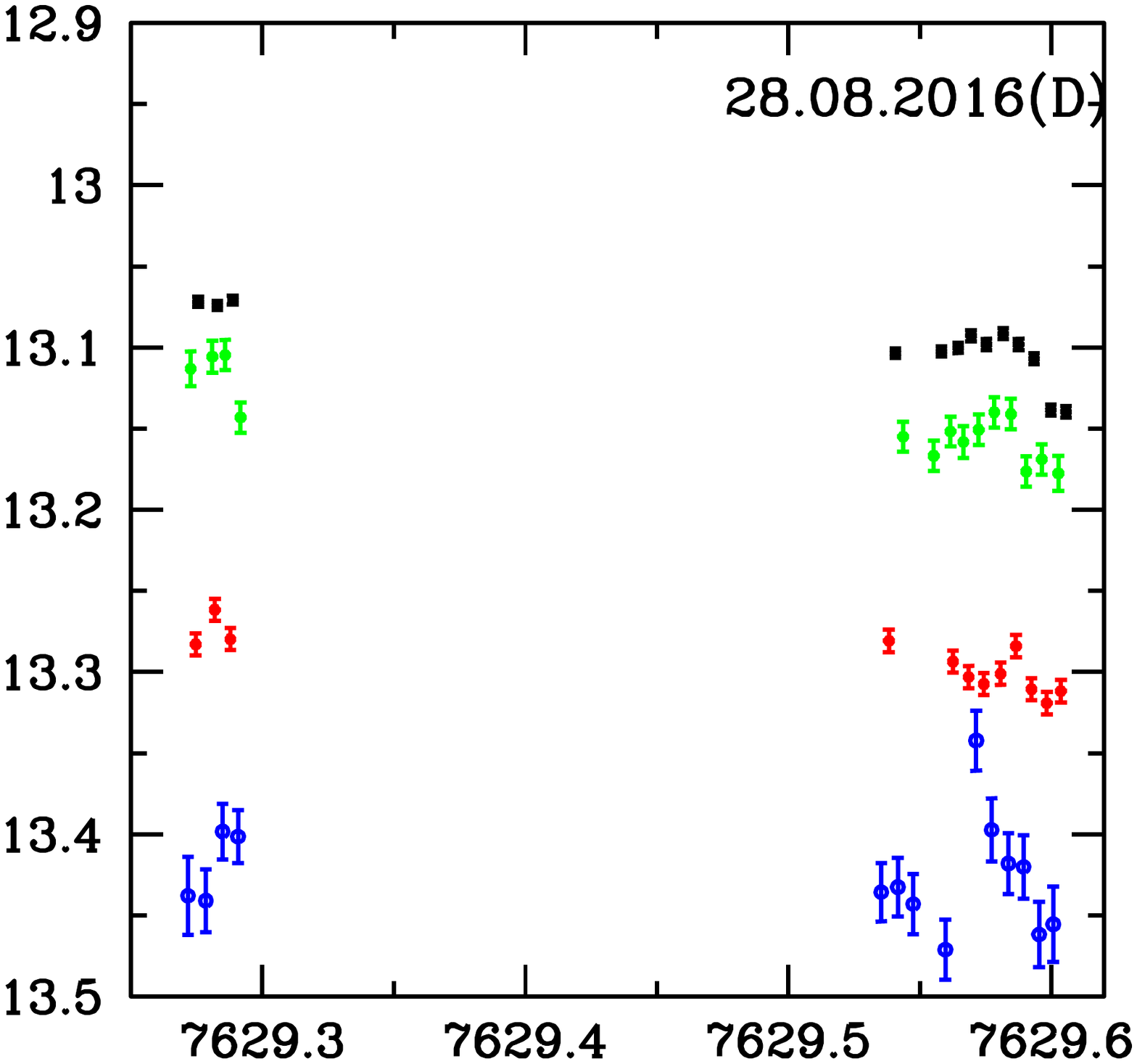}
\includegraphics[scale=0.18,trim=5 1 5 2, clip=true]{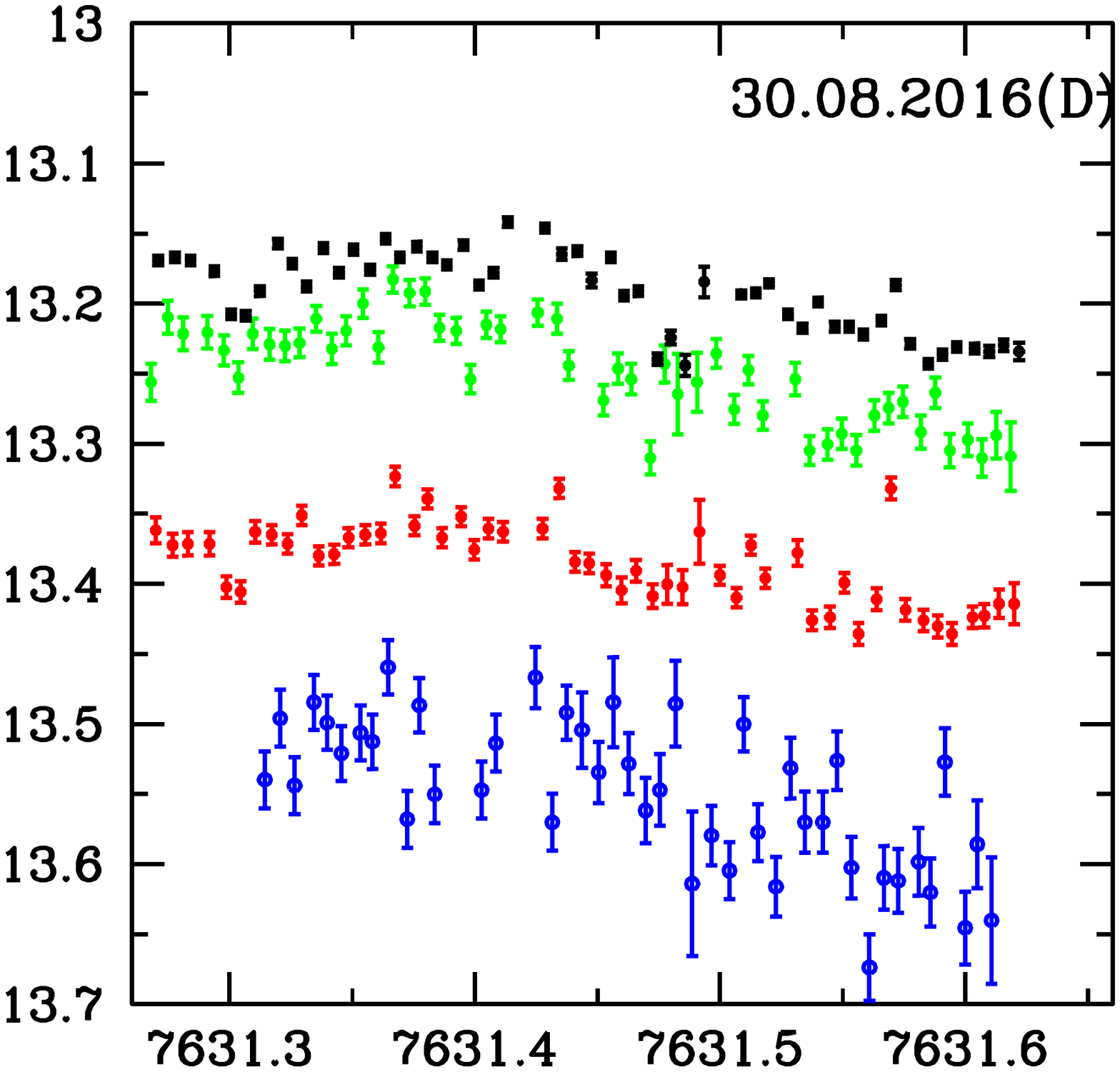}
\vspace*{-0.6in}
\includegraphics[scale=0.18,trim=5 1 5 2, clip=true]{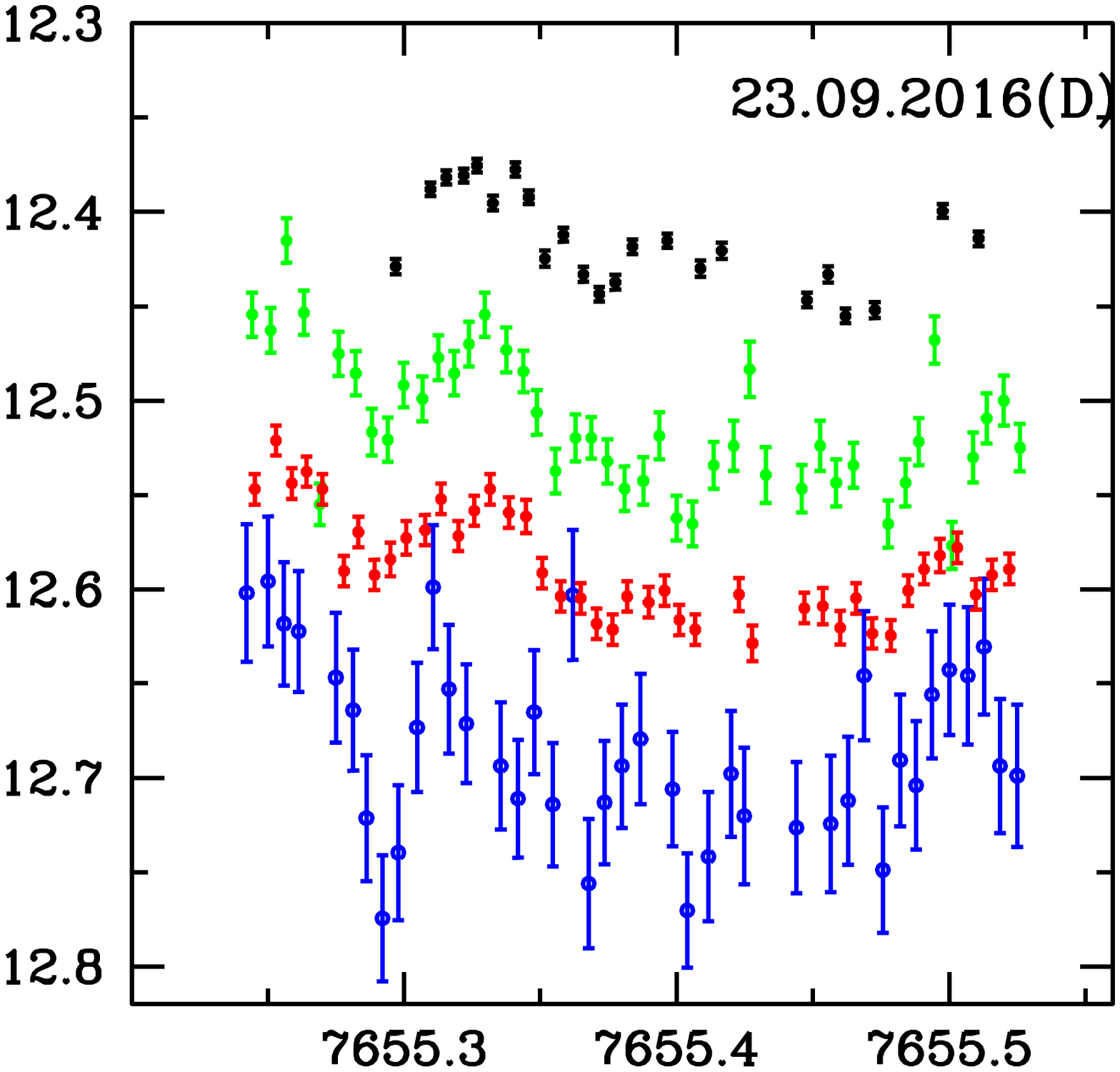}
\includegraphics[scale=0.18,trim=5 1 5 2, clip=true]{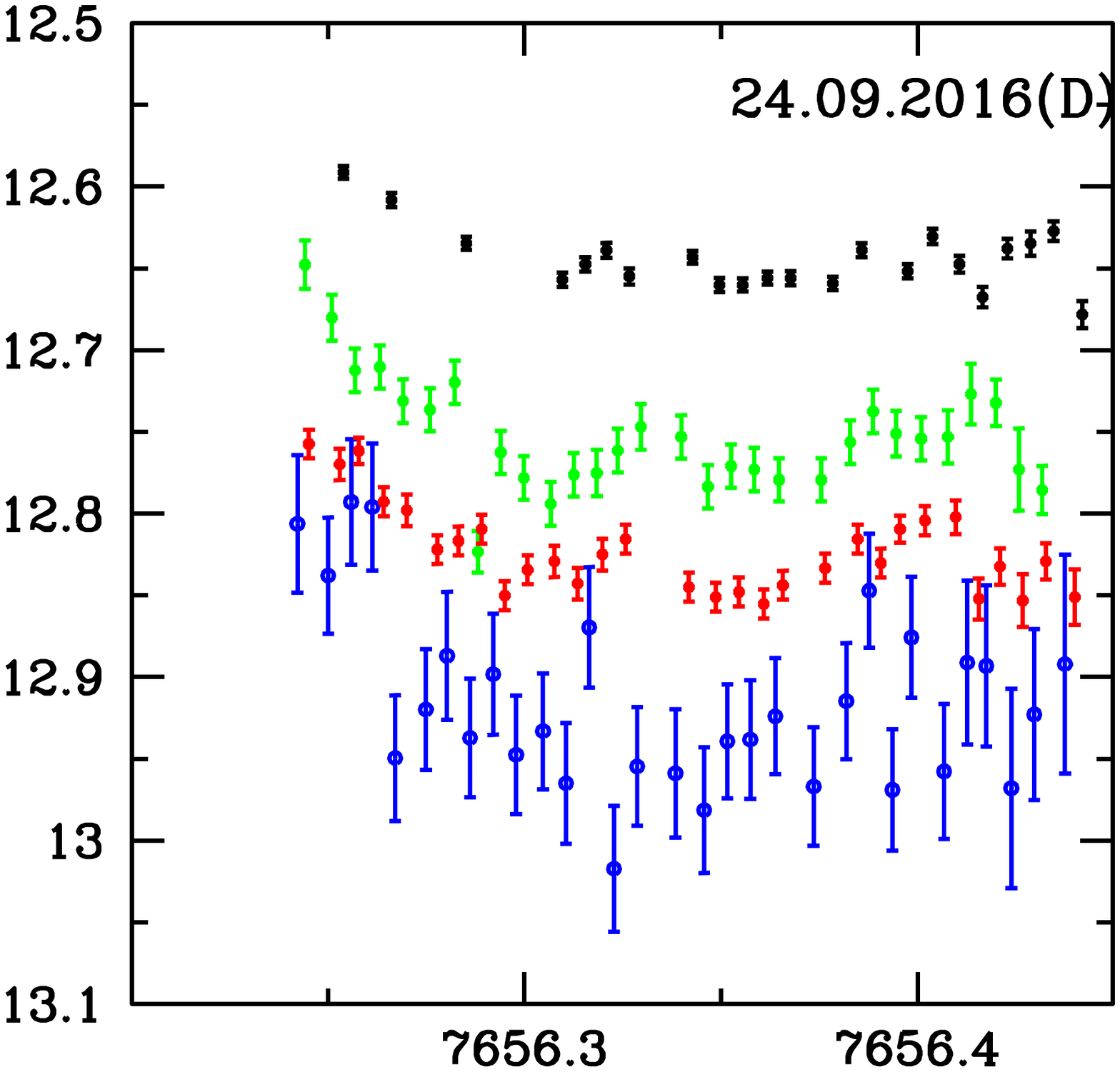}
\includegraphics[scale=0.18,trim=5 1 5 2, clip=true]{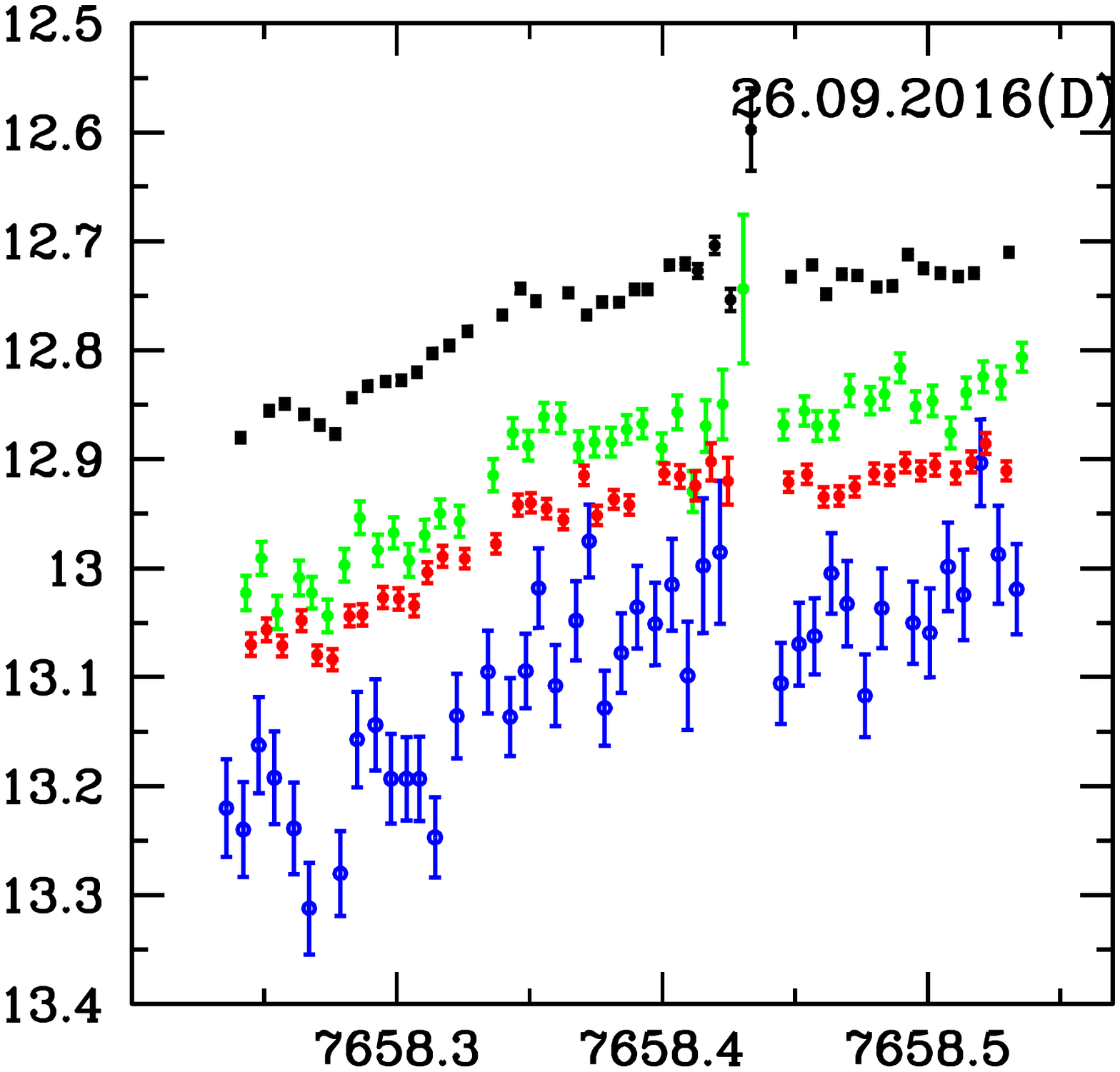}
\includegraphics[scale=0.18,trim=5 1 5 2, clip=true]{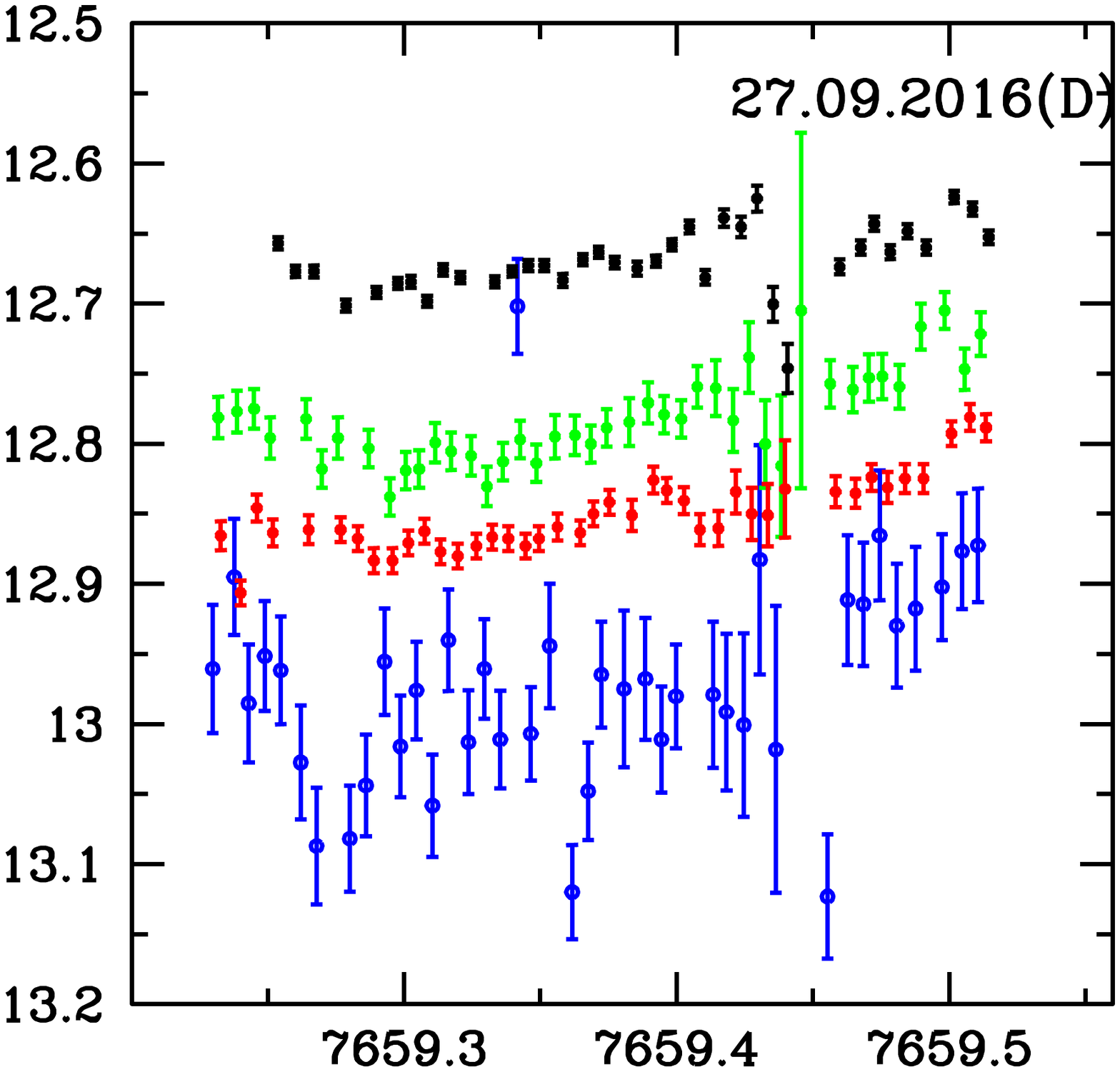}
\vspace*{-0.6in}
\includegraphics[scale=0.18,trim=5 1 5 2, clip=true]{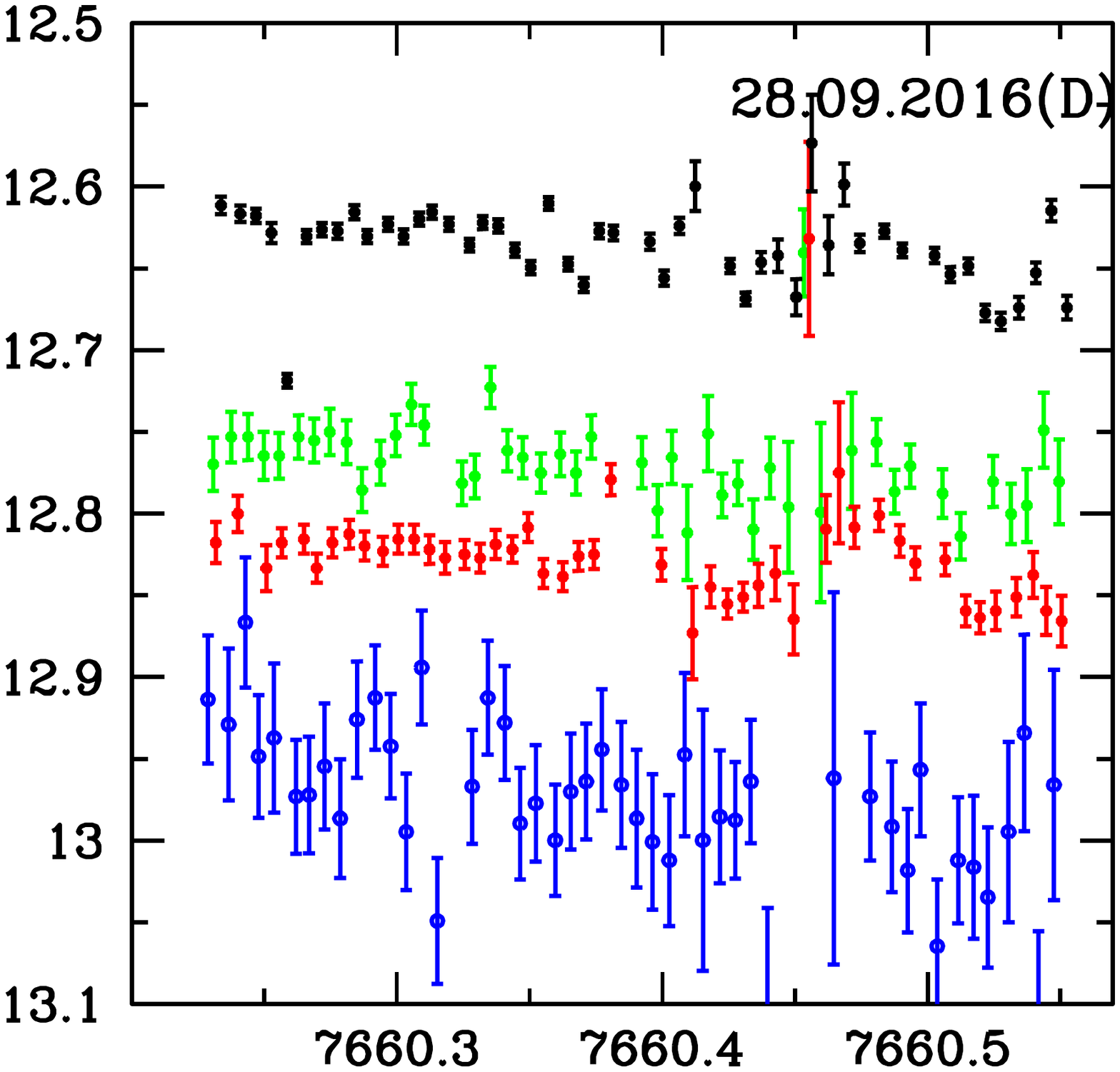}
\includegraphics[scale=0.18,trim=5 1 5 2, clip=true]{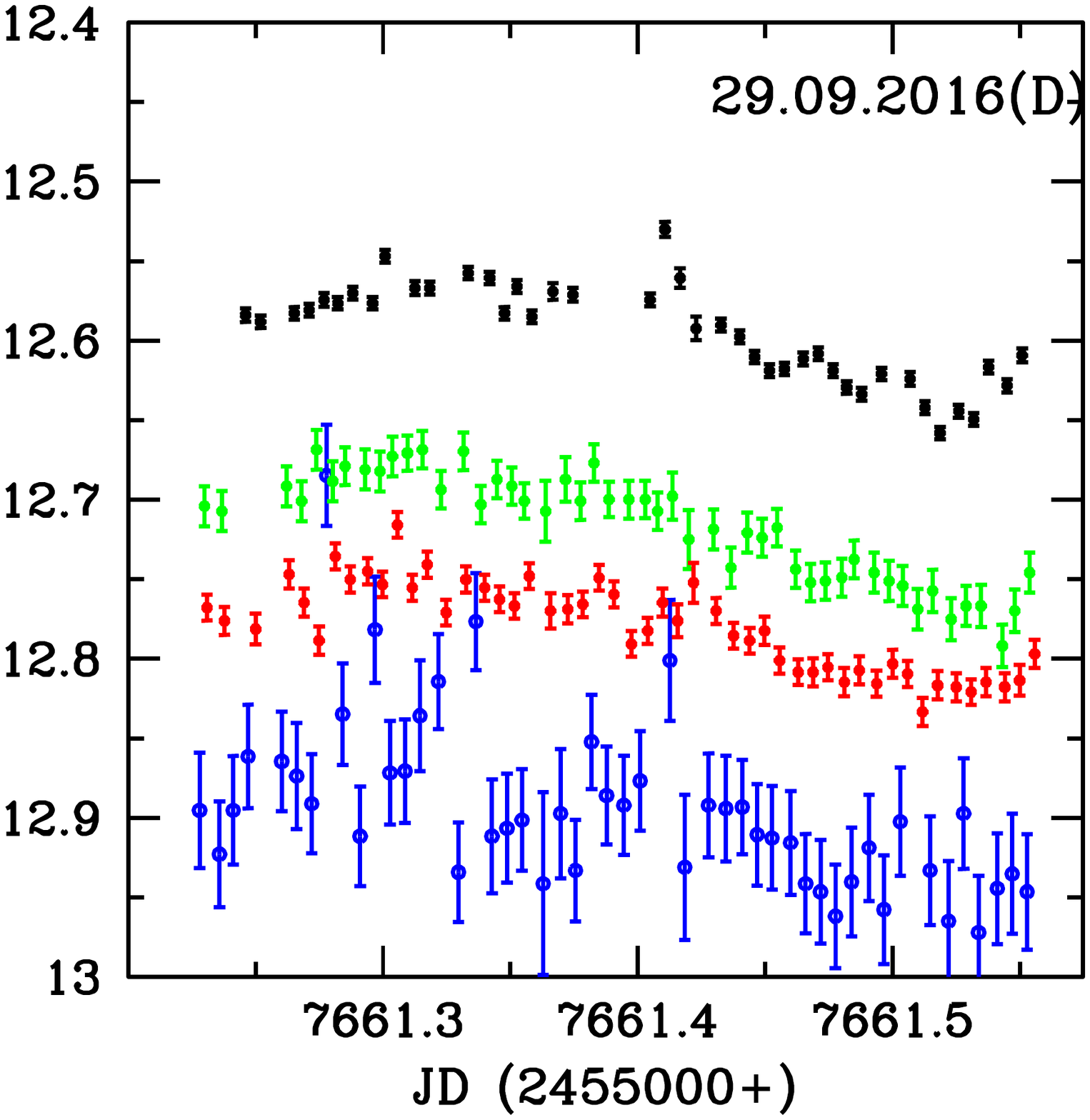}
\includegraphics[scale=0.18,trim=5 1 5 2, clip=true]{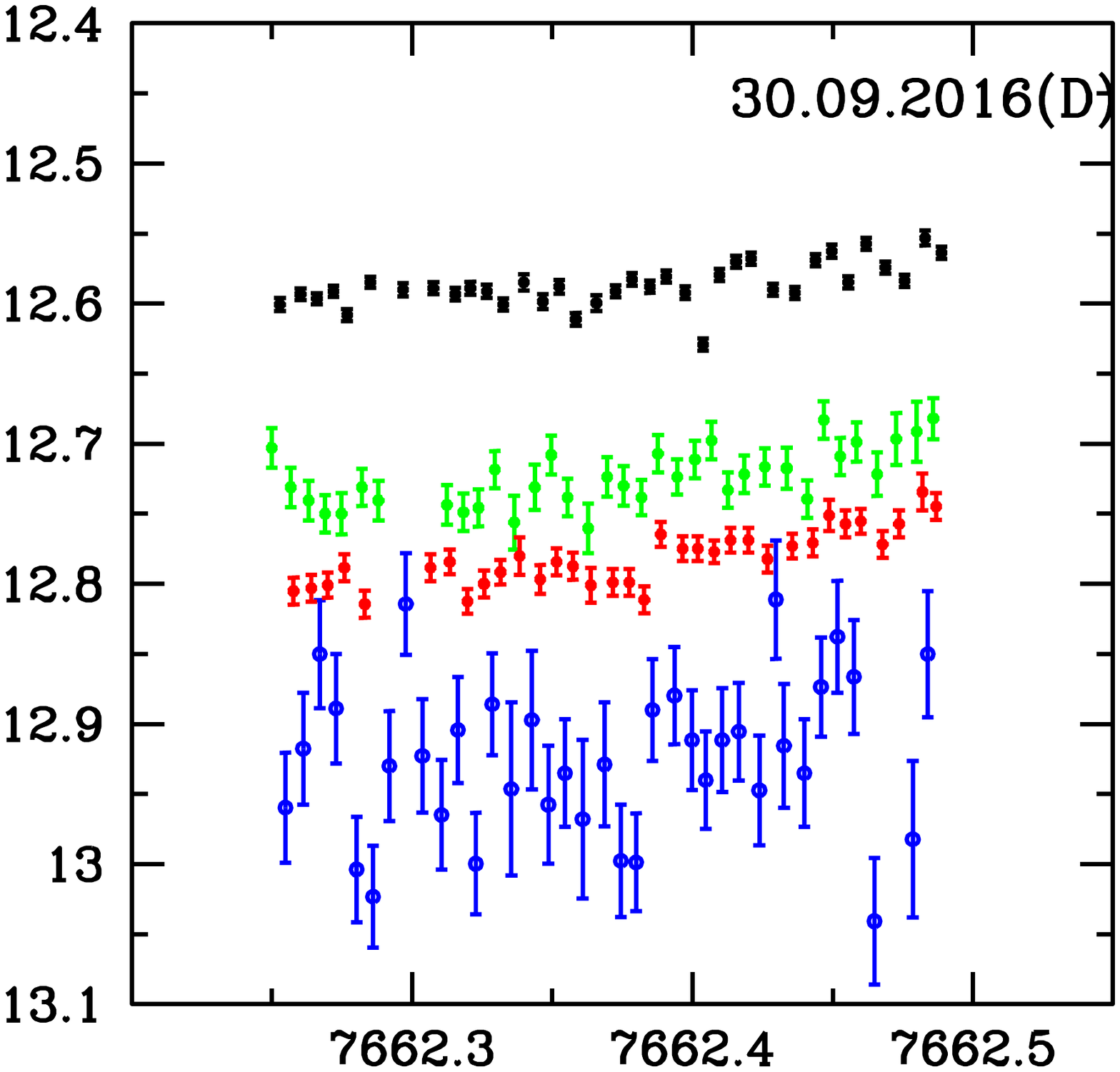}
\vspace{6pt}
\caption{{As in} Figure \ref{fig1}, for November 2015 through September 2016.}
\label{fig2}
\end{figure}

Our extensive observations of BL Lacertae were made on 43 nights during 2014--2016.  During~those nights,  BL Lac was measured for brightness quasi-simultaneously in at least two, and usually at least three, of the B, V, R and I bands, yielding a total of 143 LCs. These~LCs are displayed in Figures~\ref{fig1} and \ref{fig2} and
show that the R-band brightnesses ranged between 13.4 and 12.3 mag but were almost always brighter than $m= 13$.
Our nightly observations mostly spanned between 2 and 5 h, with typical intervals between exposures in a given band being 8 min.

\begin{figure}[H]
\centering
\includegraphics[width=0.5\textwidth,trim=0 7cm 0 5cm,clip]{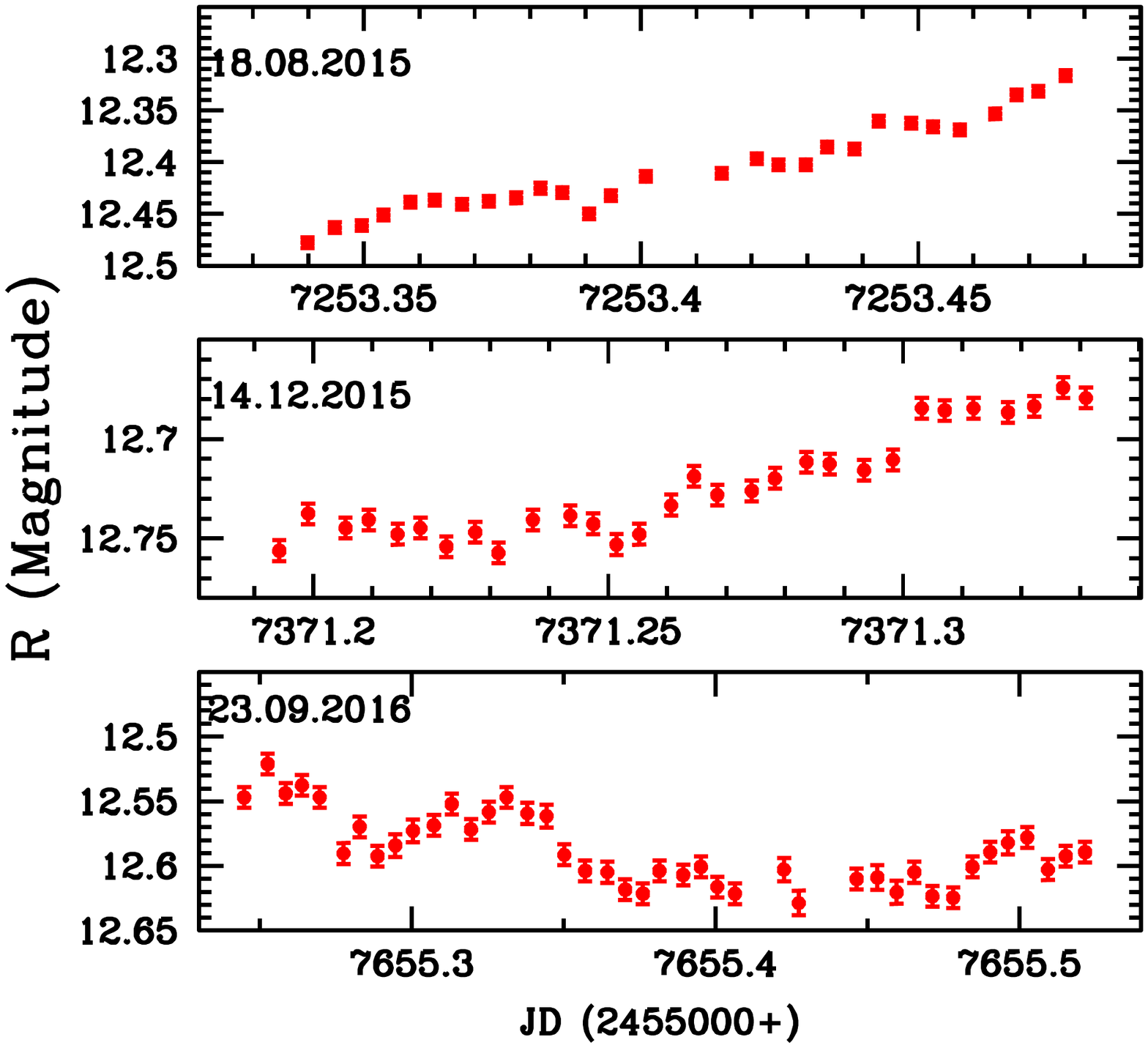}
\caption{Typical examples of LCs of BL Lacertae where $\Delta$m $>$ 0.10 in a few hours.}
\label{fig3}
 \end{figure}

It can be seen from the figures that BL Lac was variable on intra-day timescales and often showed some variability on hourly time-scales.
We used the enhanced $F$-test to search for genuine flux variations on these intra-day variability (IDV) time-scales, finding 19
LCs to be variable across the filter set over the course of 13 nights of observation.  Of these, 11 were seen in the R band,
for which we have measurements on every night we observed and which also have the smallest errors.
 These~enhanced $F$-test values are presented in Table \ref{table2}. We note that only rarely did more than one band show strong enough variability
to satisfy the stringent criterion during the same nights, mostly because of different error levels in the different bands.
In particular, the B band errors are usually significantly greater than those at other bands and essentially preclude our verifying any but the most extreme variations. The nightly LCs displayed rises and decays up to a magnitude change of  $\sim$0.18 over several hours.
A few such examples of substantial variations are shown with expanded time axes
in Figure~\ref{fig3}. In most of the observations, flickering/small subflares are superimposed on multi-hour trends that are presumably parts of larger flares.

The duty cycle (DC, in percent)  of BL Lac was found using the definition of~\cite{Romero1999} that
subsequently  was frequently employed (e.g., \cite{Stalin2004,Gaur2015a} and references therein).

Given the multi-band data we collected, we can compute the DC in different ways.
If all detections of variability are considered, we have 19 clearly variable LCs out of 143 in total, producing a
DC of around 12\%.  If we exclude the noisy B band, we have 17 variable LCs out of 102.  Perhaps the fairest approach is to only consider
the most accurate R band data, and most previous studies have just looked at one band (usually R): in this case, we have firm detections on
11 out of 43 nights, or a DC of roughly 25\%.

In our previous extensive search for IDV in BL Lac over 38 nights in the period 2010--2012, we~found 50 (out of 119) filter LCs to be variable
during 19 nights using the same enhanced $F$-test ~\cite{Gaur2015a}, or a DC = 44\%, although the B band data were usually better then.
If we restrict our computation to only R~band data, we find DC = 49\% in this earlier period.  We note that BL Lac
had a lower average brightness during 2010--2012, with R-band magnitudes ranging between 13.9 and 12.4 and it was usually fainter than $m = 13$.
These differences in the duty cycle between these two periods could be attributed to a reduction in the turbulence behind a standing shock in the jet,
which is a leading explanation for rapid, intra-day variability in blazars (e.g., \cite{Marscher2014,Calafut2015,Pollack2016}).
 This turbulence may decrease as the bulk jet speed rises and so the overall observed source flux increases. Hence, it is reasonable to
expect to see a comparatively lower DC when the source becomes brighter.

The percentage variability amplitudes are calculated using the standard method of ~\cite{Heidt1996}.
It has been found in some previous studies of BL Lacs that the variability amplitude is larger at higher frequencies,  specifically for
B (and occasionally V), on nights when variability is certainly detected in the R and I bands
~\cite{Ghisellini1997,Massaro1998,Bonning2012}; however, in other studies, the strength of the variability was  found to not be systematically larger at higher
frequencies (~\cite{Ghosh2000,Ramirez2004}). In our current observations, the B band errors are significantly higher as compared
to lower frequency bands, so that, on many occasions, our~significance thresholds in
B (also sometimes V) exceed the apparent fluctuations, even when R band variability is clearly seen and is statistically significant.
The amplitude of variability in nights with clear detections varied between 7--20\% in the R band; however,
the highest fractional amplitude of variability was seen for the night of 26 September 2016, where an amplitude of 41\% was measured for the B band.

\vspace{-12pt}

\begin{table}[H]
\centering
\caption {Results of intra-day variability.}
\label{table2}
\noindent
\scalebox{.78}{
\setlength{\tabcolsep}{0.010in}
\begin{tabular}{cccccccccccccc} \toprule
{\bf Date of}  & {\bf Band} & {\bf Number}  & {\bf $F$} &{\bf $F_{c}$(0.001)} & {\bf Variability} & {\bf Amplitude} & {\bf Date of} & {\bf Band} & {\bf Number}  &{\bf $F$}  &{\bf $F_{c}$(0.001)}  & {\bf Variability} & {\bf Amplitude}  \\
{\bf Observation}   &     &       &     &                &    &     & {\bf Observation}   &     &       &     &     &    &  \\
{\bf dd.mm.yyyy}    &     &       &     &                &    &     & {\bf dd.mm.yyyy}    &     &       &     &     &    & \\\midrule
29.07.2014       &B  &10  &4.66     &5.56   &NV   &-  &         15.09.2015       &B  &28  &0.213    &2.686  &NV   &-   \\
                 &V  &10  &3.54     &5.56   &NV   &-  &                          &R  &28  &0.163    &2.686  &NV   &-   \\
                 &R  &10  &2.81     &5.56   &NV   &-  &                          &I  &30  &0.121    &2.596  &NV   &-   \\
                 &I  &10  &1.75     &5.56   &NV   &-  &         03.11.2015       &B  &21  &0.978    &3.145  &NV   &-  \\
\midrule28.09.2014       &B  &13  &0.858    &4.393  &NV   &-  &                          &R  &21  &0.734    &3.145  &NV   &-  \\
                 &V  &17  &1.115    &3.598  &NV   &-  &                          &I  &22  &0.508    &3.06   &NV   &-  \\
                 &R  &19  &1.769    &3.345  &NV   &-  &         04.11.2015       &B  &26  &0.568    &2.79   &NV   &-   \\
                 &I  &23  &1.084    &2.983  &NV   &-  &                          &R  &28  &2.599    &2.686  &NV   &-   \\
\midrule29.09.2014       &B  &12  &0.345    &4.697  &NV   &-  &                          &I  &28  &0.749    &2.686  &NV   &-   \\
                 &V  &16  &0.277    &3.753  &NV   &-  &         05.11.2015       &B  &30  &0.568    &2.596  &NV   &-   \\
                 &R  &17  &0.346    &3.598  &NV   &-  &                          &R  &30  &1.559    &2.596  &NV   &-   \\
                 &I  &19  &0.232    &3.345  &NV   &-  &                          &I  &30  &0.833    &2.596  &NV   &-   \\
\midrule30.09.2014       &B  &19  &0.701    &3.345  &NV   &-- &         06.11.2015       &B  &31  &0.88     &2.555  &NV   &-   \\
                 &V  &20  &0.492    &3.239  &NV   &-  &                          &R  &33  &6.93     &2.481  &Var   &10.57   \\
                 &R  &22  &0.246    &3.06   &NV   &-  &                          &I  &32  &3.411    &2.517  &Var &9.33   \\
                 &I  &23  &0.355    &2.983  &NV   &-  &         07.11.2015       &B  &28  &0.245    &2.686  &NV   &-   \\
\midrule29.06.2015       &B  &10  &2.10     &5.56   &NV   &-  &                          &R  &32  &0.39     &2.517  &NV   &-   \\
                 &V  &10  &2.00     &5.56   &NV   &-  &                          &I  &32  &0.096    &2.517  &NV   &-   \\
                 &R  &10  &3.19     &5.56   &NV   &-  &         11.11.2015       &B  &46  &0.764    &2.157  &NV   &-  \\
                 &I  &10  &2.32     &5.56   &NV   &-  &                          &R  &46  &6.571    &2.157  &Var &11.30  \\
\midrule30.06.2015       &B  &10  &2.08     &5.56   &NV   &-  &                          &I  &45  &8.000    &2.176  &Var &12.39  \\
                 &V  &10  &2.14     &5.56   &NV   &-  &         12.11.2015       &B  &43  &1.212    &2.215  &NV   &-  \\
                 &R  &10  &2.76     &5.56   &NV   &-  &                          &R  &42  &0.369    &2.236  &NV   &-  \\
                 &I  &10  &1.15     &5.56   &NV   &-  &                          &I  &41  &0.879    &2.258  &NV   &-  \\
\midrule18.07.2015       &B  &33  &0.327    &2.481  &NV   &-  & 03.12.2015       &B  &15  &1.273    &3.932  &NV   &-   \\
                 &R  &33  &4.549    &2.481  &Var &7.14&                          &R  &16  &1.111    &3.753  &NV   &-   \\
                 &I  &32  &2.252    &2.517  &NV   &-  &                  &I  &18  &0.558    &3.463  &NV   &-   \\
\midrule19.07.2015       &B  &26  &0.812    &2.79   &NV   &-  & 14.12.2015       &B  &29  &0.512    &2.639  &NV   &-  \\
                 &R  &27  &3.973    &2.736  &Var &7.57&                          &R  &30  &7.858    &2.596  &Var &8.31  \\
                 &I  &27  &0.881    &2.736  &NV   &-  &                  &I  &30  &2.843    &2.596  &Var &13.09  \\
\midrule20.07.2015       &B  &22  &0.200    &3.06   &NV   &-  & 30.06.2016       &V  &20  &1.384    &3.239  &NV   &-  \\
                 &R  &24  &1.221    &2.913  &NV   &-  &                  &R  &20  &1.081    &3.239  &NV   &-  \\
                 &I  &24  &0.742    &2.913  &NV   &-  &                  &I  &20  &0.09     &3.239  &NV   &-  \\
\midrule21.07.2015       &B  &20  &1.086    &3.239  &NV   &-  & 01.07.2016       &V  &14  &0.464    &4.142  &NV   &-  \\
                 &R  &20  &6.851    &3.239  &Var &10.67&                         &R  &14  &0.708    &4.142  &NV   &-  \\
                 &I  &20  &2.226    &3.239  &NV   &-  &                  &I  &13  &0.839    &4.393  &NV   &-  \\
\midrule22.07.2015       &B  &23  &0.373    &2.983  &NV   &-  & 08.08.2016       &B  &15  &0.607    &3.932  &NV   &-  \\
                 &R  &24  &0.669    &2.913  &NV   &-  &                  &V  &15  &1.57     &3.932  &NV   &-  \\
                 &I  &24  &0.320    &2.913  &NV   &-  &                  &R  &30  &0.467    &2.596  &NV   &-  \\
\midrule23.07.2015       &B  &18  &0.460    &3.463  &NV   &-  & 09.08.2016       &B  &25  &2.594    &2.849  &NV   &-  \\
                 &R  &22  &1.337    &3.06   &NV   &-  &                  &V  &25  &4.572    &2.849  &Var &22.54  \\
                 &I  &20  &0.853    &3.239  &NV   &-  &                  &R  &51  &1.915    &2.076  &NV   &-  \\
\midrule18.08.2015       &B  &28  &1.355    &2.686  &NV   &-  & 25.08.2016       &B  &13  &0.930    &4.393  &NV   &-  \\
                 &R  &28  &13.825   &2.686  &Var &16.15&                         &V  &14  &0.791    &4.142  &NV   &-  \\
                 &I  &28  &1.924    &2.686  &NV   &-  &                  &R  &28  &1.290    &2.686  &NV   &-  \\
\midrule26.08.2015       &B  &17  &0.488    &3.598  &NV   &-  &         26.08.2016       &B  &25  &0.403    &2.849  &NV   &-  \\
                 &R  &20  &6.659    &3.239  &Var &7.67&                          &V  &26  &0.526    &2.79   &NV   &  \\
                 &I  &17  &1.125    &3.598  &NV   &-  &                          &R  &59  &0.942    &1.972  &NV   &-  \\
\midrule27.08.2015       &B  &19  &0.335    &3.345  &NV   &-  &         28.08.2016       &B  &14  &0.698    &4.142  &NV   &-  \\
                 &R  &21  &0.669    &3.145  &NV   &-  &                          &R  &12  &1.009    &4.697  &NV   &-  \\
                 &I  &20  &0.336    &3.239  &NV   &-- &                          &I  &13  &1.632    &4.393  &NV   &-  \\
\midrule28.08.2015       &B  &22  &0.633    &3.06   &NV   &-  &         30.08.2016       &B  &45  &2.218    &2.176  &Var &21.41  \\
                 &R  &23  &3.093    &2.983  &Var &10.16 &                        &V  &54  &2.498    &2.033  &Var &12.73  \\
                 &I  &21  &1.797    &3.145  &NV   &-    &                 &R  &53  &1.408    &2.047  &NV   &-  \\
\midrule08.09.2015       &B  &45  &0.928    &2.176  &NV   &-    &                 &I  &53  &1.775    &2.047  &NV   &-  \\
                 &R  &45  &1.249    &2.176  &NV   &-    &23.09.2016       &B  &44  &1.023    &2.195  &NV   &-   \\
                 &I  &45  &1.233    &2.176  &NV   &-    &                 &V  &44  &1.998    &2.195  &NV   &-   \\
\midrule14.09.2015       &B  &23  &0.422    &2.983  &NV   &-    &                 &R  &42  &3.48     &2.236  &Var &10.75   \\
                 &R  &27  &0.147    &2.736  &NV   &-    &                 &I  &23  &2.834    &2.983  &NV   &   \\
                 &I  &29  &0.16     &2.639  &NV   &-   &                  &   &    &         &       &     &  \\\bottomrule
\end{tabular} }
\end{table}

\begin{table}[H]\ContinuedFloat
\centering
\caption {{\em Cont.}}
\noindent
\begin{tabular}{ccccccc} \toprule
{\bf Date of}  & {\bf Band} & {\bf Number}  & {\bf $F$} &{\bf $F_{c}$(0.001)} & {\bf Variability} & {\bf Amplitude} \\
{\bf Observation}   &     &       &     &                &    &  \\
{\bf dd.mm.yyyy}    &     &       &     &                &    &  \\\midrule

24.09.2016       &B  &31  &0.642    &2.555  &NV   &-   \\
                 &V  &30  &1.433    &2.596  &NV   &-   \\
                 &R  &30  &2.054    &2.596  &NV   &-   \\
                 &I  &22  &0.617    &3.06   &NV   &-   \\
\midrule26.09.2016       &B  &45  &3.496    &2.176  &Var &40.87   \\
                 &V  &45  &6.274    &2.176  &Var &29.98    \\
                 &R  &43  &3.87     &2.215  &Var &19.82   \\
                 &I  &42  &1.558    &2.236  &NV  &-   \\
\midrule27.09.2016       &B  &42  &2.113    &2.236  &NV   &-   \\
                 &V  &42  &0.864    &2.236  &NV   &-   \\
                 &R  &41  &0.779    &2.258  &NV   &-   \\
                 &I  &37  &0.362    &2.358  &NV   &-   \\
\midrule28.09.2016       &B  &50  &1.284    &2.091  &NV   &-   \\
                 &V  &46  &1.002    &2.157  &NV   &-   \\
                 &R  &47  &1.387    &2.14   &NV   &-   \\
                 &I  &49  &0.567    &2.106  &NV   &-   \\
\midrule29.09.2016       &B  &49  &1.088    &2.106  &NV   &-   \\
                 &V  &49  &1.959    &2.106  &NV   &-   \\
                 &R  &50  &1.115    &2.091  &NV   &-   \\
                 &I  &42  &1.125    &2.236  &NV   &-   \\
\midrule30.09.2016       &B  &38  &1.898    &2.331  &NV   &--   \\
                 &V  &35  &0.330    &2.416  &NV   &-   \\
                 &R  &34  &0.429    &2.448  &NV   &-  \\
                 &I  &36  &0.853    &2.386  &NV   &-  \\\bottomrule
\end{tabular} \\
\begin{tabular}{@{}c@{}}
\multicolumn{1}{p{\textwidth -.88in}}{\footnotesize $F$ = Enhanced \emph{F}-test values;
$F_{c}$(0.001) =  Critical Values of \emph{F} Distribution at 0.1\%;~Amplitude: Variability~Amplitude; Var/NV: Variable/Non-variable.}
\end{tabular}
\end{table}
\unskip

\begin{figure}[H]
\centering
\includegraphics[width=0.5\textwidth,trim=0 0 0 5cm,clip]{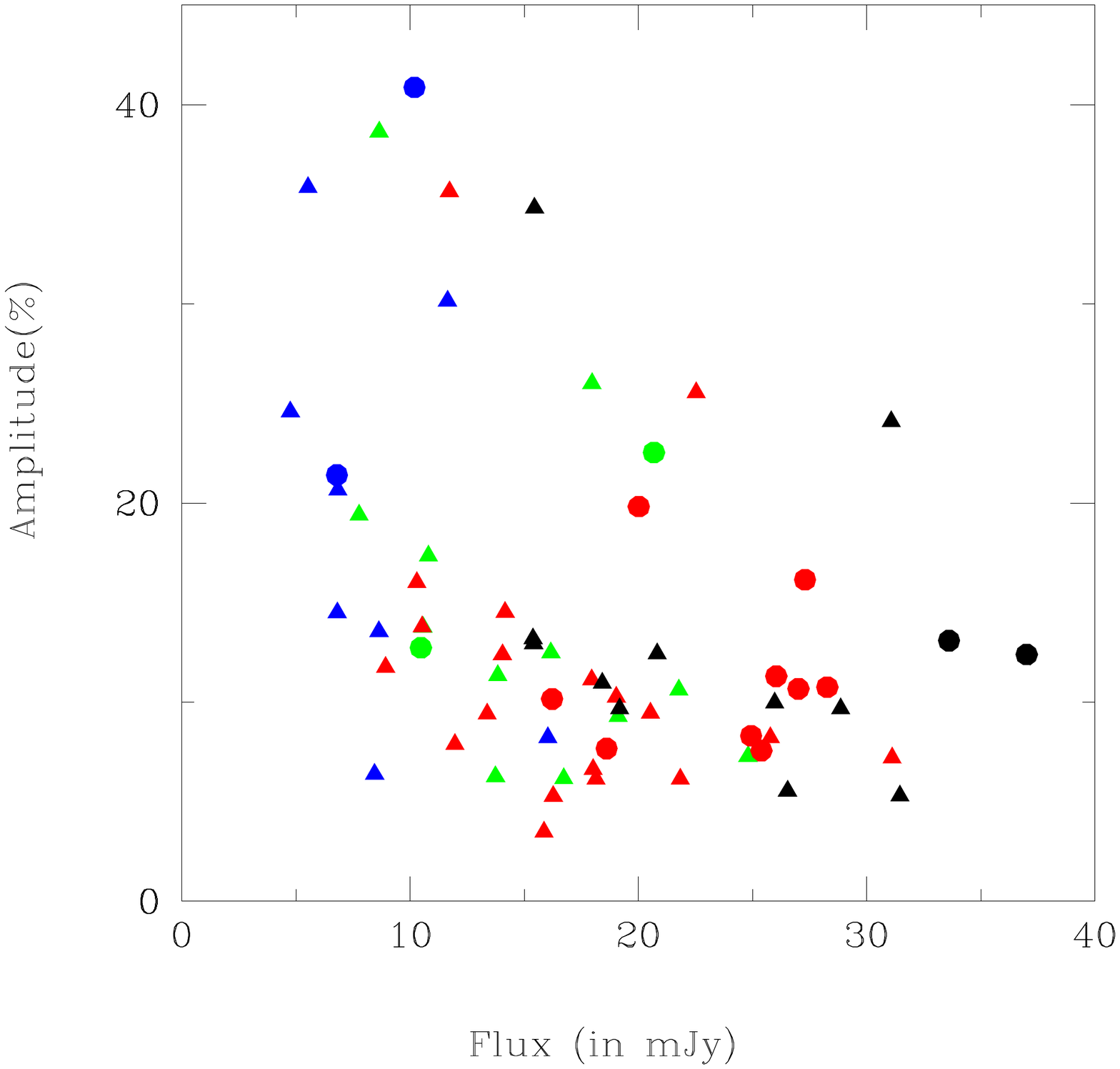}
\vspace*{-0.85in}
\caption{Amplitude of variability versus flux. Again, blue, green, red and black colors represent B, V, R and I bands, respectively, triangles give data from~\cite{Gaur2015a} and circles provide these new data. The~Spearman correlation
coefficient is $-0.42$ with $p > 0.9996$.}
\label{fig4}
 \end{figure}

The correlation between the variability strength and source flux is shown in Figure~\ref{fig4}; we see that the
 variability amplitude drops  as the source flux rises ($\rho=-0.42$ with $p=0.9996$ where $\rho$ is the Spearman correlation
coefficient). Here, we included our data from~\cite{Gaur2015a} for the 2010--2012 period. As~for our above discussion of the DC,
this trend can be explained if, as the source's flux  increases, the~irregularities in the jet flow, i.e., turbulence near shocks/knots, decrease.
As this turbulence is quite likely to be responsible for the intra-night optical
variability (~\cite{Marscher2014,Calafut2015,Pollack2016}), an overall increased flux rise would lead to a reduction in the fractional
amplitude (e.g., \cite{Gupta2008}). Part of the scatter in the plot could arise because observed
 amplitudes of variability seen for similar luminosities normally go up for longer observations, and there is a significant spread in the
observation times in our data.

The inter-band time lags between the B, V, R and I bands for all of the LCs presented here were computed using
Discrete Correlation Functions (DCF; details are given in~\cite{Gaur2015a}).~
 All the lags are consistent with zero, as expected because the optical bands are very close to each other; if lags are present,
 they must be shorter than the resolution in our LCs, which is $\sim$8 min.

\subsection{Color Indices}

We searched for spectral variations in BL Lac by looking at  how color varied with respect to the source brightness.
We calculated color indices (CIs)  by combining nearly simultaneous (within~8~min)
V, R and I measurements to produce  CIs: V$-$R and R$-$I.

It should be noted that the light from the host galaxy or the accretion disk in the AGN can also affect the color variations of
BL Lac~\cite{Hawkins2004}. Hence, in calculating the color indices, we have used data that is corrected for the host galaxy contribution.
In addition, since the source is in a bright state during our observations, the accretion disk component can be ignored, as the Doppler
boosted flux from the relativistic jet would dominate the overall emission.

\begin{figure}[H]
\centering
\includegraphics[width=0.5\textwidth,trim=0 6cm 0 4cm,clip]{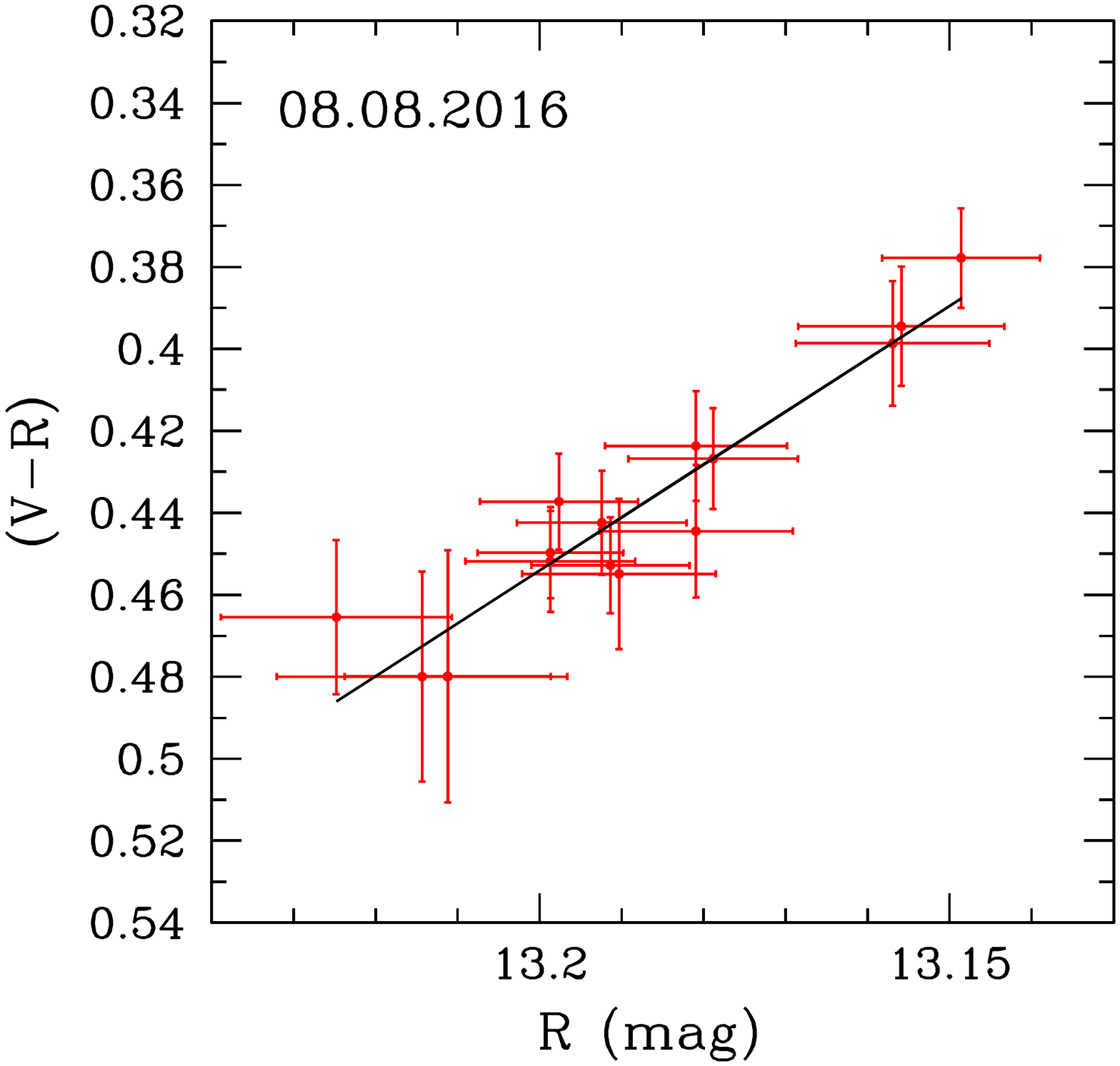}
\vspace{-6pt}
\caption{Sample color-magnitude diagram for BL Lacertae.}
\label{fig5}
 \end{figure}

We found significant
correlations between CI and magnitude in 24 of our observations. One~example is shown in Figure~\ref{fig5}.
To quantify this, we follow~\cite{Gaur2015a} and fit the color-R-magnitude diagrams with a
linear model of the form CI = $m \times $ mag $+$~c.  Pearson correlation coefficients ($r$) and
their $p$-values (here, the null hypothesis probability, where we take a confirmed CI correlation with the R magnitude when $p < 0.05$)
as well as the values of slopes, $m$, and intercepts, $c$ are given in Table \ref{table4}.

In these plots of 24 of our observations, the CIs positively correlate with the source brightness,
which shows hardening of the spectrum as BL Lac gets brighter (or a bluer-when-brighter behavior).
In the remaining 19 observations where we had collected sufficient data
to search for CI-brightness trends, we did not find any significant
correlations. These could be explained by the lighthouse effect in the shock-in-jet model (e.g., \cite{Gopal1992}),
where relativistic electrons are moving towards an observer on helical trajectories in the jets and
the flares will be produced by the sweeping beam whose direction varies with time but no differences in variations can be seen over such
limited ranges in frequency.

 \unskip
 \begin{table}[H]
\centering
\caption{Results of linear fits to color index-flux diagrams.}
\label{table4}
\setlength{\tabcolsep}{0.020in}
\begin{tabular}{cccccc} \toprule
{\bf Date of}    & {\bf Band}      & {\bf $r$}   & {\bf $p$}    &{\bf $m$}  & {\bf $c$}     \\
{\bf Observation}      &                 &       &                    & {\bf Slope}   & {\bf Intercept} \\
{\bf dd.mm.yyyy}       &                 &       &                    &               &                 \\\midrule
19.07.2015   &(R$-$I)                &0.65 &0.0002                &0.42     &$-4.76$ \\
20.07.2015   &(R$-$I)                &0.51 &0.011                 &0.39     &$-4.31$  \\
21.07.2015   &(R$-$I)                &0.53 &0.015                 &0.22     &$-2.18$    \\
22.07.2015   &(R$-$I)                &0.41 &0.047                 &0.41     &$-4.54$  \\
18.08.2015   &(R$-$I)                &0.62 &0.0005                &0.39     &$-4.23$  \\
27.08.2015   &(R$-$I)                &0.88 &3.74$\times10^{-7}$   &1.13     &$-14.10$  \\
08.09.2015   &(R$-$I)                &0.42 &0.004                 &0.29     &$-3.10$    \\
14.09.2015   &(R$-$I)                &0.73 &3.51$\times10^{-5}$   &0.87     &$-10.52$  \\
15.09.2015   &(R$-$I)                &0.66 &0.00012               &0.42     &$-4.60$  \\
05.11.2015   &(R$-$I)                &0.41 &0.026                 &0.47     &$-5.26$      \\
06.11.2015   &(R$-$I)                &0.46 &0.009                 &0.15     &$-1.23$     \\
07.11.2015   &(R$-$I)                &0.60 &0.0002                &0.80     &$-9.38$    \\
11.11.2015   &(R$-$I)                &0.33 &0.028                 &0.14     &$-1.12$    \\
12.11.2015   &(R$-$I)                &0.35 &0.023                 &0.76     &$-8.92$   \\
03.12.2015   &(R$-$I)                &0.72 &0.004                 &0.73     &$-8.60$     \\
30.06.2016   &(R$-$I)                &0.45 &0.045                 &0.66     &$-7.72$  \\
08.08.2016   &(V$-$R)                &0.94 &1.83$\times10^{-7}$   &1.29     &$-16.58$  \\
26.08.2016   &(V$-$R)                &0.65 &0.0004                &0.54     &$-6.94$   \\
30.08.2016   &(R$-$I)                &0.33 &0.017                 &0.21     &$-2.26$   \\
22.09.2016   &(R$-$I)                &0.68 &0.016                 &0.46     &$-5.21$   \\
24.09.2016   &(R$-$I)                &0.66 &0.0009                &0.44     &$-5.02$   \\
27.09.2016   &(R$-$I)                &0.55 &0.0004                &0.47     &$-5.51$    \\
28.09.2016   &(R$-$I)                &0.72 &8.30$\times10^{-9}$   &0.50     &$-5.79$    \\
30.09.2016   &(R$-$I)                &0.65 &3.03$\times10^{-5}$   &0.51     &$-5.96$  \\  \bottomrule
\end{tabular} \\
\begin{tabular}{ccc}
\multicolumn{1}{c}{\footnotesize$r$ and $p$: Pearson Correlation Coefficient and its null hypothesis probability values, respectively.}
\end{tabular}
\end{table}

 BL Lacertae does sometimes exhibit bluer-when-brighter trends with a range of regression slopes but no redder-when-brighter behavior.
This bluer-when-brighter dominance for BL Lac has been reported earlier on both long-term and short-term time-scales
(e.g., \cite{ Villata2004,Stalin2006,Gu2006,Larionov2010,Gaur2012,Wierzcholska2015}).
Villata 2004~\cite{ Villata2004} characterized  intra-day flares to be strongly bluer-when-brighter chromatic events with a
slope of $\sim$0.4.  They noted that these could be  modeled if there was  one variable component with a flatter slope that was dominant during
the flaring states as well as a steeper sloped component that was more stable and produces the bulk of the longer term, basically achromatic, emission.
  In our current study, we~found bluer-when-brighter changes in CI with slopes that were in the range 0.14--1.13 (Table~\ref{table4}).
Hence, we~find that the intra-night flares seen during 2014--2016 are usually more chromatic as compared to our
earlier studies for the period 2010--2012 during  where significant positive slopes  only ranged between 0.14--0.29 ~\cite{Gaur2015a}.

In some of the light curves, significant color variations are seen, but the slopes are relatively flat, such as for the
 observations performed on 11 November 2015, 18 August 2015, 21 July 2015, and~\mbox{30 August 2016}. This could be explained by  differences beween flux states
during these particular observations. It can be seen from the LCs of these observations that they exhibit more small sub-flares
on top of the long-term rising or decaying trends. Hence, the superposition of multiple
components can produce an overall weakening of the color-magnitude relations.
Similar findings were made by~\cite{Bonning2012}, where they found that their sample of BL Lacs showed complicated color-magnitude diagram behavior
including hysteresis tracks, as well as achromatic flares; these indicate that  different jet components dominate
at different times.  Dai~\cite{Dai2013} found intra-day spectral
hysteresis loop patterns for the blazar S5 $0716+714$ in its color-magnitude diagram.
These loop-like patterns are attributed to the differences in the amplitudes and cadences at different wavelengths~\cite{Hu2006}.
 We did not find any hints of such loop patterns during our observations, given the absence of any detectable time lags between~bands.

\section{Discussion}\label{sec5}

We have presented  our photometric monitoring of BL Lacertae during three seasons in  2014--2016 for a total of 43 nights
in the B, V, R and I bands in order to further study its flux and spectral variability. We found genuine IDV in a total of 14
nights of observation using the power enhanced $F$-test at a stringent level. The LCs frequently exhibit gradual rises and decays,
 sometimes with superimposed sub-flares. We found no evidence for periodicity or other characteristic time scales.
Various models that can explain optical IDV of blazars involve: changes in the electron energy density distribution
of the relativistic particles that produce variable synchrotron emission; shocks accelerating  particles in the bulk relativistic
plasma jet; turbulence behind an outgoing shock in a jet; irregularities in the jet flows, such as  ``mini-jets''  (e.g., \cite{Marscher1992,Giannios2009,Marscher2014,Gaur2014,Calafut2015,Pollack2016}).
 As we are focused on intra-night flux variations that involve color variations, these are very likely to
be associated with models centered on shocks moving through  a turbulent plasma jet.

 We found the duty cycle of the source between 2014 and 2016, based on the light curves of all the bands to be $\sim$12\%,
which is low compared to the DC (44\%) found in our previous campaign during 2010--2012 over 38 nights of observations.  If we
restrict consideration to the R band, only the current DC is $\sim$25\% while that for the earlier observations was $\sim$49\%.
BL Lacertae is an LBL and previous studies have indicated that
LBLs typically display stronger IDV than do HBLs (high frequency peaked blazars).
The DCs have been claimed to be as high as  $\sim$70\% for LBLs and $\sim$30--50\% for HBLs (~\cite{Heidt1998,Romero2002})
However, Ref. \cite{Gopal2011} found that the overall DC was 22\% for HBLs
and 50\% for LBLs in a large sample of blazars. Recently, ~\cite{Gupta2016} considered a group of LBLs in X-ray bands and found
that the since the peaks of the synchrotron component for such sources lie in the near IR/optical, more energetic electrons are
available for the synchrotron emission there and hence they display more DC in these bands. Hence, a~DC of $\sim$44\% for BL Lacertae
is in agreement with  previous studies, but the current $\sim$12\% value is low. This could well be related to the overall high state
of BL Lac during our recent observations. However, it has been found that, during this period, BL Lacertae has not
shown any flaring activity at radio frequencies (~\cite{Tsujimoto2017,Kim2017}).

The variability amplitude is typically larger at higher frequencies. This is generally
explained in terms of higher energy electrons only being able to emit  closer to the shock front as those more quickly lose energy while
producing higher frequency synchrotron radiation~\cite{Marscher1985}. Due to the closeness of the B, V, R and I bands,
the time when a flare is initiated and their lengths are nearly the same for all optical bands.
The errors in our B band measurements for this campaign are higher than those in  lower frequency bands, hence on many occasions the
amplitude of variability is not seen at sufficient statistical strength, and generally not seen as
stronger,  at higher frequencies. We found significant positive correlations between the observed variability amplitude
 with respect to the length of the nightly observations. It can be seen from Figure \ref{fig1} and \ref{fig2} that the duration of the
monitoring of our observations varies between 1.5--7 h.

We found that the probability of seeing a significant intra-day variability
usually increases if the source is continuously observed for more than 4 h, but most of our measurements were shorter than 4~h.
 We also searched for  correlations between the variability amplitude with respect to flux and saw that the amplitude drops as the
source flux rises. This trend could be explained if, as the source flux increases, the fluctuations in the turbulent jet decrease~\cite{Marscher2014} and thus this more uniform jet flow  leads to a decline in the variability amplitude.

We looked for  correlations between  magnitude and color-index, finding  significant ($>$95\%) positive correlations
 in 24 observations out of the total of 43. BL Lac exhibited bluer-when-brighter trends with~slopes that varied between 0.14--1.13 as shown in Table \ref{table4}.~We never saw redder-when-brighter behavior.

The structure of  color-magnitude diagrams gives information on the amplitude
differences and time-lags between the observed bands. As the optical bands have close proximity, it is often hard to measure
time lags between them, as was the case for these observations.  The simulations of~\cite{Dai2011} indicated that both the
amplitude differences and time delays between temporal changes at different wavelengths typically result in spectral hardening
while a flare rises. They showed that the amplitude difference between two light curves
resulted in a diagonal path in the color-magnitude diagram, whereas a
time-lag combined with amplitude variations resulted in a counter-clockwise loop pattern on a color index-magnitude plot.
As we were not able to determine any significant lags between these bands, we can not address this possibility and can only  conclude that
any lags, if present, are probably shorter than the sampling time of our observations.

\section{Conclusions}\label{sec6}

Our conclusions can be summarized as follows:

\begin{itemize}
\item We observed BL Lacertae quasi-simultaneously in B, V, R and I bands during the period 2014--2016 in 43 nights to search
for intra-day variability and found strong variability in at least one band on a total of 14  nights.
\item BL Lac was moderately variable with changes well correlated across the bands.
The variations usually were smooth, with gradual rises/decays. Magnitude changes of over $\sim$0.10 over a few hours were seen
on several nights.
\item The amplitude of variability is positively correlated  with the duration of the observation, but it
 decreases as the  source becomes brighter.
\item We did not find any significant time delays between the B, V, R and I bands in these observations. This indicates that the variations are nearly simultaneous across the optical bands and any possible time lags are shorter
than our data sampling interval of $\sim$8 min.
\item The flux changes are accompanied by spectral variations on intra-day time-scales. In 24 of the 43 observations, the
optical spectrum showed the overall bluer-when-brighter trend, probably associated with very variable jet emission.
\item The color vs.\ magnitude diagrams are highly chromatic on the intra-night timescales and are likely associated with the rapid
 injection or acceleration of relativistic particles, subsequently followed by  rapid  synchrotron cooling.

\end{itemize}

\vspace{6pt}


\acknowledgments{We thank the referees for constructive comments that have improved the presentation of~our work.
H.G. is sponsored by the Chinese Academy of Sciences (CAS) Visiting Fellowship for Researchers from Developing
Countries; CAS Presidents International Fellowship Initiative (PIFI) (Grant No. 2014FFJB0005); supported by the National Natural
Science Foundation of China (NSFC) Research Fund for International Young Scientists (Grant Nos. 11450110398, 11650110434) and
supported by a Special Financial Grant from the China Postdoctoral Science Foundation (Grant No. 2016T90393).
This research was partially supported by the Bulgarian National Science Fund of the Ministry of Education and Science under
grants DN 08-1/2016 and DM08-2/2016 and by funds of the project RD-08-102 of Shumen University. The authors thank the Director
of Skinakas Observatory Ioannis Papamastorakis and Iossif Papadakis for the award of telescope time.
The Skinakas Observatory is a collaborative project of the University of Crete, the Foundation for Research
and Technology-Hellas, and the Max-Planck-Institut f\"ur Extraterrestrische Physik. M.F.G. acknowledges
support from the National Science Foundation of China (Grant 11473054 and U1531245) and by the Science
and Technology Commission of Shanghai Municipality (14ZR1447100)}. 

\authorcontributions{H.G. and A.C.G. planned the project. H.G. did major work e.g., further analysis of~the~intra-day
light curves, made the plots, and wrote the manuscript. A.C.G., P.J.W. and M.F.G. contributed to the writing and interpretation
of the results of the manuscript. R.B., A.S., E.S., and S.I. performed observations and~photometric analysis of the data.}

\conflictsofinterest{The authors declare no conflict of interest.}

\reftitle{References}

\end{document}